\documentclass[apj]{emulateapj}
\usepackage{apjfonts,amsmath,natbib}
\makeatletter

\newcommand{\Rmnum}[1]{\expandafter\@slowromancap\romannumeral #1@}
\makeatother

\newcommand{\gps}{\ensuremath{g_{\rm P1}}}
\newcommand{\rps}{\ensuremath{r_{\rm P1}}}
\newcommand{\ips}{\ensuremath{i_{\rm P1}}}
\newcommand{\zps}{\ensuremath{z_{\rm P1}}}
\newcommand{\yps}{\ensuremath{y_{\rm P1}}}

\newcommand{\grizy}{\gps\rps\ips\zps\yps}
\newcommand{\griz}{\gps\rps\ips\zps}

\slugcomment{Accepted by the Astrophysical Journal}

\shorttitle{TDSS}
\shortauthors{Morganson et al.}

\begin{document}


\title{The Time Domain Spectroscopic Survey: Variable Object Selection and Anticipated Results}


\author{Eric Morganson\altaffilmark{1}}
\author{Paul J. Green\altaffilmark{1}}
\author{Scott F. Anderson\altaffilmark{2}}
\author{John J. Ruan\altaffilmark{2}}
\author{Adam D. Myers\altaffilmark{3}}
\author{Michael Eracleous\altaffilmark{4}}
\author{Brandon Kelly\altaffilmark{5}}
\author{Carlos Badenes\altaffilmark{6}}
\author{Eduardo Ba\~{n}ados\altaffilmark{7}}
\author{Michael R. Blanton\altaffilmark{8}}
\author{Matthew A. Bershady\altaffilmark{9}}
\author{Jura Borissova\altaffilmark{10}}
\author{William Nielsen Brandt\altaffilmark{4}}
\author{William S. Burgett\altaffilmark{11}}
\author{Kenneth Chambers\altaffilmark{12}}
\author{Peter W. Draper\altaffilmark{13}}
\author{James R.~A. Davenport\altaffilmark{2}}
\author{Heather Flewelling\altaffilmark{12}}
\author{Peter Garnavich\altaffilmark{14}}
\author{Suzanne L. Hawley\altaffilmark{2}}
\author{Klaus W. Hodapp\altaffilmark{12}}
\author{Jedidah C. Isler\altaffilmark{15, 1}}
\author{Nick Kaiser\altaffilmark{11}}
\author{Karen Kinemuchi\altaffilmark{16}}
\author{Rolf P. Kudritzki\altaffilmark{12}}
\author{Nigel Metcalfe\altaffilmark{12}}
\author{Jeffrey S. Morgan\altaffilmark{11}}
\author{Isabelle P\^aris\altaffilmark{17}}
\author{Mahmoud Parvizi\altaffilmark{18}}
\author{Rados\l{}aw Poleski\altaffilmark{19}}
\author{Paul A. Price\altaffilmark{20}}
\author{Mara Salvato\altaffilmark{21}}
\author{Tom Shanks\altaffilmark{13}}
\author{Eddie F. Schlafly\altaffilmark{7}}
\author{Donald P. Schneider\altaffilmark{4}}
\author{Yue Shen\altaffilmark{22, 23}}
\author{Keivan Stassun\altaffilmark{18}}
\author{John T. Tonry\altaffilmark{12}}
\author{Fabian Walter\altaffilmark{7}}
\author{Chris Z. Waters\altaffilmark{12}}

\altaffiltext{1}{Harvard Smithsonian Center for Astrophysics, 60 Garden St, Cambridge, MA 02138, USA}
\altaffiltext{2}{Department of Astronomy, University of Washington, Box 351580, Seattle, WA 98195, USA}
\altaffiltext{3}{Department of Physics and Astronomy, University of Wyoming, Laramie, WY 82071, USA}
\altaffiltext{4}{Department of Astronomy \& Astrophysics, 525 Davey Laboratory, The Pennsylvania State University, University Park, PA 16802, USA}
\altaffiltext{5}{Department of Physics, Broida Hall, University of California, Santa Barbara, CA 93106-9530, USA}
\altaffiltext{6}{Department of Physics and Astronomy and Pittsburgh Particle Physics, Astrophysics and Cosmology Center (PITT PACC), University of Pittsburgh, 3941 O'Hara St, Pittsburgh, PA 15260, USA}
\altaffiltext{7}{Max-Planck-Institut f\"ur Astronomie, K\"onigstuhl 17, 69117 Heidelberg, Germany}
\altaffiltext{8}{Center for Cosmology and Particle Physics, Department of Physics, New York University, 4 Washington Place, New York, NY 10003, USA}
\altaffiltext{9}{Department of Astronomy, University of Wisconsin, 475 N. Charter St., Madison, WI 53706, USA}
\altaffiltext{10}{Instituto de F\'isica y Astronom\'ia, Universidad de Valpara\'iso, Av. Gran Breta\~na 1111, Playa Ancha, Casilla 5030, and Millennium Institute of Astrophysics (MAS), Santiago, Chile}
\altaffiltext{11}{GMTO Corp, Suite 300, 251 S. Lake Ave, Pasadena, CA 91101, USA}
\altaffiltext{12}{Institute for Astronomy, University of Hawaii at Manoa, Honolulu, HI 96822, USA}
\altaffiltext{13}{Department of Physics, University of Durham Science Laboratories, South Road Durham DH1 3LE, UK}
\altaffiltext{14}{Department of Physics, University of Notre Dame, Notre Dame, IN 46556,USA}
\altaffiltext{15}{Syracuse University, Syracuse, NY 13244, USA}
\altaffiltext{16}{Apache Point Observatory, P.O. Box 59, Sunspot, NM 88349, USA}
\altaffiltext{17}{INAF - Osservatorio Astronomico di Trieste, Via G. B. Tiepolo 11, I-34131 Trieste, Italy}
\altaffiltext{18}{Department of Physics \& Astronomy, Vanderbilt University, VU Station B 1807, Nashville, TN, USA}
\altaffiltext{19}{Department of Astronomy, Ohio State University, 140 West 18th Avenue, Columbus, OH 43210, USA}
\altaffiltext{20}{Department of Astrophysical Sciences, Princeton University, Princeton, NJ 08544, USA}
\altaffiltext{21}{Max Planck institute for extraterrestrial Physics, Giessenbachstr. 1, Garching, D-85748, Germany} 
\altaffiltext{22}{Carnegie Observatories, 813 Santa Barbara Street, Pasadena, CA 91101, USA}
\altaffiltext{23}{Kavli Institute for Astronomy and Astrophysics, Peking University, Beijing 100871, China}

\email{emorganson@cfa.harvard.edu}



\begin{abstract} 

We present the selection algorithm and anticipated results for the Time Domain Spectroscopic Survey (TDSS). TDSS is an SDSS-IV eBOSS subproject that will provide initial identification spectra of approximately 220{,}000 luminosity-variable objects (variable stars and AGN) across 7{,}500 deg$^2$ selected from a combination of SDSS and multi-epoch Pan-STARRS1 photometry. TDSS will be the largest spectroscopic survey to explicitly target variable objects, avoiding pre-selection on the basis of colors or detailed modeling of specific variability characteristics. Kernel Density Estimate (KDE) analysis of our target population performed on SDSS Stripe 82 data suggests our target sample will be 95\% pure (meaning 95\% of objects we select have genuine luminosity variability of a few magnitudes or more). Our final spectroscopic sample will contain roughly 135{,}000 quasars and 85{,}000 stellar variables, approximately 4{,}000 of which will be RR Lyrae stars which may be used as outer Milky Way probes. The variability-selected quasar population has a smoother redshift distribution than a color-selected sample, and variability measurements similar to those we develop here may be used to make more uniform quasar samples in large surveys. The stellar variable targets are distributed fairly uniformly across color space, indicating that TDSS will obtain spectra for a wide variety of stellar variables including pulsating variables, stars with significant chromospheric activity, cataclysmic variables and eclipsing binaries. TDSS will serve as a pathfinder mission to identify and characterize the multitude of variable objects that will be detected photometrically in even larger variability surveys such as LSST.

\end{abstract}





\section{Introduction}\label{sect:intro}

Variability in optical luminosity is an important behavior in many astronomical objects, and enhancing our understanding of the physics of a variety of systems. In this paper, we discuss objects whose optical luminosity varies by a tenth of a magnitude or more on time scales of a year or less, a level that can be easily measured with ground-based observations, and refer to them as "variable objects". This term encompasses both stellar variables and AGN. The large majority of AGN (especially quasars) and approximately one percent of stars satisfy this definition. Quasars and other AGN generally vary stochastically in optical bands by up to several tenths of a magnitude over  months and years \citep{GIVE++99,VAND++04}. The main cause of quasar variability in the optical continuum is instability in the accretion disk \citep{REES84,KAWA++98,PERE++06,RUAN++14}. Blazars, generally accepted to be AGN whose relativistic jets point along the line of sight \citep{ANTO93,URRY++95}, vary due to Doppler beaming of their jet emission \citep{ULRI++97}. Microlensing by stars in intervening lensing galaxies \citep{WAMB06,MORG++10} can also contribute to AGN variability in some cases. 

Stellar variability is produced by a large variety of physical processes. Chromospheric magnetic fields cause flaring stellar activity \citep{SCHA62,WILS63,BALI++95,HALL++09,MATH++14} that produces significant optical variability, particularly in younger late-type stars. Periodically pulsating variable stars exhibit large amplitude variability caused by the $\kappa$ mechanism in which a star's atmospheric opacity varies periodically \citep{ZHEV59}. These are more likely to appear as early-type stars, and the most famous pulsators, RR Lyrae and Cepheid variables, are commonly used as "standard candle" distance probes \citep{HUBB29,RODG57,P&V87,SMIT95,FREE++01,SESA++10}. Cataclysmic variables (CVs) are binaries in which a white dwarf accretes material from its companion producing occasional outbursts that can generate several magnitudes of variability \citep{MUMF63,SMIT07,KNIG11}. CV donor stars can appear as a wide variety of stellar types although most CVs involve a red dwarf or giant. Eclipsing binaries can also produce significant periodic variability \citep{STEP60,DEBO++11,BECK++14} across all stellar types. 

Because of its astrophysical importance, variability has become the focus of many recent and upcoming photometric surveys in which the same region of sky is imaged multiple times. A series of small (20-100 deg$^2$) surveys including the Faint Sky Variability Survey \citep{GROO++03}  and MACHO \citep{ALCO++01} have obtained hundreds to thousands of photometric measurement epochs. The Kepler Mission \citep{BORU++10} is probing similarly sized areas with much greater photometric precision and tens of thousands of observation epochs. OGLE I-OGLE IV \citep{UDAL++08, WYRZ++14}, the QUEST RR-Lyrae Survey \citep{VIVA++04}, the Sloan Digital Sky Survey \citep[SDSS,][]{YORK++00} Stripe 82 \citep{SESA++07} and the VISTA Variables in the V\'ia L\'actea ESO Public Survey \citep{CATE++11} cover 2{,}000, 700, 290 and 560 deg$^2$ respectively with each providing of order 100 measurement epochs per source. Recently, a number of "full sky" (at least 10{,}000 deg$^2$) variability surveys have been completed. ROTSE-I \citep{AKER++00,WOZN++04a}, The La Silla-QUEST Variability Survey in the Southern Hemisphere \citep{HADI++12}, the Catalina Sky Survey \citep{DRAK++09}, the Palomar Transient Factory \citep[PTF,][]{LAW++09}, All-Sky Automated Survey \citep[ASAS,][]{POJM++02}, the Lincoln Near-Earth Asteroid Research survey \citep[LINEAR,][]{PALA++13} and Pan-STARRS1 \citep[PS1,][]{KAIS++02,KAIS++10} obtain between 50 and 400 measurements per object. Of these, PS1 is the deepest and covers the largest area, and PS1 data will be the focus of this paper. In the near future, the Gaia mission \citep{LIND++08} and the Large Synoptic Survey Telescope \citep{ABEL++09} will extend full sky surveys to greater precision, more rapid cadences and much fainter limits. 

These photometric surveys have been accompanied by many large spectroscopic surveys. The SDSS-III Baryon Oscillation Spectroscopic Survey \citep[BOSS,][]{DAWS++13}, its SDSS-IV extension eBOSS (eBOSS, Dawson et al. 2015, \textit{in preparation}) and the LAMOST ExtraGAlactic Surveys \citep[LEGAS,][]{WANG++09} will eventually take $1.3\times10^6$ spectra of quasars, which are generally variable. SDSS has also taken $2.4\times10^5$ optical stellar spectra in the Sloan Extension for Galactic Understanding and Exploration \citep[SEGUE,][]{YANN++09} and will take $10^5$ high resolution infrared spectra with the APO Galactic Evolution Experiment \citep[APOGEE,][]{ZASO++13}. The Bulge Radial Velocity Assay \citep[BRAVA,][]{KUND++12}, the Radial Velocity Experiment \citep[RAVE,][]{KORD++13}, the LAMOST Experiment for Galactic Understanding and Exploration \citep[LEGUE,][]{DENG++12} and the GALactic Archaeology with HERMES survey \citep[GALAH,][]{ZUCK++12} will obtain between $10^4$ and $2.5 \times 10^6$ stellar spectra each. We expect roughly 1\% of the stars in each of these surveys to satisfy our definition of variable. Finally, the Gaia mission will obtain high resolution ($R \approx 11{,}500$) narrow filter (8{,}470 \AA\ $< \lambda <$ 8{,}740) and low resolution (10 $< R <$ 200) broad filter (3{,}300 \AA\ $< \lambda <$ 10{,}000) spectroscopy of $10^8$ $V < 17$ objects. These spectra will provide precise radial velocity measurements and generally characterize a wide variety of astrophysical objects, but for the broad variety of galactic and extragalactic variable objects we target may be less useful at characterizing e.g., specific absorption and emission lines that fall outside the narrow high resolution spectra.  
 
Despite these dedicated photometric variability surveys and similarly large spectroscopic surveys, large spectroscopic surveys of variable objects are somewhat lacking. There have been variability-selected samples of quasars \citep[e.g.][]{PALA++11} and RR Lyrae stars \citep[e.g.][]{DRAK++13} as well as relatively small SDSS spectroscopic variability studies of subdwarfs \citep{GEIE++11}, white dwarf main-sequence binaries \citep{REBA++11}, white dwarfs \citep{BADE++09,MULL++09,BADE++13} and field stars more generally \citep{POUR++05}. But these surveys have been relatively small in size and have used color information, spectra or specific light curve character to target specific types of variables.
 
The Time Domain Spectroscopic Survey (TDSS) has been designed to widen the scope of spectroscopic surveys of variable objects and will soon become the largest medium resolution ($R \approx 2{,}000$), broad wavelength (3{,}600 \AA\ $< \lambda <$ 10{,}400 \AA)  spectroscopic survey of variable objects. This survey, a subproject of the SDSS-IV Extended Baryon Oscillation Spectroscopic Survey (eBOSS), will cover 7{,}500 deg$^2$ and include 220{,}000 variability-selected targets with no focus on any specific variability or photometric type in target selection. TDSS is not well-suited to spectroscopic identification of rapid transients, because plug plates to accommodate the 1,000 spectroscopic fibers must
be drilled well in advance of observations. 

Roughly 90\% of TDSS targets will be TDSS's Single Epoch Spectroscopy (SES) targets, for which TDSS will produce a single discovery (identification/classification) spectrum. For bright, quickly varying targets within this sample, we will also be able to study spectroscopic variability by examining the spectroscopic sub-exposures taken over hours or sometimes several nights. This TDSS SES sample is designed to be a  probe of general optical variability and will be the subject of this paper. 

The remaining 10\% of TDSS spectra will be drawn from one of TDSS's nine Few Epoch Spectroscopy (FES) projects, a series of smaller ($\approx$ 1{,}000 targets each) samples with previous SDSS spectroscopy for which TDSS will obtain another spectrum for two and occasionally three-epoch comparison. The FES projects are each designed to probe a specific type of variable object and science topic. These nine current projects are devoted to:
\begin{itemize}
\item Radial velocity variation in dwarf carbon stars
\item M-dwarf white dwarf binaries 
\item Activity in ultracool dwarfs on decadal time scales
\item Stars with more than 0.2 magnitudes of variability
\item Broad absorption line trough variability \citep[as in][]{AK++13} in quasars
\item Balmer line variability in high signal to noise quasars
\item Double-peaked broad emission line quasars
\item Searching for binary black hole quasars via Mg II line velocity shifts
\item Quasars with more than 0.7 magnitudes of variability
\end{itemize} 
The details of these FES projects will be addressed in future papers.

In this paper, we describe how the TDSS SES Project (subsequently referred to as simply TDSS) produces a large sample of photometric variable objects with a broad range of variability types while avoiding spurious, non-astrophysical ``variability'' in its target selection. In Section \ref{sect:eboss} we outline TDSS's role in eBOSS, the larger SDSS-IV optical spectroscopy project. In Section \ref{sect:data} we demonstrate how the combination of SDSS and Pan-STARRS1 photometry allows the construction of a 7{,}500 deg$^2$, relatively uniform sample, and we describe our algorithm for quantifying variability into a single metric in Section \ref{sect:varmeas}. In Section \ref{sect:prioritization} we present our ultimate target prioritization. We estimate our survey purity (fraction of candidates that genuinely vary by a few tenths of magnitudes) and show how it varies across the sky in Section \ref{sect:purity}. We describe the selection of a small subsample of $i$-band dropouts that would have been missed by our algorithm without special effort in Section \ref{sect:idrops}. We statistically classify our complete list of targets by their colors in Section \ref{sect:phot} and discuss how our selection percentage varies as a function of color in Section \ref{sect:fraction}. Finally, we compare the targets selected by our algorithm using our dataset to small sets of known variable objects and objects with existing SDSS spectra from SDSS Stripe 82 in Section \ref{sect:spec}. 

\section{TDSS and eBOSS}\label{sect:eboss}

\begin{figure}[ht]
\includegraphics[width=0.98\columnwidth]{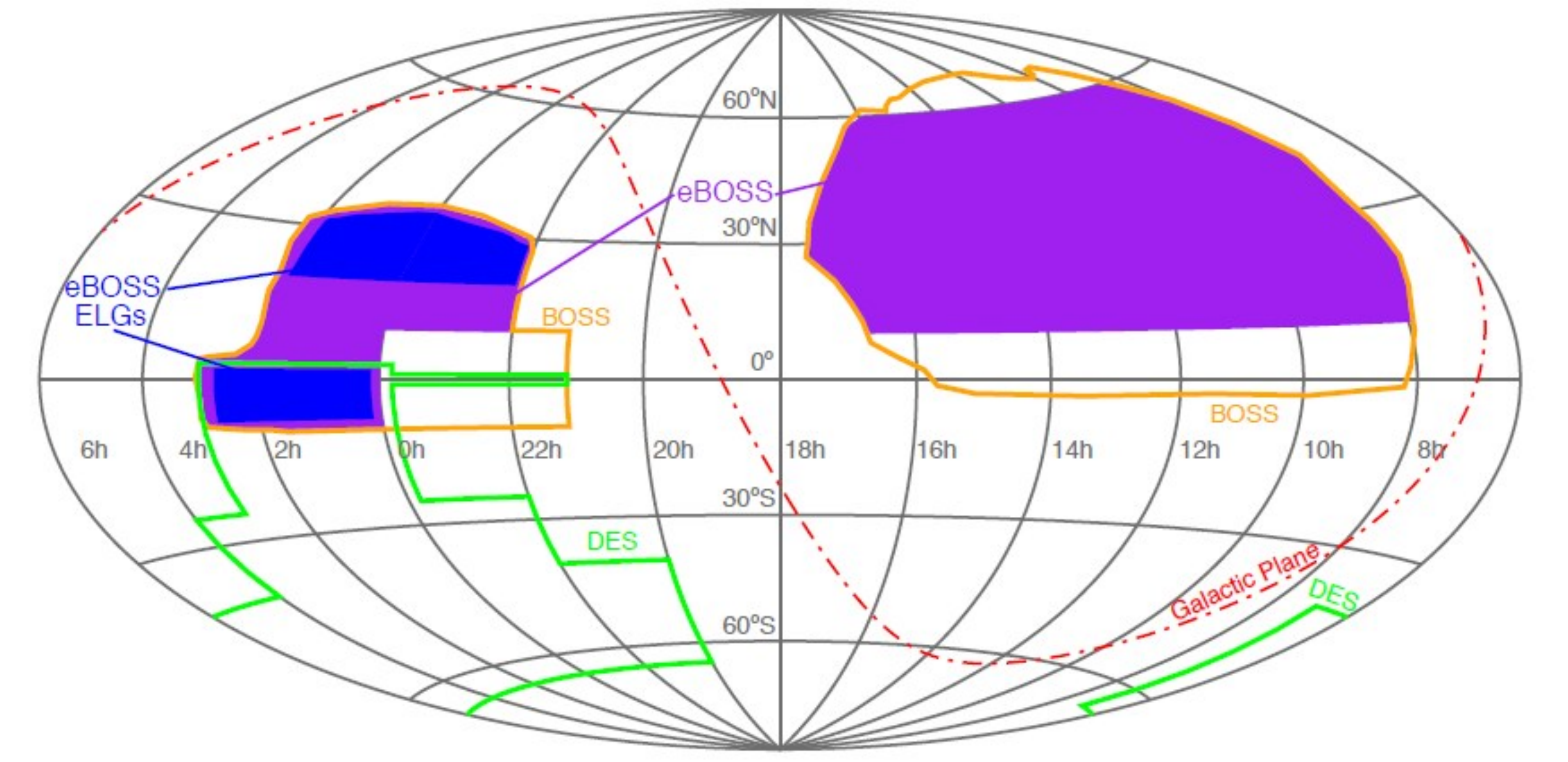}
\caption{\rm{The planned eBOSS (and by extension TDSS) area is shown in blue and purple. The blue area may be sampled twice for the eBOSS Emission Line Galaxy (ELG) project. The eBOSS predecessor, BOSS, is outlined in orange, and the Dark Energy Survey, which may be of interest in ELG targeting, is outlined in green.}}
\label{fig:eBOSS}\end{figure}

TDSS is a subprogram of the Extended Baryon Oscillation Spectroscopic Survey (eBOSS). eBOSS is an SDSS-IV project designed to perform a variety of cosmological measurements with spectroscopy of quasars (Myers et al. 2015, \textit{in preparation}), luminous red galaxies (Prakash et al. 2015, \textit{in preparation}), X-ray emitting quasars and cluster galaxies (Menzel et al. 2015 \textit{in preparation}, Finoguenov et al. 2015, \textit{in preparation} and Clerc et al. 2015, \textit{in preparation}) and emission line galaxies \citep{COMP++13a}. TDSS will be paired with the main eBOSS survey (shown in Fig.\ \ref{fig:eBOSS}) and is planned to cover a total of 7{,}500 deg$^2$ in the Northern and Southern Galactic Caps. eBOSS devotes 10 fibers deg$^{-2}$ to TDSS-only targets. But TDSS also selects an additional 23 TDSS-joint targets deg$^{-2}$ that have previous SDSS spectroscopy or are part of the main eBOSS quasar target list most of which is selected using colors alone with the $XDQSOz$ algorithm \citep{BOVY++12}. A small number of eBOSS quasars are also selected using a combination of colors and optical variability from the Palomar Transient Factory \citep{PELA++11}. The full TDSS sample will thus include 33 objects deg$^{-2}$. See Section \ref{sect:purity} for more details. 

Spectroscopy for the main TDSS sample will be obtained as part of the eBOSS schedule on the BOSS spectrograph \citep{SMEE++13}. At the TDSS $i = 21$ magnitude limit, we will obtain per pixel signal to noise ratios of 5 or better \citep{DAWS++13} (typical pixel size is roughly 1 \AA). We use an $i = 17$ bright limit to prevent saturation and signal leaking between adjacent fibers. The spectra cover 3{,}700 \AA\ $< \lambda <$ 10{,}400 \AA\ in two channels (red and blue). The spectrograph's resolution runs from $R = $1{,}560 at 3{,}700 \AA\ to $R = $2{,}270 at 6{,}000 \AA\ (blue channel), and from $R = $1{,}850 at 6{,}000 \AA\ to $R = $2{,}650 at 9{,}000 \AA\ (reds channel). These spectra will be easily good enough measure continua, major absorption and emission features, quasar redshifts and stellar velocities (to better than 50 km s$^{-1}$). 
s

\section{The SDSS-PS1 Dataset}\label{sect:data}

In order to measure optical variability, TDSS uses a combination of single-epoch SDSS and multi-epoch PS1 photometry. We use SDSS photometry from SDSS Data Release 9 \citep{GUNN++98,YORK++00,GUNN++06,AIHA++11,EISE++11,AHN++12}. SDSS DR9 covers 14{,}555 square degrees in the $u$, $g$, $r$, $i$ and $z$ filters which span the 3{,}000 \AA\ $< \lambda <$ 10{,}000 \AA\ spectral range \citep{FUKU++96}. The imaging footprint covers most of the high Galactic longitude area north of declination $-10^\circ$. Throughout this paper, we use $u$, $g$, $r$, $i$ and $z$ to refer to the SDSS magnitudes and not the (very similar) PS1 analogs. 

The PS1 3$\pi$ survey \citep{KAIS++02,KAIS++10,CHAM11} covers its 30{,}000 deg$^2$ area north of declination $-30^\circ$. This region includes the entire SDSS survey imaging footprint. The PS1 $\gps$, $\rps$, $\ips$ and $\zps$ filters cover the 4000 \AA\ $< \lambda <$ 9200 \AA\ spectral range similarly to the corresponding SDSS $g$, $r$, $i$ and $z$ filters. PS1 also has a $\yps$ filter which, including the spectral response of the camera, covers 9200 \AA\ $< \lambda <$ 10500 \AA. These PS1 filters are described in detail in \citet{TONR++12}. The PS1 survey takes four exposures per year for 3.5 years with each of the $\grizy$ filters (non-simultaneously) and fills approximately 90\% of the 30{,}000 deg$^2$ area in each band. The missing area is mostly due to non-detection areas on the camera plane and weather restricting the survey to two or rarely zero exposures per filter in some areas of the sky. Individual PS1 exposures are generally shallower than analogous SDSS images. However, the $10\sigma$ limiting PSF magnitudes of the PS1 average catalogs, produced by taking a weighted average of individual detections rather than stacking the images (PS1 image stacking is still being developed), are well-matched to the SDSS single-exposure limits as summarized in Table \ref{tab:limmag}.

In this work, we use an updated version of the "\"ubercalibrated" PS1 data from \citet{SCHL++12}, which includes the PS1 data up through July 2013 (using PV1 of the PS1 pipeline) and is calibrated absolutely to 0.02 magnitudes or better. This database excludes detections flagged by PS1 as cosmic rays, edge effects and other defects.

\begin{table}
\begin{tabular}{ccccc}
        \hline
Filter &  SDSS    &   PS1 Exposure & PS1 Mean & PS1 Stack  \\
        \hline
$u$   & 21.2 & --   & --   & --   \\
$g$   & 22.3 & 21.2 & 21.7 & 22.4 \\
$r$   & 21.8 & 21.0 & 21.7 & 22.2 \\
$i$   & 21.4 & 20.8 & 21.5 & 22.0 \\
$z$   & 19.9 & 20.1 & 20.7 & 21.3 \\
$y$   & --   & 19.1 & 19.7 & 20.3 \\
        \hline
\end{tabular}
\caption{\rm{Median 10$\sigma$ Limiting AB PSF Magnitudes of SDSS, PS1 3$\pi$ single exposures, the PS1 3$\pi$ mean catalog and the current PS1 3$\pi$ stack. Similarly-named filters from different surveys are not exactly the same.}}\label{tab:limmag}
\end{table}

To convert between SDSS and PS1 magnitudes, we use the conversions from \citet{FINK++14} which follow the equation
\begin{eqnarray} 
m_{\rm{P1}}-m_{\rm{SDSS}} &=& \rm{a}_0 + \rm{a}_1\ \mathit{gi} + \rm{a}_2\ \mathit{gi}^2 + \rm{a}_3\ \mathit{gi}^3,\label{eq:spcon}\\ 
gi &=& g-i.\nonumber
\end{eqnarray}
where $m = griz$ and a$_{0123}$ are in Table \ref{tab:coeff}. \citet{TONR++12} also provide a similar conversion from SDSS to PS1 calculated from PS1 filter curves, but we use the Finkbeiner equations because they are optimized to be accurate for a broad stellar population, and because they are calculated within the \citet{SCHL++12} \"ubercalibrated system. For the non-varying stars for which these coefficients were fit, these conversions are accurate to 0.01 magnitudes or better. We add this 0.01 mag in quadrature to our statistical error. When comparing SDSS and PS1 magnitudes, we convert them to standard logarithmic magnitudes, rather than the default asinh-based "Luptitudes" that SDSS reports \citep{LUPT++99}.

\begin{table}
\begin{tabular}{ccccc}
        \hline
Filter &  a$_0$   & a$_1$    & a$_2$    & a$_3$    \\
        \hline
$g$    & $0.00128$  & $-0.10699$ & $0.00392$  & $0.00152$  \\
$r$    & $-0.00518$ & $-0.03561$ & $0.02359$  & $-0.00447$ \\
$i$    & $0.00585$  & $-0.01287$ & $0.00707$  & $-0.00178$ \\
$z$    & $0.00144$  & $0.07379$  & $-0.03366$ & $0.00765$  \\
        \hline
\end{tabular}
\caption{\rm{The coefficients used to convert from SDSS magnitude to PS1 magnitudes in Eq.\ \ref{eq:spcon}. Ensemble error bars are insignificant; for individual stars, these conversions are good to 0.01 magnitudes.}}\label{tab:coeff}
\end{table}

All database analysis and cross-matching of surveys is performed with the Large Survey Database software \citep[LSD][]{JURI11}. LSD is a versatile, parallelized, python-based database module optimized for astronomical querying and cross-matching. We compare PS1 and SDSS PSF magnitudes in all cases, only work with objects that are unresolved in SDSS (morphology type "star") and match PS1 and SDSS objects with a radius of 1.5''.

\begin{figure}[ht]
\includegraphics[width=0.98\columnwidth]{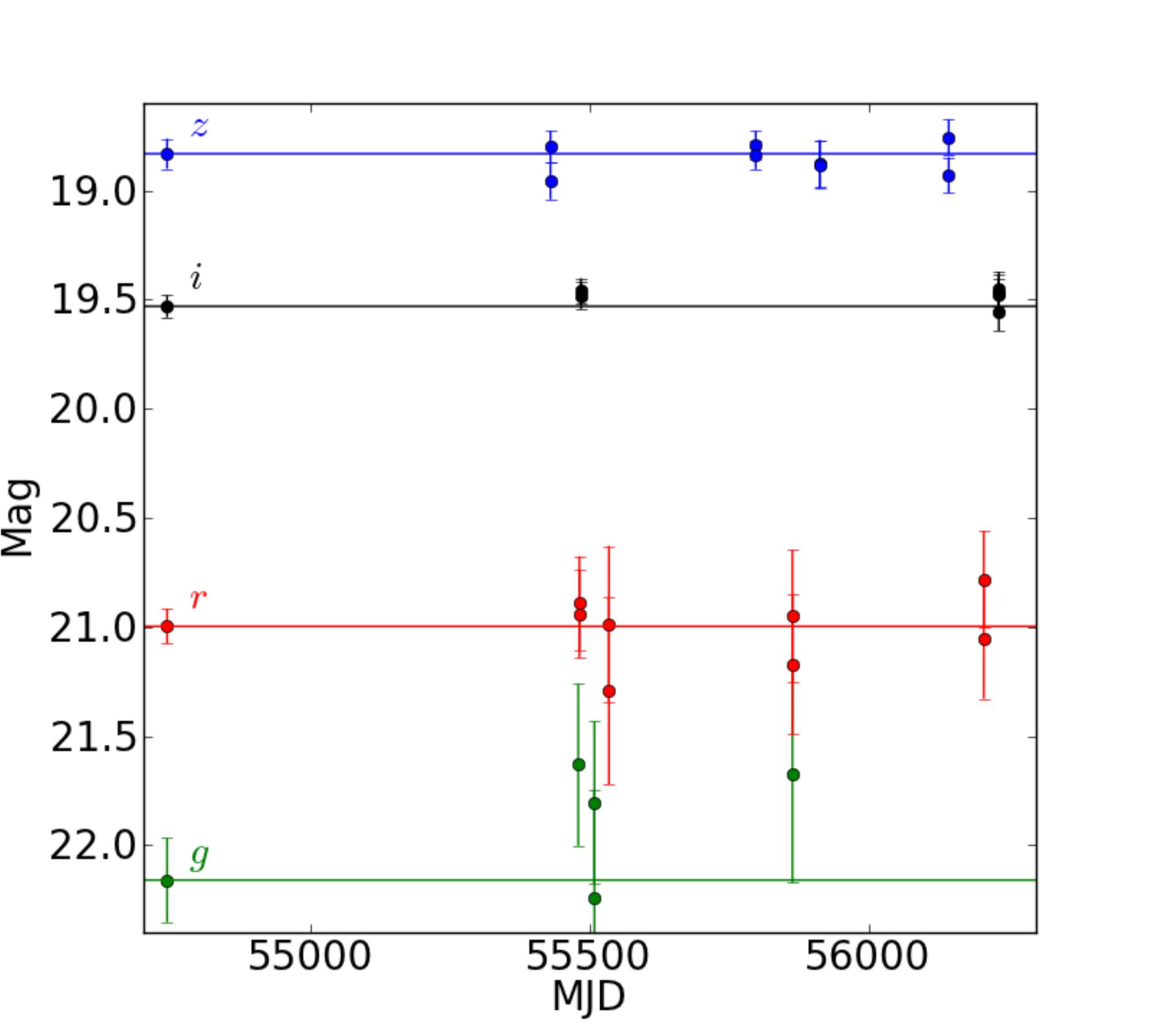}
\includegraphics[width=0.98\columnwidth]{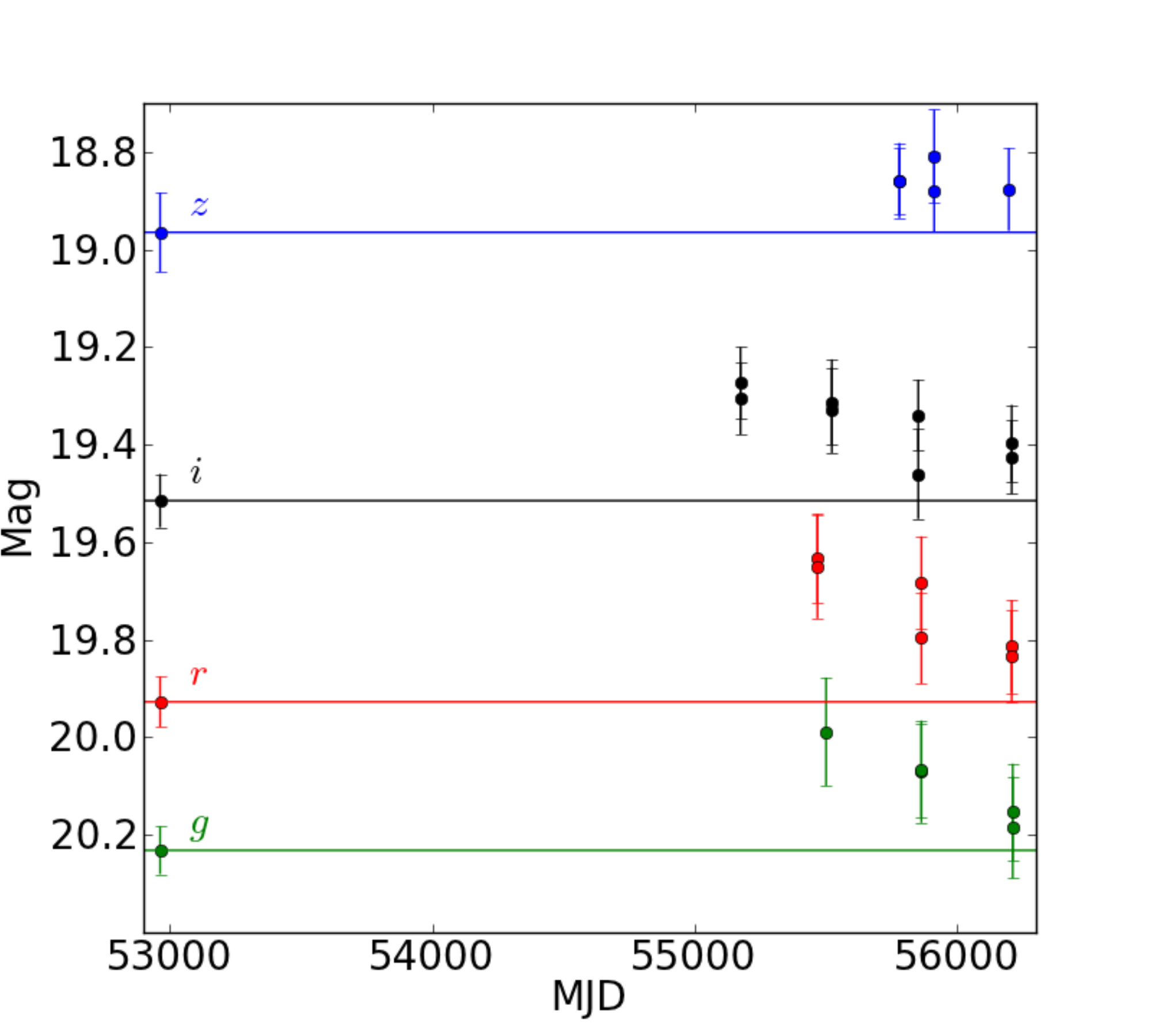}
\caption{\rm{Typical SDSS-PS1 light curves from a Stripe 82 photometric standards (top) and variable objects (bottom). The four lines in each figure represent, from bottom to top, the light curve from $g$ (in green), $r$ (in red), $i$ (in black) and $z$ (in blue) filters. The first data point on each light curve and the horizontal line are taken from (PS1-converted) SDSS, and all other datapoints are from PS1.} }
\label{fig:lightcurve}\end{figure}

Fig.\ \ref{fig:lightcurve} shows a typical SDSS-PS1 light curve for a non-variable and a variable object. The SDSS and PS1 magnitudes are consistent for the non-variable, confirming at least the approximate validity of the Finkbeiner conversions in this case. It is easily discernible that the variable object is varying in the PS1 data, but it is also clear that the very sparse PS1 sampling prevents a detailed characterization of a single object's variability (e.g. determining a period). This limitation, combined with a desire to avoid biasing our sample to any specific variability type, led to the relatively simple variability criteria described in Section \ref{sect:varmeas}. 

\subsection{PS1 Photometric Uncertainties}\label{sect:errors}
Accurate photometric uncertainties are very important for variability measurements. If we overestimate photometric error, we will underestimate variability and vice versa. To assess the level of spurious variability induced by incorrect error bars in PS1, we draw from the PS1 catalog a population of 4{,}032{,}258 (theoretically) constant photometric F stars which satisfy the following SDSS criteria (all magnitudes PSF magnitudes and are dereddened using the \citet{SCHL++98} extinction map):

\begin{eqnarray}
16 &<& r < 20,\label{eq:fstar}\\
(u-g-0.82)^2 &+& (g-r-0.30)^2 + (r-i-0.09)^2 + \nonumber\\
(i-z-0.02)^2 &<& 0.04,\nonumber\\
\rm{Type_{SDSS}} &=& 6\ \rm{(star)}.\nonumber
\end{eqnarray}
This selection volume is essentially a 0.2 magnitude four color sphere around the position of F stars in color space \citep{IVEZ++07}. F stars are useful standards because they are common, and because their luminosity peaks roughly in the middle of our \griz wavelength range. 

\begin{figure}[ht]
\includegraphics[width=0.49\columnwidth]{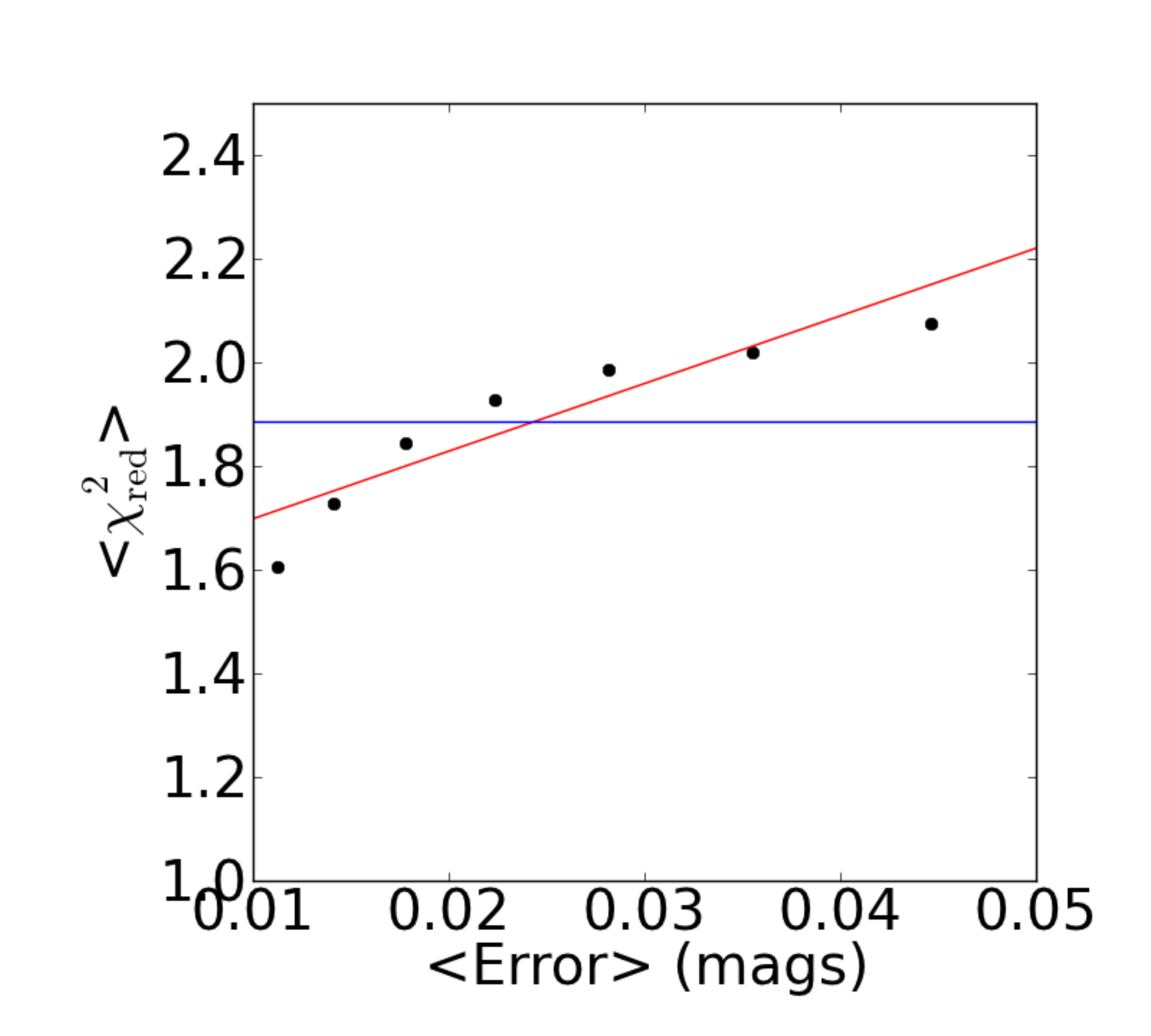}
\includegraphics[width=0.49\columnwidth]{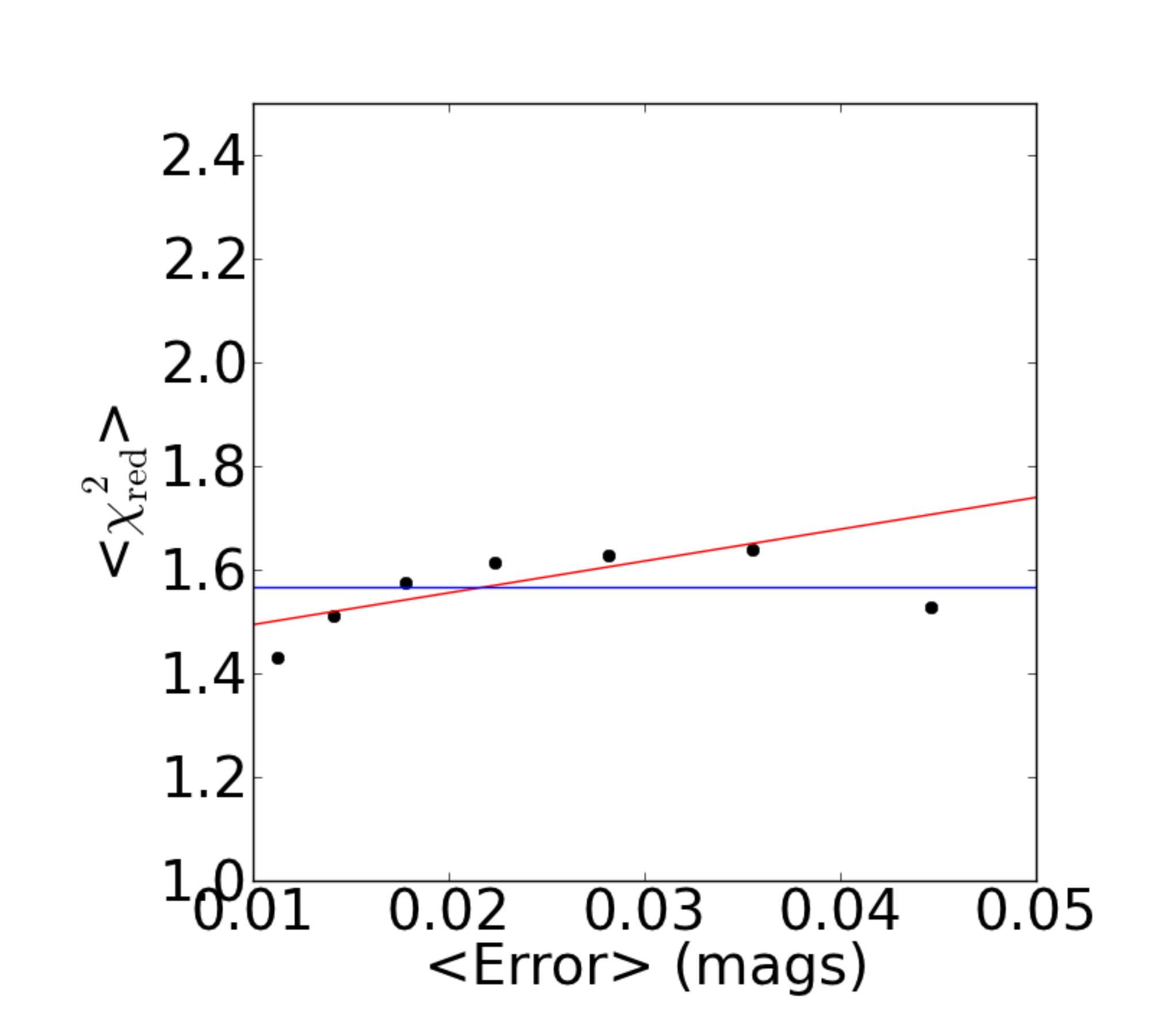}
\includegraphics[width=0.49\columnwidth]{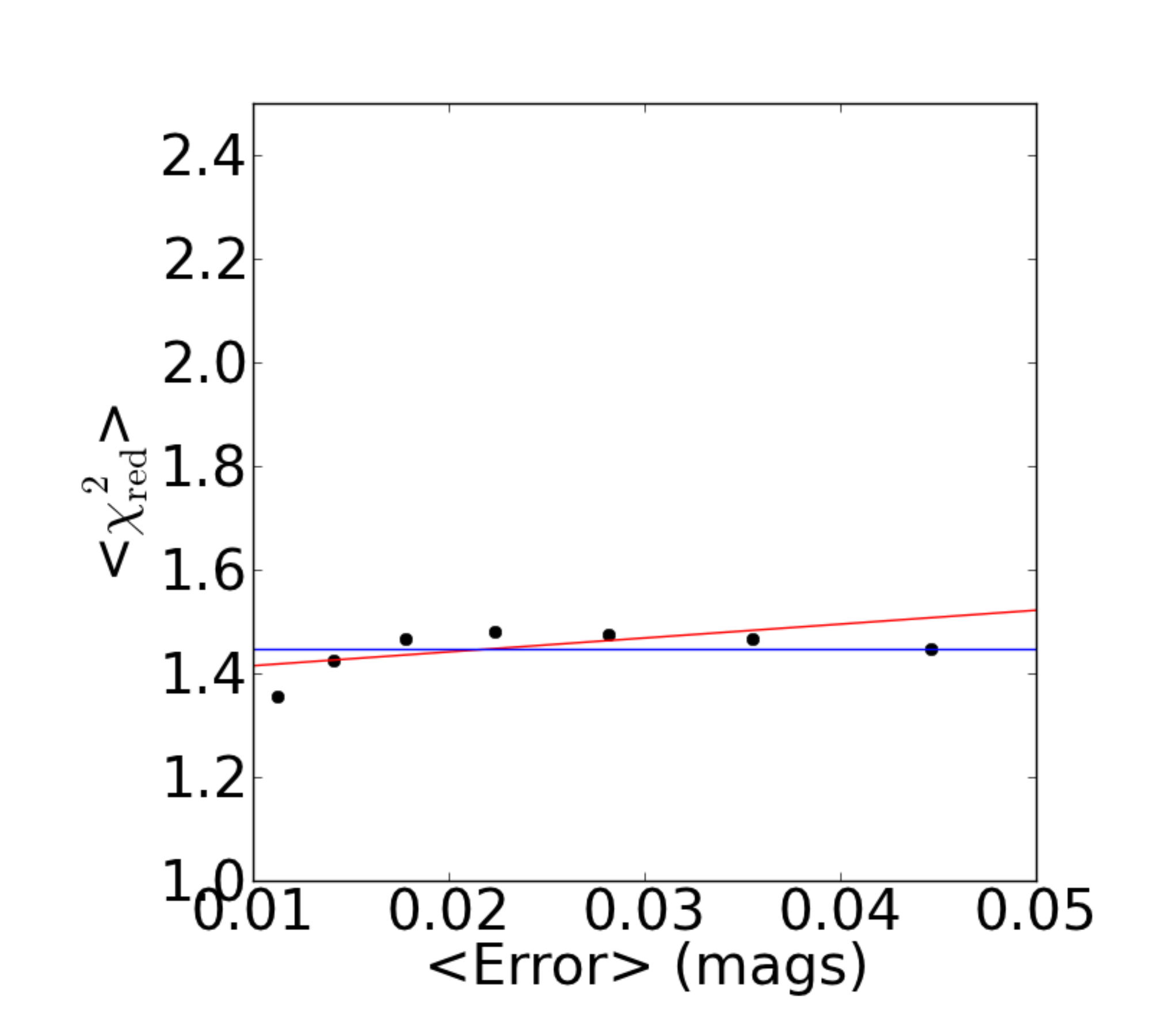}
\includegraphics[width=0.49\columnwidth]{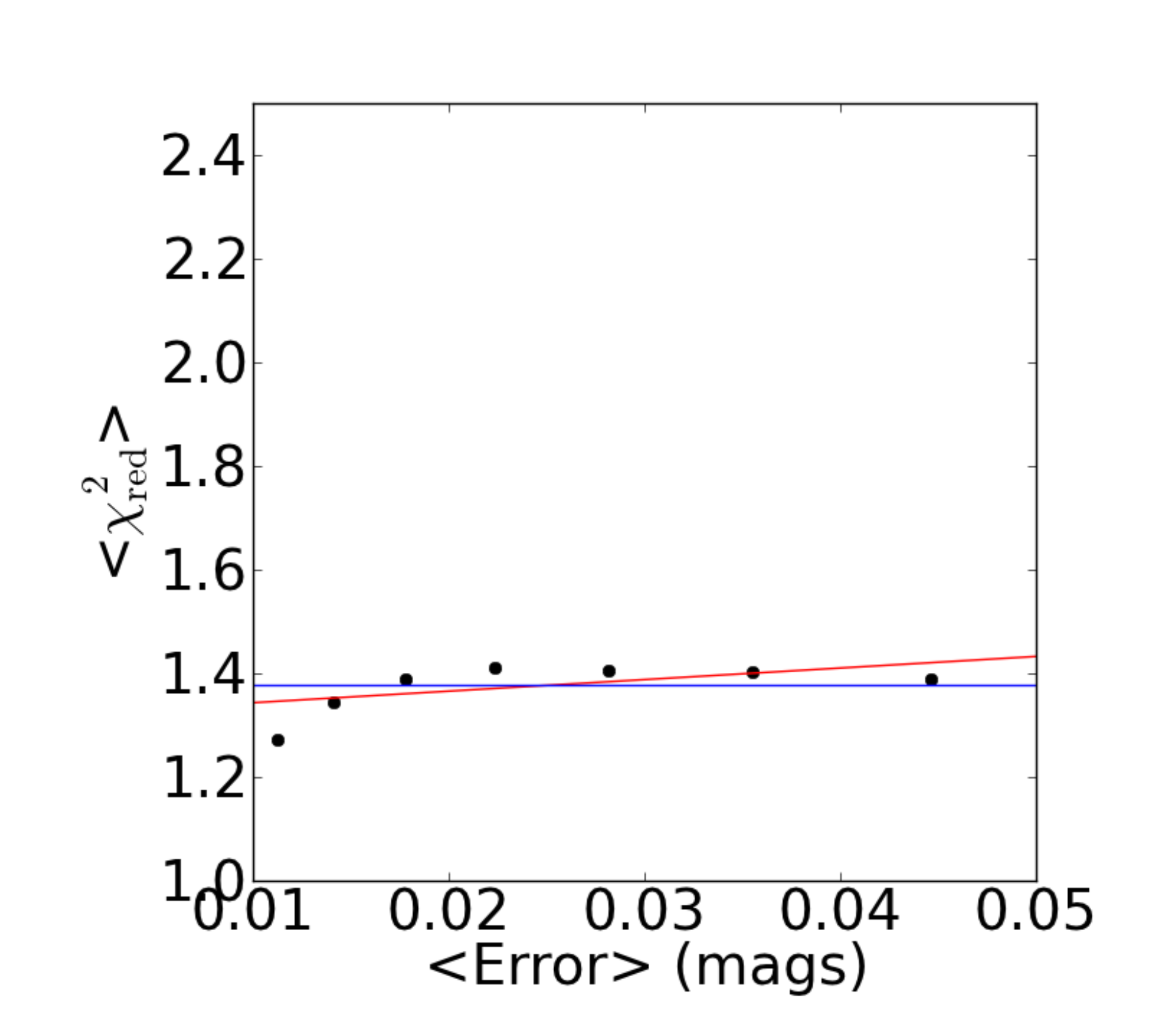}
\caption{\rm{$\chi^2_{\rm{red}}$ versus average photometric measurement error for F stars that satisfy Eq.\ \ref{eq:fstar} in \gps (upper left), \rps (upper right), \ips (lower left) and \zps (lower right). Each is fit as a constant (blue) and as a line (red). The data are very roughly consistent with a constant model, with a different constant for each filter.} }
\label{fig:chierr}\end{figure}

We examine the reduced $\chi^2$ distribution for single filter F star light curves, assuming a constant luminosity model, i.e.:
\begin{eqnarray}
\chi^2_{\rm{red}} &=& \frac{1}{n-1} \sum \frac{(m_i-\bar{m})^2}{\sigma_i^2}\\
\bar{m} &=& \frac{\sum m_i/\sigma_i^2}{\sum 1/\sigma_i^2 }.\nonumber
\end{eqnarray}
The quantity $\chi^2_{\rm{red}}$ should approach unity for large ensembles of constant sources, implying that the variation in the mean magnitude is consistent with the error bars. We plot the median $\chi^2_{\rm{red}}$ versus the average of the error bars from different measurements in Fig.\ \ref{fig:chierr}. $\chi^2_{\rm{red}}$ is never 1, but it is fairly constant with respect to the size of the error bars (although there is a small positive correlation between the two). The square root of this constant is 1.387, 1.327, 1.249, 1.228 and 1.170 in \gps, \rps, \ips, \zps and \yps filters, respectively. We multiply the standard PS1 error bars by these constants in our work. 



\section{TDSS Variability Measurement}\label{sect:varmeas}

\begin{figure}[ht]
\includegraphics[width=0.98\columnwidth]{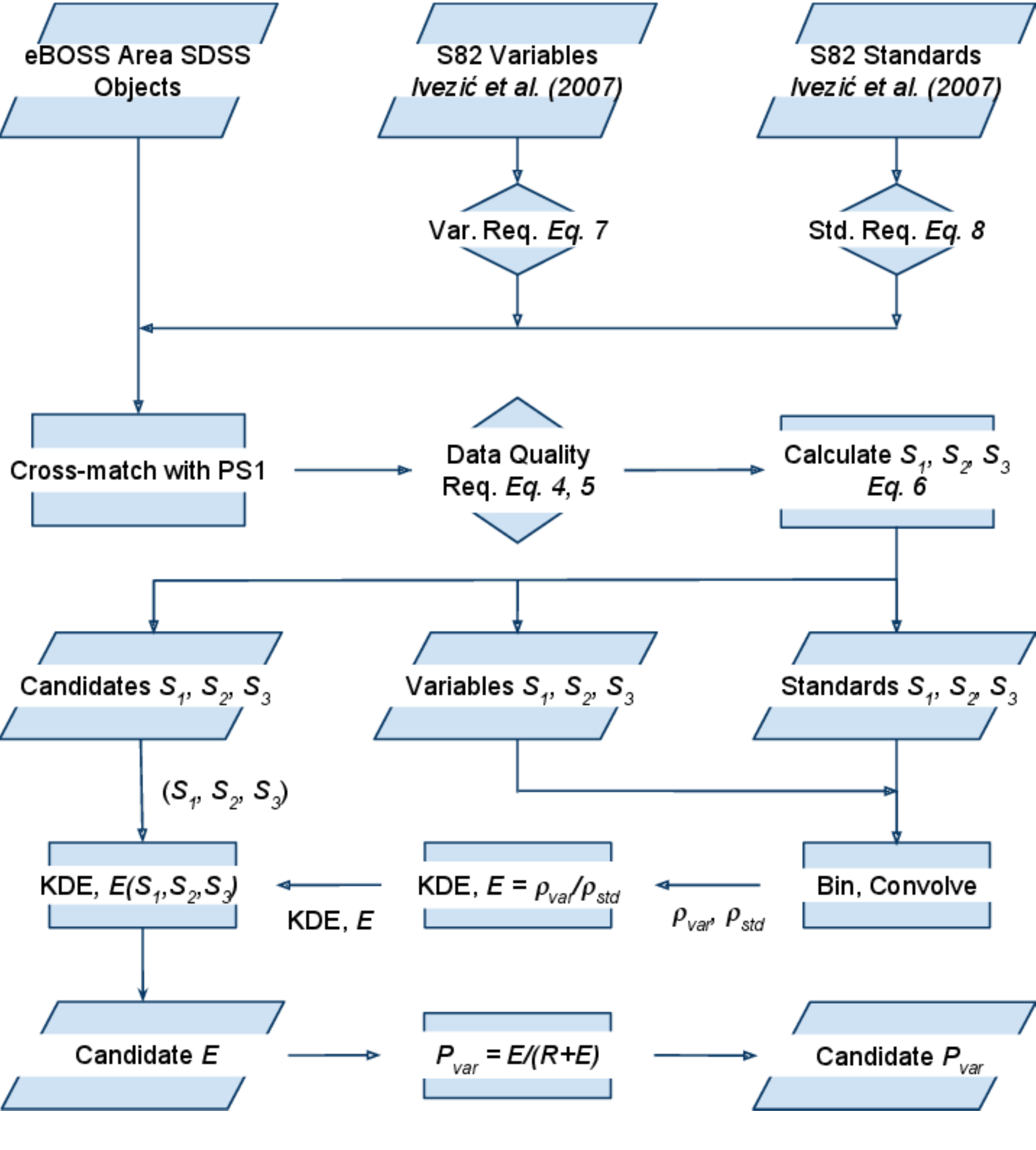}
\caption{\rm{A flowchart describing the process by which we determine $P_{\rm{variable}}$, the probability that a given candidate is a variable object. Parallelograms represent data objects. Diamonds are conditional statements that reject some of the data. Rectangles are functions. We start with all SDSS objects in the eBOSS area as well as two Stripe 82 training sets, Variables (both stellar and AGN) and Standards. Each of the three data sets pass a set of data quality cuts and are cross-matched with PS1 data. Using PS1 data and single epoch SDSS data, we calculate our variability parameters, ($S_1$, $S_2$, $S_3$) for all three datasets. We use the variability parameters from the training sets to produce a Kernel Density Estimate, $E(S_1, S_2, S_3)$. Using this function, which is static across the sky, we can assign every potential TDSS candidate an $E$ value, which is easily converted to  $P_{\rm{variable}}$, via $R$, the (assumed constant) ratio of variable objects to nonvariables.} }
\label{fig:flowchart}\end{figure}

TDSS aims to take full advantage of the SDSS-PS1 combined dataset to select a highly pure sample of variable objects without any overt bias with regard to color or variability pattern. To achieve this goal we preselect targets whose variability can be robustly measured. We then combine data across filters in which a given source is well-measured into a single three dimensional parameter space. Finally, we use a kernel density estimator \citep[KDE,][]{ROSE56,PARZ62} and a Stripe 82 training set to assign each object a probability of being a true variable object based on its location within this 3D KDE space. Fig.\ \ref{fig:flowchart} outlines this process.

Among objects detected in both SDSS an PS1, we preselect a subset with good data quality to avoid wasting computational resources on sources for which we could not reliably measure variability at the 0.1 magnitude level by requiring that 
\begin{eqnarray}
17 &<& i < 21,\label{eq:dataq}\\
g,\ r,\ z &>& 16,\nonumber\\
\rm{type_{SDSS}} &=& 6\ \rm{(star)},\nonumber\\
r_{22} &>& 5'',\nonumber\\
r_{17} &>& 10'',\nonumber\\
r_{15} &>& 20'',\nonumber\\
r_{13} &>& 30'',\nonumber\\
n_{PS1\ griz} &>& 10.\nonumber
\end{eqnarray}
Here, all magnitudes are SDSS PSF magnitudes. We find that 95\% of such objects have SDSS $i$ band errors and PS1 \ips\ mean errors of less than 0.1 magnitudes at $i = 21$. The $i > 17$ and $g,\ r,\ z > 16$ requirements prevent selection of very bright sources whose flux would bleed into neighboring spectroscopic fibers. We are obtaining spectra of $16 < i < 17$ targets with smaller telescopes and will discuss this bright extension of TDSS in a future paper. We restrict ourselves to unresolved objects (SDSS morphological type "star"), because it is difficult to perform consistent measurements of extended sources in varying observation conditions at the precision we need. Based on our experience with visual inspection, we also require that the sources not have an $i < 22$ neighbor within $5''$ as this can confuse the photometry ($r_{22} > 5''$). Similarly, we require no $i < 17,\ 15,\ 13$ neighbors within $10''$, $20''$, $30''$, respectively. Finally, we require PS1 detections at more than 10 epochs across the \gps\rps\ips\zps filters ($n_{PS1\ griz} > 10$) for each object to ensure that we have a significant amount of variability information. This last requirement is the most restrictive and is met by approximately 85\% of $17 < i < 21$ SDSS sources with PS1 matches.

We also examine each source in every filter to determine in which filters we can reliably measure variability. For a given source, we only measure variability in filters in which 

\begin{eqnarray}
\rm{err_{SDSS}} &<& 0.1,\label{eq:req}\\
\rm{err_{PS1}} &<& 0.1,\nonumber\\
n_{\rm{PS1}} &>& 1.\nonumber
\end{eqnarray}
Here, $\rm{err_{SDSS}}$ and $\rm{err_{PS1}}$ are the SDSS and PS1 mean magnitude errors, respectively. $n_{PS1}$ is the number of detections in a single PS1 filter. Because PS1 lacks a $u$ filter and SDSS lacks a $y$ filter, we only examine variability across the $griz$ filters. To eliminate some obvious artifacts, we ignore filters in which the PS1-SDSS difference is greater than 3 magnitudes, unless the PS1-SDSS difference is greater than 1.2 magnitudes in another filter. For a given source, we designate filters which pass these criteria as "good" and only examine the variability of sources with at least two good filters.

Many groups have demonstrated that advanced machine learning algorithms are highly effective at selecting variable objects \citep{WOZN++04b,RICH++11}. These algorithms are generally optimized to accept a large number of inputs to distinguish between variable objects and nonvariables with fairly complex routines that can be difficult to assess. We opted for a simpler 3D KDE estimator for two main reasons. First, TDSS aims to be a variability-only survey, and it is difficult to ensure that machine learning algorithms (which work best with many input parameters) are primarily using variability to select astrophysical variables. For instance, a boosted decision tree, given the $griz$ magnitudes of a set of variable objects (including many quasars) and non-variables (with mainly stars), can locate quasars clustered in color space, which may reproduce quasar color selection and ignore actual variability entirely. Second, the depth and number of observations per source varies significantly across the PS1 survey, and it is difficult to ensure that a complicated machine learning algorithm is operating efficiently across the whole sky when its inputs change. Furthermore, when we restricted a boosted decision tree to a small number of robust parameters that contained no color information, we found that the KDE detected more variable objects at a similar threshold. We discuss our boosted decision tree results in detail in the appendix. 

Through extensive testing, we have settled on a simple 3D ($S_1,S_2,S_3$) KDE parameter space:
\begin{eqnarray}
S_1 &=& \rm{median}(|\rm{mag}_{\rm{PS1}}-\rm{mag}_{\rm{SDSS}}|),\label{eq:xyz}\\
\rm{Var}_{\rm{PS1}} &=& \rm{Variance}_{\rm{PS1}}-\rm{Err}_{\rm{PS1}}^2(n_{\rm{PS1}}-1),\nonumber\\
S_2 &=& \rm{median}(\rm{sign}(\rm{Var}_{\rm{PS1}}) |\rm{Var}_{\rm{PS1}}|^{1/2}),\nonumber\\
S_3 &=& \rm{median}(\rm{mag}_{\rm{PS1}}).\nonumber
\end{eqnarray}
Qualitatively, $S_1$ is the PS1 SDSS difference and represents long term (multi-year) variability. $S_2$ is the PS1 only variability and represents short term (days to a few years) variability. $S_3$ is just an apparent magnitude. The word ``median'' refers to the median magnitude value across all good filters (Eq. \ref{eq:req}) for a given source. If there are only two good filters, $S_1$ and $S_2$ become minima to prevent individual outlier filters from creating false positive variable targets. All magnitudes used are Point Spread Function (PSF) magnitudes. The PS1 magnitudes, mag$_{\rm{PS1}}$, are median magnitudes, used to improve robustness due to the non-simultaneous nature of the PS1 measurements in different filters. Var$_{\rm{PS1}}$ is an estimate of true PS1 magnitude variability above the expected random variability given the error bars. Var is negative for sources whose photometry randomly varies less than their error bars would indicate. The variable $S_2$ is a simple function of Var that accounts for this possible negativity while also converting Var into units of magnitudes (where the distribution is more useful for KDE analysis). The variable $S_3$ is the median PS1 magnitude across good filters. While there is no obvious trend of variability with magnitude, our ability to accurately measure variability decreases as objects get fainter, and using $S_3$ in our selection allows us to adjust our threshold accordingly.

To assess which bins are the most likely to contain true variable objects, we use a set of confirmed Stripe 82 variable and standard (non-variable) objects. Both catalogs are from \citet{IVEZ++07} and are made with Stripe 82 light curves. Often, our error bars are of order 0.1 magnitudes, so to maintain high purity of the Stripe 82 variable object catalog, we require
\begin{eqnarray}
g\rm{Ampl} &>& 0.1,\\
r\rm{Ampl} &>& 0.1,\nonumber\\
i\rm{Ampl} &>& 0.05,\nonumber
\end{eqnarray}
where Ampl is the estimated amplitude of variation in magnitudes from SDSS. We use a lower threshold in the $i$ band, because both stellar variables and quasars tend to vary less in redder bands, and because SDSS is shallower and less sensitive to variability in the $i$ band. A total of 89\% of Stripe 82 variable objects from \citet{IVEZ++07} satisfy this requirement. To increase the purity of the Stripe 82 standards catalog, we require
\begin{eqnarray}
n_{\rm{SDSS}} &>& 7,\\
\chi^2_{g\ \rm{red}} &<& 2,\nonumber\\
\chi^2_{r\ \rm{red}} &<& 2,\nonumber\\
\chi^2_{i\ \rm{red}} &<& 2,\nonumber
\end{eqnarray}
where $n_{\rm{SDSS}}$ is the number of SDSS measurements and the $\chi^2_{\rm{red}}$ values are fits assuming a constant magnitude in each filter. Approximately 66\% of Stripe 82 standards from \citet{IVEZ++07} satisfy this requirement. To obtain a sample that is  similar to TDSS, we use the region of Stripe 82 where $RA > 315^\circ$ or $RA < 60^\circ$. This avoids the $300^\circ < RA < 315^\circ$ region that is at low Galactic latitude and has a stellar density well above that typical in TDSS. Our variable object and standard catalogs have 12{,}523 and 411{,}219 sources, respectively. 

We divide our standard and variable object KDE spaces into 200 equally spaced bins in each of the three dimensions (i.e., 200$^3$ total bins). We set the bounds along each dimension to include the middle 99.8\% of our variable object training set. The remaining sources are placed in either the minimum or maximum bin as appropriate. We convolve our binned parameter space with a normalized, symmetric Gaussian filter with $\sigma$ = 5 bins (0.02 $\times$ 0.008 $\times$ 0.1 mags in ($S_1$, $S_2$, $S_3$)-space) so that regions with a small number of sources are filled uniformly as a continuous function. We normalize each KDE density so that it is effectively a probability density. 

To prioritize targets, we examine the smoothed, continuous, normalized (so that it integrates to unity) KDE density of Stripe 82 variable objects and standards, which we designate $\rho_{\rm{var}}$ and $\rho_{\rm{stan}}$. We assign each bin in ($S_1$, $S_2$, $S_3$)-space a KDE value, $E(S_1, S_2, S_3)$, defined as
\begin{equation}
E = \frac{\rho_{\rm{var}}}{\rho_{\rm{stan}}}.
\end{equation}
Areas of parameter space with the highest values of $E$ are the most efficient places to find variable objects and are initially assigned the highest priority. In Section \ref{sect:prioritization} we will discuss how our final target list does not strictly follow the $E$ value above. This quantity is, in principle, simple to relate to the probability of an object being a variable object:
\begin{equation}
P_{\rm{variable}} = \frac{E}{R+E},
\label{eq:PE}
\end{equation}
Where $R$ is the ratio of nonvariables to variable objects. In practice, $R$ depends on Galactic latitude and longitude, survey depth, observation cadence and the chosen threshold for variability. Different variability surveys could thus have wildly different values of $R$. Consulting color-based quasar selection, we estimate that an average of 2.4\% of sources which pass our data quality preselection in Eqs.\ \ref{eq:dataq} and \ref{eq:req} are quasars, which we generally assume to be variable objects. 58\% of objects in our Stripe 82 variable object catalog are quasars. We combine these numbers to estimate that approximately 4\% of objects which pass our preselection are variable objects. This leads to an estimate $R=25$, which we use in every region of the sky. While inaccuracies in $R$ will moderately affect our estimates of purity, they do not directly affect the actual targets we select.

\section{Prioritization of TDSS Variable Objects}\label{sect:prioritization}

Given our allotted fiber density across the sky, we seek a statistically uniformly selected target list of 10 TDSS-only targets deg$^{-2}$ across the entire TDSS area. To move from our 3D "efficiency" space defined in Stripe 82 to this uniform density target list, we divide the sky into equal area "pixels", determine a sensible threshold for our value $E$ (defined in Section \ref{sect:varmeas}), accept all targets that cross that threshold in the 20\% lowest target density pixels and randomly subsample targets which cross that threshold in the 80\% higher target density pixels. Our final sample is then uniform in the sense that objects everywhere pass the same $E$ threshold, but we use more subsampling in denser, low Galactic latitude areas. 

We start by dividing the sky into $2\times2$ degree square pixels. In each pixel, we assign an $E$ threshold that selects exactly 10 TDSS-only targets deg$^{-2}$ after removing the numerous targets shared with the eBOSS CORE quasar program, targets with previous SDSS spectroscopy and a small set of targets selected from the Palomar Transient Factory. We use "TDSS-only targets" to refer to objects selected for observation exclusively by TDSS and refer to the complete set of objects which satisfy our selection criteria as "total targets". We do not formally exclude the objects we share with the eBOSS CORE quasar sample, and they are part of the final TDSS survey. The distinction between these samples is made in our targeting procedure, because TDSS targets that are also in the eBOSS CORE quasar sample are not charged to our survey fiber allotment of 10 targets deg$^{-2}$. 

\begin{figure}[ht]
\includegraphics[width=0.98\columnwidth]{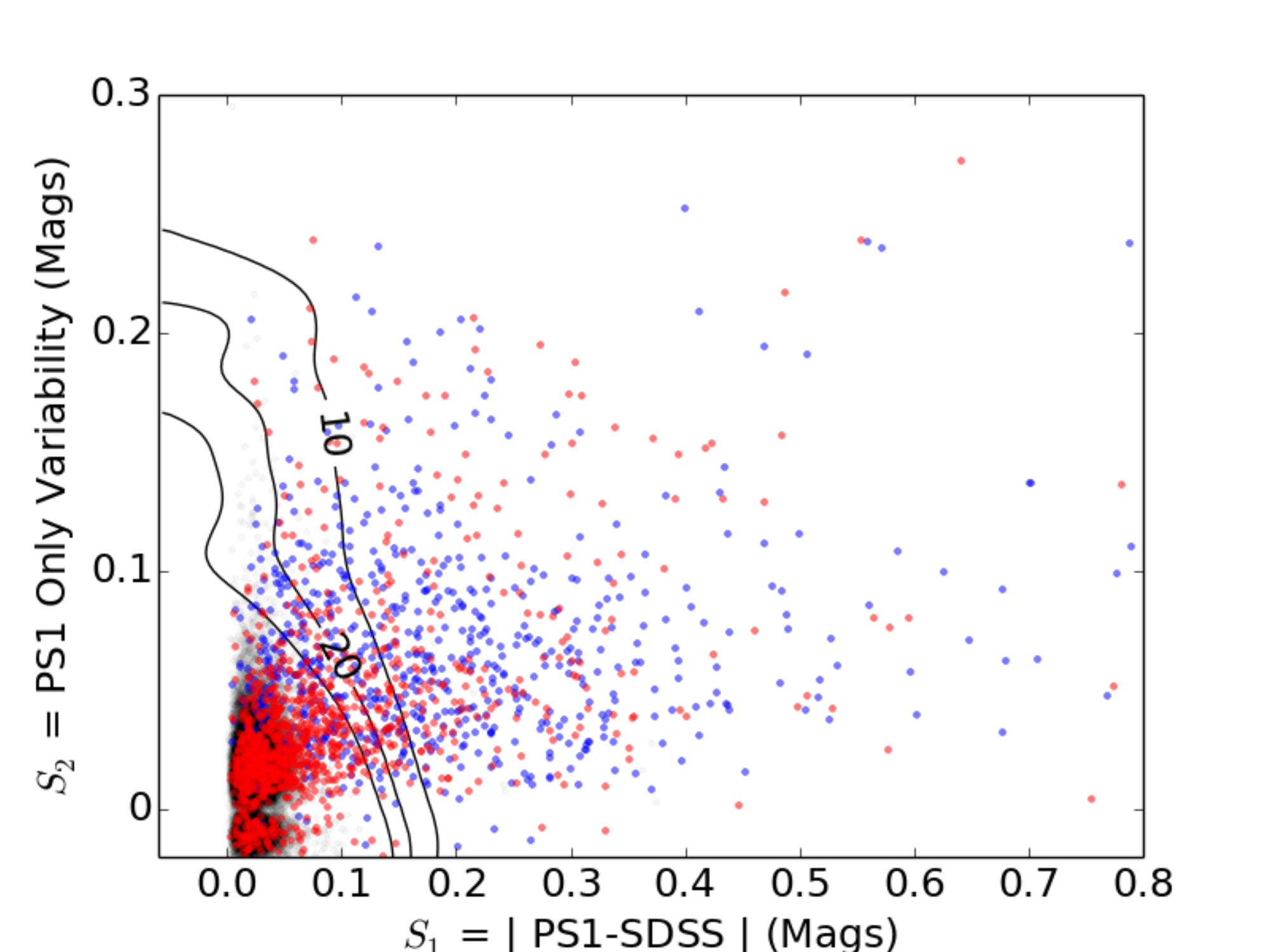}
\includegraphics[width=0.98\columnwidth]{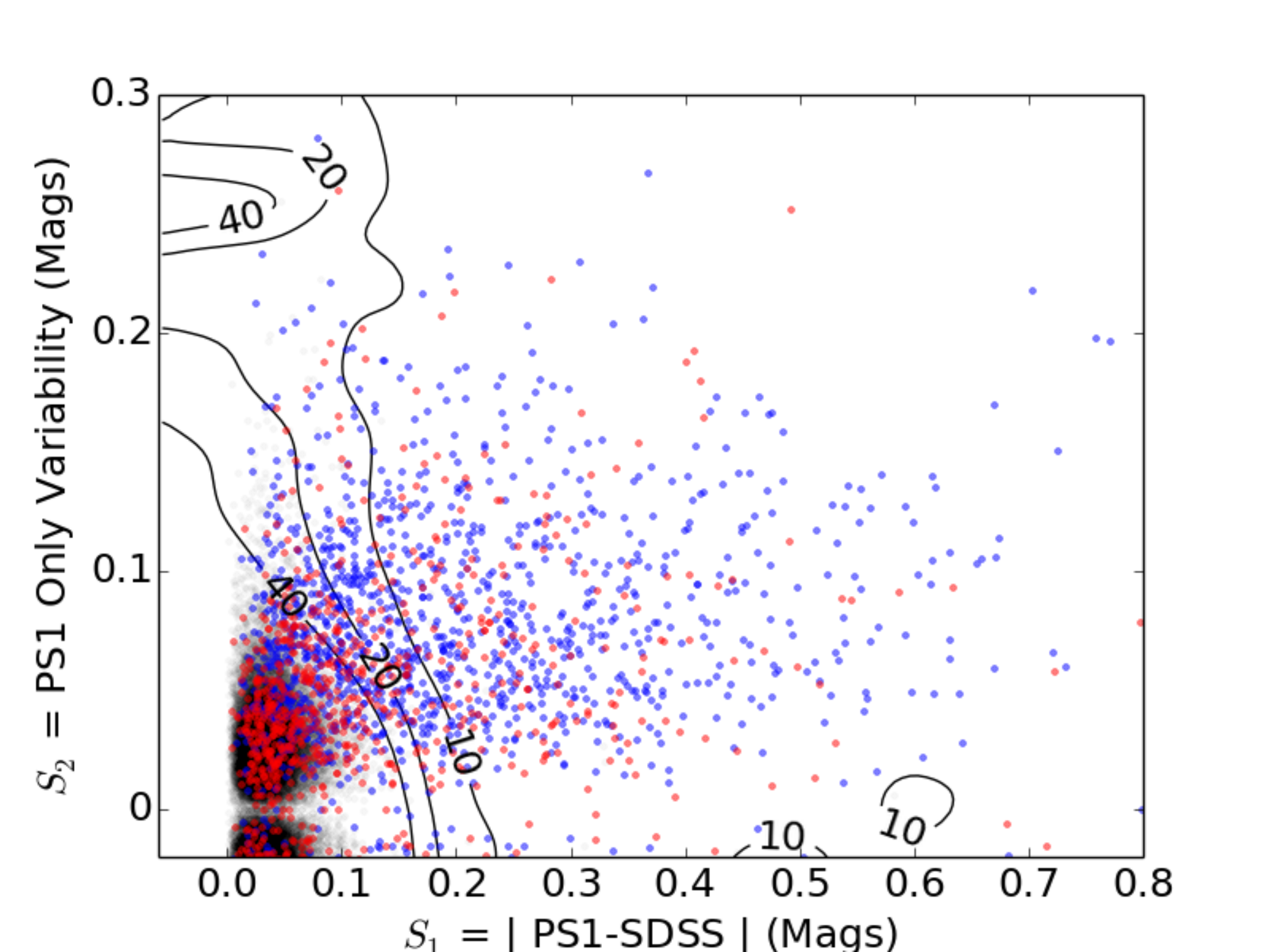}
\includegraphics[width=0.98\columnwidth]{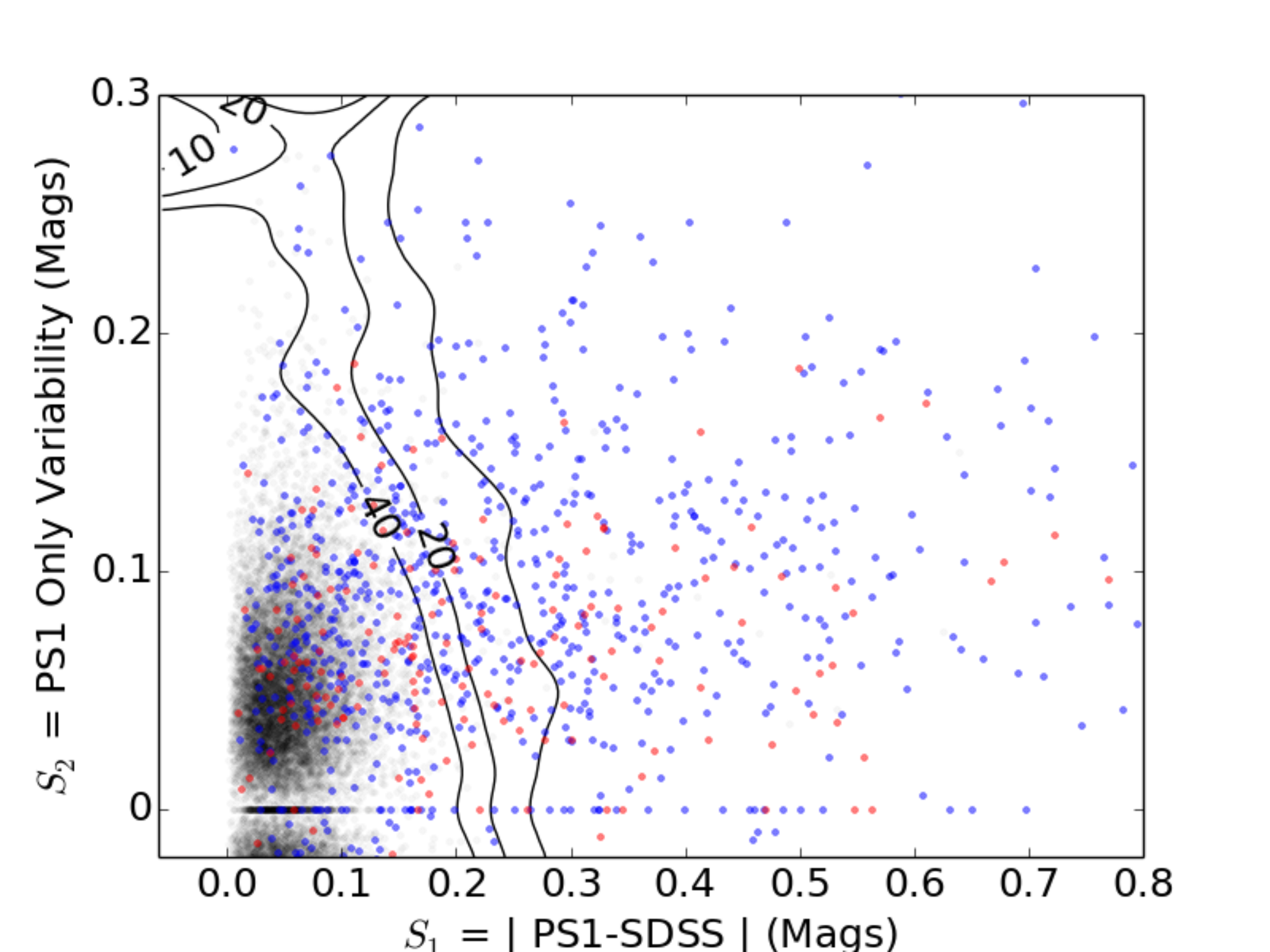}
\caption{\rm{2D cross section from our 3D KDE. The top, middle and bottom panels are cross sections centered around median magnitude, $S_3$ = 17.5, 18.5, 20, respectively. The contour labels specify the number of TDSS-only targets deg$^{-2}$ obtained with each cut. We show the Stripe 82 standards (black), variable quasars (blue) and variable non-quasars (red). Here quasars either have SDSS spectral type 'q' or $P(\rm{qso}) > 0.5$ according to the eBOSS CORE photometric quasar selection algorithm. The lack of points near the $S_2 = 0$ axis is due to the square root in the definition of $S_2$ and does not significantly affect the binned KDE.} }
\label{fig:contours}\end{figure}

In Fig.\ \ref{fig:contours} we show three cross sections of our 3D KDE taken from a large region ($135^\circ < RA < 150^\circ,\ 45^\circ < DEC < 60^\circ$) for statistical robustness. These cross sections demonstrate how selection varies in $S_1$ (|PS1-SDSS|) and $S_2$ (PS1 Variability) at different values of $S_3$ (median magnitude) where $S_1$, $S_2$ and $S_3$ are defined in Eq.\ \ref{eq:xyz}. The three density contours represent the cutoffs we use to obtain 10, 20 and 40 TDSS-only targets deg$^{-2}$. Our threshold in $S_1$ and $S_2$ expands outward at fainter magnitudes indicating, sensibly, that we require stronger variability to observe fainter objects, since they have larger error bars. Objects are generally required to vary by approximately 0.2 magnitudes to meet a 10 target deg$^{-2}$ limitation across most of the sky. Our KDE can fail in regions near the edge of our KDE parameter space where the density of both variable objects and standards is small. To avoid this problem, we assign any object with $S_1 > 0.5$ or $S_2 >  0.25$ a value of $E = 100$ if its $E$ does not already exceed 100. Only 15\% of our TDSS-only targets and 8\% of our total targets have $E$ assigned to 100.

\begin{figure}[ht]
\includegraphics[width=0.98\columnwidth]{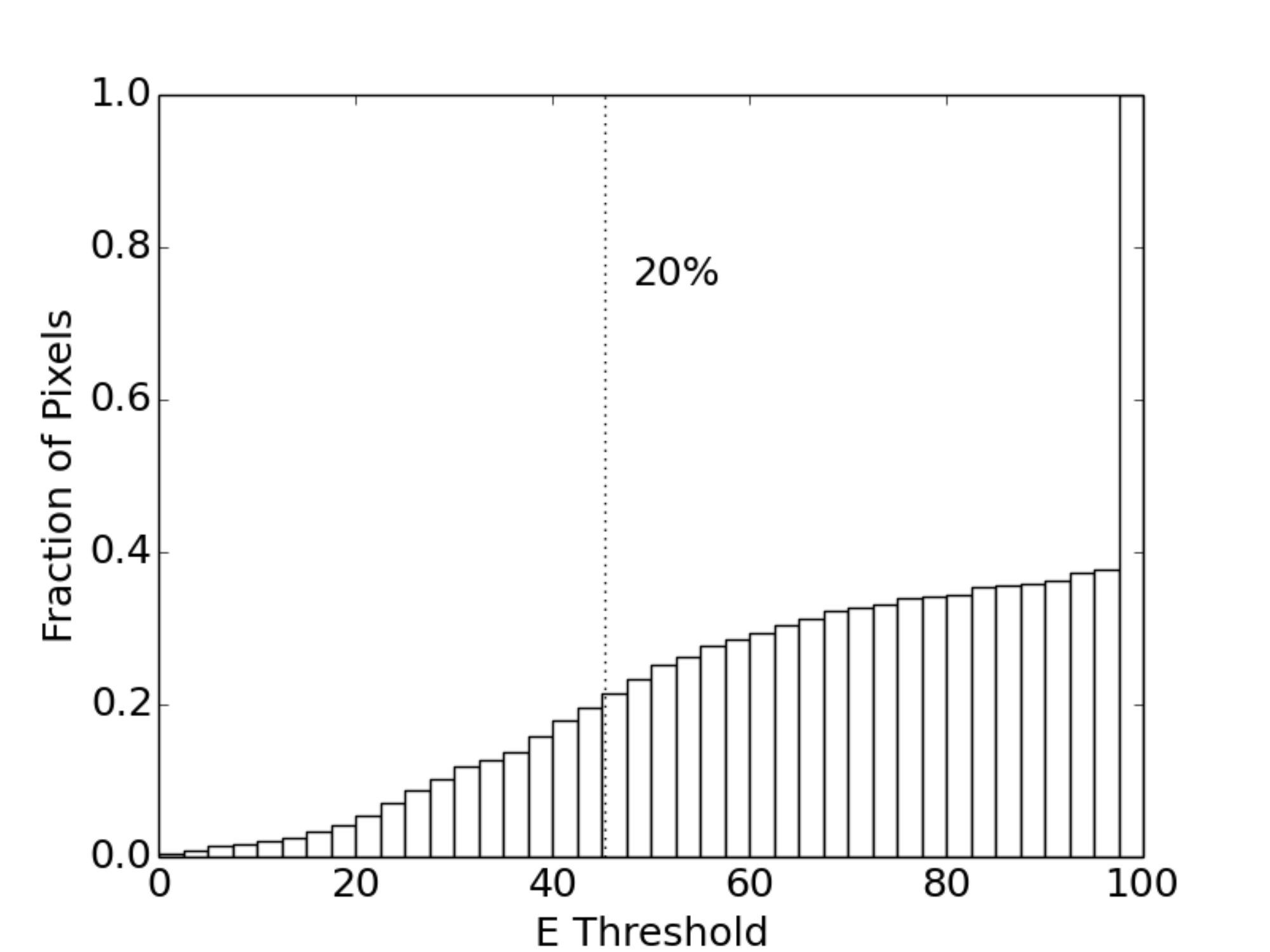}
\caption{\rm{The fraction of pixels with a 10 TDSS-only targets deg$^{-2}$ $E$ threshold less than a given value. Our global $E$ threshold, 45.4, is marked with a dotted line. Only 20\% of pixels have a 10 targets deg$^{-2}$ threshold less than 45.4. As noted in the text, 15\% of TDSS-only targets have their $E$ manually set to 100 which leads to the jump at $E = 100$.} }
\label{fig:thresh}\end{figure}

The KDE that underlies Fig.\ \ref{fig:contours} is derived exclusively from a fixed set of Stripe 82 standards and variable objects and can thus be applied to any area of the sky. However, the positions of the contours in Fig.\ \ref{fig:contours} corresponding to a particular target density are only applicable to a specific 135 deg$^2$ area of the sky. Different pixels across the sky will have different 10 targets deg$^{-2}$ thresholds (contour positions) corresponding to the variation in density of stellar variables (and stars more generally) across the sky. Fig \ref{fig:thresh} shows the distribution of 10 target deg$^{-2}$ $E$ thresholds across our pixels. A total of 80\% of pixels have a 10 targets deg$^{-2}$ $E$ threshold greater than 45.4, and we adopt this value as our nominal global $E$ threshold. Again, E($S_1$, $S_2$, $S_3$) is a static function defined by Stripe 82, so the only thing that changes across the sky is the density of objects with $E > 45.4$. Eq.\ \ref{eq:PE} states that the expected variable object purity of targets with $E = 45.4$ is 65\%. This is a lower bound on our sample purity, and our estimated purity (Section \ref{sect:purity}) is significantly higher.

\begin{figure}[ht]
\includegraphics[width=0.98\columnwidth]{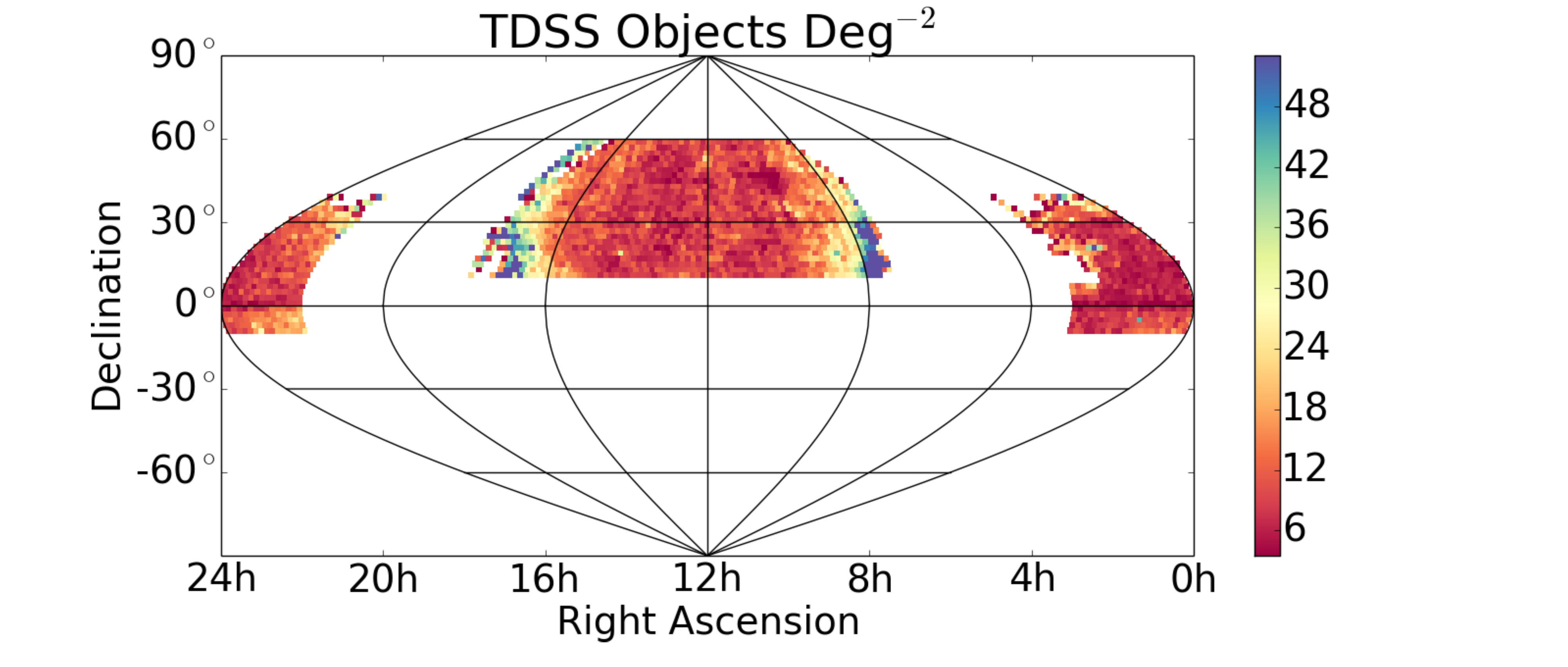}
\includegraphics[width=0.98\columnwidth]{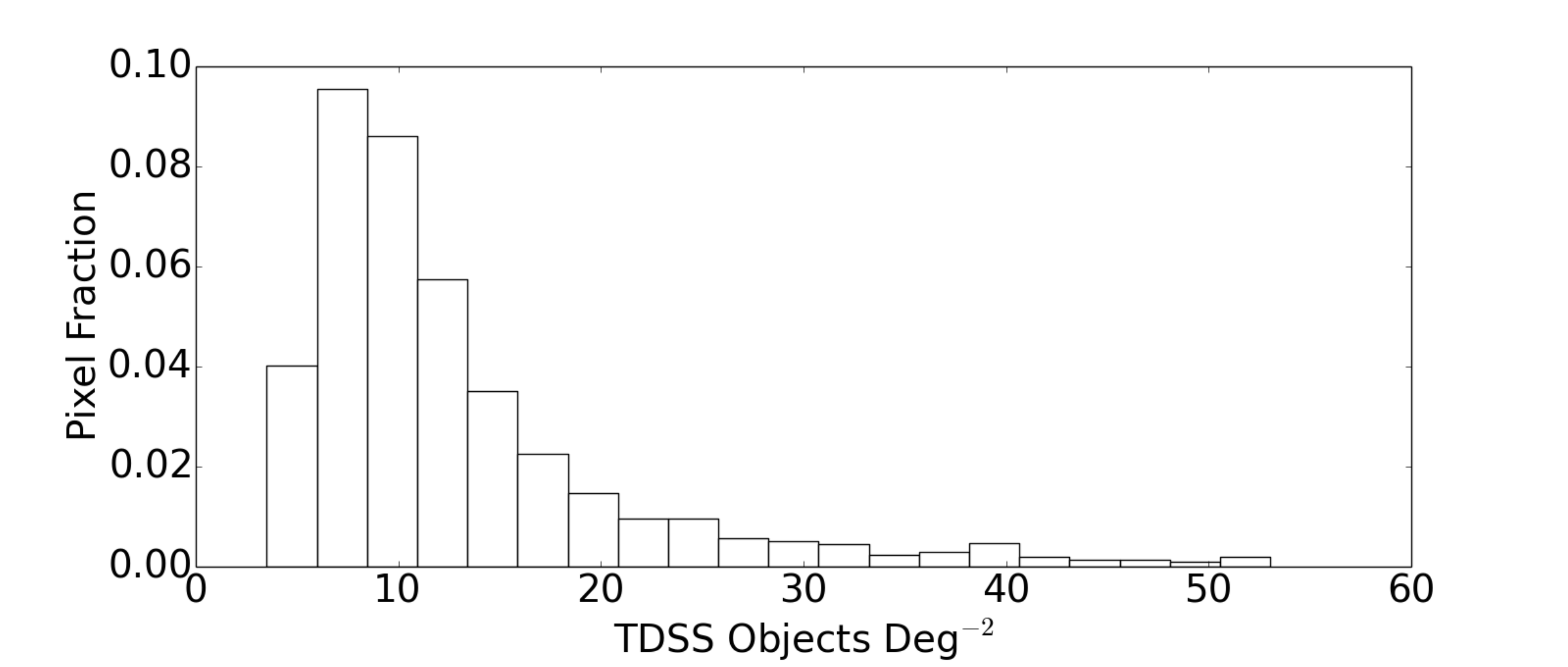}
\includegraphics[width=0.98\columnwidth]{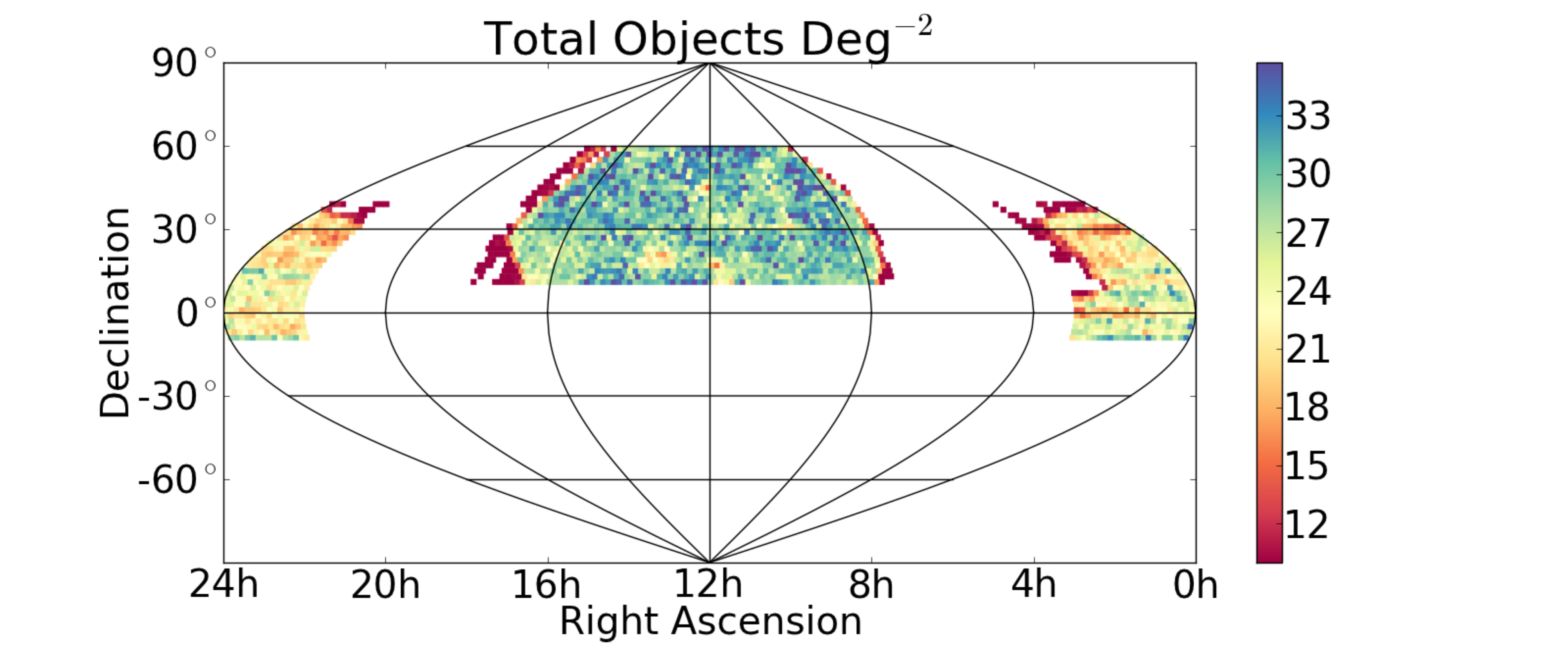}
\includegraphics[width=0.98\columnwidth]{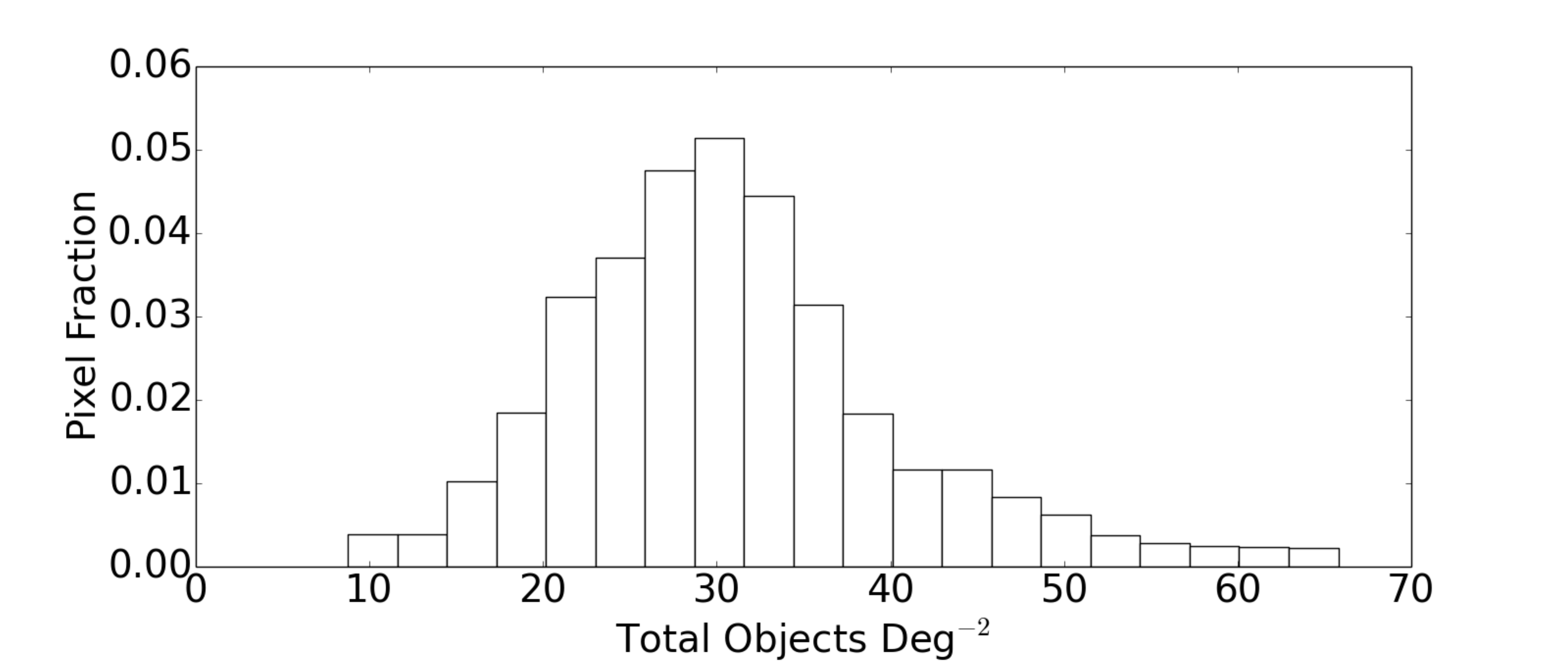}
\caption{\rm{Map showing the density of TDSS-only objects (excluding CORE quasars and objects with previous spectroscopy) that exceed the $E = 45.4$ threshold in each $2\times2$ degree pixel and the distribution of these densities (top 2 panels). Map showing the total density of all objects that exceed the $E = 45.4$ threshold in each pixel, including objects shared with the eBOSS CORE quasar sample and objects with previous spectra, and the corresponding distribution of these densities (bottom two panels). } }
\label{fig:maps}\end{figure}

Fig.\ \ref{fig:maps} presents the distribution of the density of TDSS-only objects (those objects not selected as part of the eBOSS CORE quasar sample and not having previous spectroscopy) and of all objects that cross the $E = 45.4$ threshold in each pixel. The density rises precipitously in the low Galactic latitude regions at the edges of our survey. This result simply implies that a significant fraction of our variable objects are stars that become more common at lower Galactic latitude. Some of the pixels with low target density are near the very edge of our estimated survey bounds. Many of these pixels will not be included in the actual spectroscopic survey. Globally, the average density of TDSS-only targets with $E> 45.4$ is 14 deg$^{-2}$, and we are sparse sampling 70\% of these sources. The majority of pixels with fewer than 10 targets deg$^{-2}$ with $E > 45.4$, have at least 8 targets deg$^{-2}$, so only a small number of spectra of $E < 45.4$ objects will be taken.

Having established a threshold, we must still determine how to make a uniform target list with 40 TDSS-only targets (10 deg$^{-2}$) in each 4 deg$^2$ pixel. In the 20\% of pixels with 40 or fewer TDSS-only targets that cross the $E$ threshold, we simply select the 40 targets with the highest $E$ estimate (a small fraction of which have $E < 45.4$). In the 80\% of pixels with more than 40 targets, we prioritize a small number of hypervariable targets (described in Subsection \ref{sect:hyper}) and then assign a random priority to the remaining targets with $E > 45.4$, choosing the targets with the highest priority until we reach our 10 deg$^{-2}$ target quota.

Our final step to produce a target list is to visually inspect every object's SDSS image. Visual inspection was performed by authors Morganson, Green, Anderson and Ruan. Objects judged to have significant flux from nearby neighbors, objects with unflagged processing errors or objects within approximately $30''$ of a diffraction spike are removed. Lower priority objects rise in the queue naturally, with $E < 45.4$ objects being prioritized directly by $E$ value. The fraction of objects removed by visual inspection ranges between 5\% and (rarely) 30\%. The rejection fraction is highest at low Galactic latitudes where there are many very bright stars (that can influence photometry over distances of several arcminutes)  and close stellar pairs (unresolved in the SDSS catalog). Fortunately, these regions also have an abundance of high $E$ targets. 

\subsection{TDSS Prioritization of Hypervariables}\label{sect:hyper}

While the main goal of TDSS is to provide a statistically uniform sample of variable objects, TDSS also provides a unique opportunity to obtain a statistical sample of the most variable objects in the sky, which we designate as hypervariables. While most of these hypervariables would be observed naturally as part of the survey, we wish to ensure that hypervariables which vary above a particular threshold are all observed, regardless of the local target density. Our KDE method is not well-designed to select hypervariables, since the extreme regions of variability space are poorly populated by either variable objects or standards. Instead we reduce our variability parameters from Eq.\ \ref{eq:xyz} to a single parameter:
\begin{eqnarray}
V &=& \left(\rm{median}(|\rm{mag}_{\rm{PS1}}-\rm{mag}_{\rm{SDSS}}|)^2+4\ \rm{median}(Var_{\rm{PS1}})^2\right)^{1/2},\label{eq:V}\\
  &=& \left(S_1^2+4 S_2^2\right)^{1/2}.\nonumber
\end{eqnarray}
This is an elliptical contour of approximately constant density in our ($S_1$, $S_2$) variability space. The factor of 4 accounts for the fact that $x$, the SDSS-PS1 difference, is generally of order twice $y$, the PS1 only variability. This ratio is not exact and is specific to this data. It is likely due to the longer time scales of the SDSS-PS1 difference. 

\begin{table}
\begin{tabular}{cccccccc}
        \hline
$V_{\rm{Thresh}}$ & $H$ & $H_{\rm{QSO}}$ & $H_{*}$ & $H_{\rm{CORE}}$ & $H_{\rm{prev}}$ & $H\ \rm{deg}^{-2}$ & $H_{\rm{low}}\ \rm{deg}^{-2}$ \\
        \hline
1.0 & 8411 & 7 & 6489 & 1274 & 640 & 1.05 & 1.54\\
1.2 & 4784 & 3 & 4046 & 481 & 253 & 0.60 & 0.91\\
1.4 & 3069 & 1 & 2725 & 224 & 118 & 0.38 & 0.55\\
1.6 & 2071 & 0 & 1899 & 120 & 51 & 0.26 & 0.42\\
1.8 & 1492 & 0 & 1375 & 80 & 36 & 0.19 & 0.24\\
2.0 & 1108 & 0 & 1033 & 52 & 22 & 0.14 & 0.19\\
2.2 & 823 & 0 & 778 & 29 & 15 & 0.10 & 0.15\\
2.4 & 629 & 0 & 595 & 21 & 12 & 0.08 & 0.10\\
2.6 & 483 & 0 & 463 & 12 & 8 & 0.06 & 0.08\\
2.8 & 401 & 0 & 385 & 10 & 6 & 0.05 & 0.06\\
3.0 & 338 & 0 & 323 & 9 & 6 & 0.04 & 0.06\\
        \hline
\end{tabular}
\caption{\rm{Estimated total number of hypervariables ($H$) that pass different thresholds of V ($V_{\rm{Thresh}}$ from Eq.\ \ref{eq:V}). We also show the expected number of TDSS-only quasars ($H_{\rm{QSO}}$), TDSS-only non-quasars ($H_{*}$), quasars shared with the CORE quasar group ($H_{\rm{CORE}}$) and targets with previous SDSS spectra ($H_{\rm{prev}}$). The last two columns are the total density (deg$^{-2}$) of hypervariables across the whole survey and the density in a low Galactic latitude region ($120^\circ < \rm{RA} < 130^\circ, 10^\circ < \rm{DEC} < 20^\circ$).}}\label{tab:hyper}
\end{table}

\begin{figure}[ht]
\includegraphics[width=0.98\columnwidth]{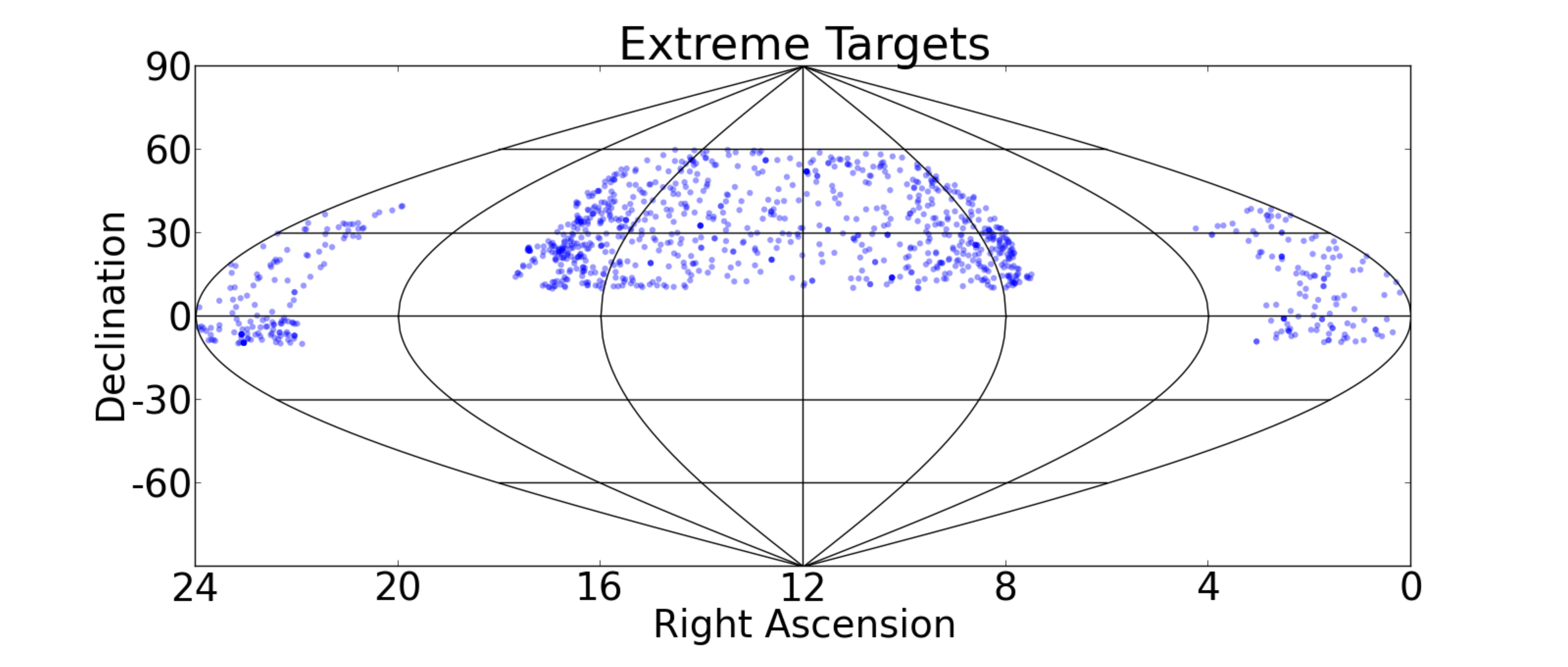}
\caption{\rm{Map showing the locations of $V > 2.0$ hypervariables across the sky in equatorial coordinates. } }
\label{fig:hypermap}\end{figure}

Table \ref{tab:hyper} lists the number of hypervariables as a function of different thresholds of $V$ (Eq.\ \ref{eq:V}). Using the densities presented here and the density map shown in Fig.\ \ref{fig:hypermap}, we set our $V$ threshold to 2.0 magnitudes. This choice yields 1{,}108 targets, most of which would have likely been observed naturally by our KDE selection method. Globally, this population density is 0.14 deg$^{-2}$ but in a representative low Galactic latitude region ($120^\circ < \rm{RA} < 130^\circ, 10^\circ < \rm{DEC} < 20^\circ$), the density of hypervariables is 0.19 deg$^{-2}$, 2\% of our low latitude targets. While the majority of targets TDSS selects have colors consistent with being quasars, approximately 95\% of our hypervariables do not. We will briefly investigate the likely identities of hypervariables in Section \ref{sect:hypercolor}.

\section{Anticipated Purity}\label{sect:purity}

When evaluating our selection criteria, we were primarily concerned with the purity of all targets selected at a given threshold,  $P_{\rm{tot}}$, and the purity of our TDSS-only target list after eBOSS CORE quasars and targets with previous SDSS spectroscopy are removed, $P_{\rm{tar}}$. We estimate the anticipated purity of our total sample using our results in Stripe 82. Specifically, building on Eq.\ \ref{eq:PE}, we define:
\begin{eqnarray} 
P_{\rm{tot}} &=& \frac{f_{\rm{var\ S82}}}{ R\ f_{\rm{stan\ S82}}+f_{\rm{var\ S82}}},\\
R &=& 25,\nonumber\\
P_{\rm{tar}} &=& \frac{f_{\rm{tar}} f_{\rm{var\ S82}}}{ R\ f_{\rm{stan\ S82}}+ f_{\rm{tar}} f_{\rm{var\ S82}}},\label{eq:purity}\\
f_{\rm{tar}} &=& \frac{N_{\rm{tar}}} {N_{\rm{tar}}+N_{\rm{CORE}}+N_{\rm{prev}}}.\nonumber
\end{eqnarray} 
Here, $f_{\rm{var\ S82}}$ and $f_{\rm{stan\ S82}}$ are the fraction of Stripe 82 variable objects and standards (as defined in section \ref{sect:varmeas}) that pass a given threshold. The quantity $R$ is the expected ratio of nonvariables to variable objects discussed in Section \ref{sect:varmeas}, and $f_{\rm{tar}}$ is the fraction of objects which pass a given threshold that will be TDSS-only targets. $N_{\rm{tar}}$, $N_{\rm{CORE}}$ and $N_{\rm{prev}}$ are the numbers of TDSS-only targets, eBOSS CORE quasar targets and objects with previous SDSS spectroscopy that pass a given threshold, respectively. The sum $N_{\rm{tar}}+N_{\rm{CORE}}+N_{\rm{prev}}$ is the total number of objects that pass a given threshold. 

\begin{table*}
\begin{tabular}{cccccccccc}
        \hline
$N_{\rm{tar\ 20}}$ & $N_{\rm{tar\ test}}$ & $N_{\rm{QSO}}$ & $N_{*}$ & $N_{\rm{lovar}}$ & $N_{\rm{CORE}}$ & $N_{\rm{prev}}$ & $N_{\rm{tot}}$ & $P_{\rm{tar}}$ & $P_{\rm{tot}}$ \\
        \hline
60 & 67.8 & 3.3 & 16.7 & 47.7 & 18.6 & 27.4 & 113.7 & 45.2 & 58.0 \\
50 & 56.5 & 3.0 & 16.6 & 36.8 & 18.0 & 25.8 & 100.3 & 49.2 & 63.3 \\
40 & 45.4 & 2.7 & 16.4 & 26.4 & 17.2 & 24.2 & 86.8 & 54.5 & 69.6 \\
30 & 35.2 & 2.4 & 15.8 & 17.0 & 16.2 & 22.4 & 73.8 & 61.4 & 76.9 \\
20 & 23.7 & 2.1 & 14.1 & 7.5 & 14.6 & 19.8 & 58.1 & 73.3 & 87.1 \\
10 & 11.3 & 1.4 & 8.3 & 1.7 & 10.8 & 14.6 & 36.7 & 86.4 & 95.4 \\
        \hline
\end{tabular}
\caption{\rm{Estimated target counts and purities from Stripe 82 tests at different variability cutoffs. All counts are deg$^{-2}$. All purities are percentages. $N_{\rm{tar\ 20}}$ is the number of targets in the 20th percentile pixel for a given threshold while $N_{\rm{tar\ test}}$ is the number of targets in our test field. $N_{\rm{QSO}}$, $N_{*}$ and $N_{\rm{lovar}}$ are the estimated numbers of TDSS-unique quasars, stars and low-variability objects, respectively. $N_{\rm{CORE}}$ and $N_{\rm{prev}}$ are the estimated numbers of objects we share with the CORE quasar sample or have previous SDSS spectroscopy. $N_{\rm{tot}}$ is the total number of candidates.  $P_{\rm{tar}}$ and $P_{\rm{tot}}$} are the estimated purities of our TDSS-only targets and our total targets, respectively.}\label{tab:s82res}
\end{table*}

To assess the performance of our variable object selection algorithm, we tested it on a large, representative patch of sky, the $135^\circ < RA < 150^\circ,\ 45^\circ < DEC < 60^\circ$ region previously mentioned in Section \ref{sect:varmeas}. We used this region to set our selection threshold to obtain 10, 20 ... 60 TDSS-only targets deg$^{-2}$ in Table \ref{tab:s82res}. Having set a threshold to obtain a known density of targets, cross-matched with eBOSS and SDSS databases to remove eBOSS CORE quasars and objects with previous spectra and calculated purity with Eq. \ref{eq:purity}, we can estimate the number of low variability sources that scatter into our selection space:
\begin{equation}
N_{\rm{lovar}}=N_{\rm{tar}}(1-P_{\rm{tar}}).
\end{equation}

To estimate our quasar fraction, we assume that everything in the color box:
\begin{eqnarray}
u_{\rm{SDSS}} - g_{\rm{SDSS}} &<& 0.8,\nonumber\\
g_{\rm{SDSS}} - r_{\rm{SDSS}} &<& 0.65\label{eq:qsobox}
\end{eqnarray}
is a quasar.

There are
\begin{equation}
N_*=N_{\rm{tar}}-N_{\rm{QSO}}-N_{\rm{lovar}}
\end{equation}
remaining objects which are expect to be mostly stellar variable objects. We also calculate $N_{\rm{CORE}}$, $N_{\rm{prev}}$ and $N_{\rm{tot}}$. 

Table \ref{tab:s82res} demonstrates how various quantities change as we lower our target selection threshold. The thresholds are set so that the 20th percentile pixel (as described in Section \ref{sect:prioritization}) has $N_{\rm{tar\ 20}}$. We actually derive our statistics from our larger test region, which is quite similar to the 20th percentile pixel (the density of targets in this region is $N_{\rm{tar\ test}}$). We acquire 10 TDSS-only spectra deg$^{-2}$, so the final line is most useful. The TDSS selection algorithm at that surface density produces a target list that is $P_{\rm{tot}}$ = 95\% pure. Many of these targets are shared with the eBOSS CORE quasar sample or have previous SDSS spectra. After these targets are removed, the TDSS-only targets are $P_{\rm{tar}}$ = 88\% pure. Note that "low variability" sources are sources that did not vary in the Stripe 82 data. Some unknown, but likely significant, fraction of these sources are true variable objects that varied during the PS1 epochs or between SDSS and PS1. In addition, our visual inspection removes a significant fraction of non-variable objects that is not accounted for in this analysis.

Since most of the quasars we select are shared with the eBOSS CORE quasar sample, the TDSS-only targets are approximately 90\% non-quasars (mostly variable stars). In practice, we expect to find a significant fraction of unusual quasars with colors not described by Eq.\ \ref{eq:qsobox}, so precise estimates of the quasar fraction will require spectra. Our data and selection method are optimized for 10 TDSS targets deg$^{-2}$. If we were to expand our target list to the 20 targets deg$^{-2}$ threshold, we would be selecting 6.5 additional variable objects and 5.6 additional standards in our test field (roughly 5.4 variable objects and 4.6 standards in our 20\% field). Our additional targets would be only 54\% pure. Selecting a "deeper" set of variable objects with high purity likely requires higher precision PS1 data or significantly better-sampled light curves. 


\section{TDSS Selection of $i$-Band Dropouts}\label{sect:idrops}

TDSS strives to produce a sample of variable objects that is unbiased in color space. We make one small exception for $i$-dropouts, objects that are observed in the $z$ band but are either not observed in any bluer bands or have extremely large $i-z$ colors. Among known astrophysical $i$-dropouts are late M and L-type dwarfs and $z \approx 6$ quasars, all of which are rarely detected and may have interesting variability properties. Our two filter requirement would exclude these objects if we did not create a separate pipeline to identify them.

Our $i$-dropout selection method closely follows the main selection method described in Section \ref{sect:varmeas}. First, we make an initial database level cut: 
\begin{eqnarray}
i_{\rm{SDSS}}-z_{\rm{SDSS}} &>& 1.0\label{eq:idreq}\\
\rm{err}_{z_{\rm{SDSS}}} &<& 0.1,\nonumber\\
\rm{err}_{i_{\rm{SDSS}}},\ \rm{err}_{r_{\rm{SDSS}}},\ \rm{err}_{g_{\rm{SDSS}}} &>& 0.1,\nonumber\\
r_{22} &>& 5'',\nonumber\\
r_{17} &>& 10'',\nonumber\\
r_{15} &>& 20'',\nonumber\\
r_{13} &>& 30'',\nonumber\\
n_{\rm{PS1\ z}},\ n_{\rm{PS1\ y}} &>& 3.\nonumber
\end{eqnarray}
The first three requirements are all purely SDSS-based and are designed to find $i$-dropouts while excluding any sources found in the main sample. The next four requirements remove objects whose photometry has likely been altered by a nearby bright object. The final requirement ensures that we have sufficient PS1 data to make a variability measurement. These criteria yield 11{,}594 sources with typical limiting magnitudes of $z < 19.9$, $y < 19.7$. These requirements also ensure that the objects are real and not just cosmic rays or other artifacts in a single $z$ band image, which can be problematic for $i$-dropout searches \citep{FAN++01,MORG++12}.

\begin{figure}[ht]
\includegraphics[width=0.98\columnwidth]{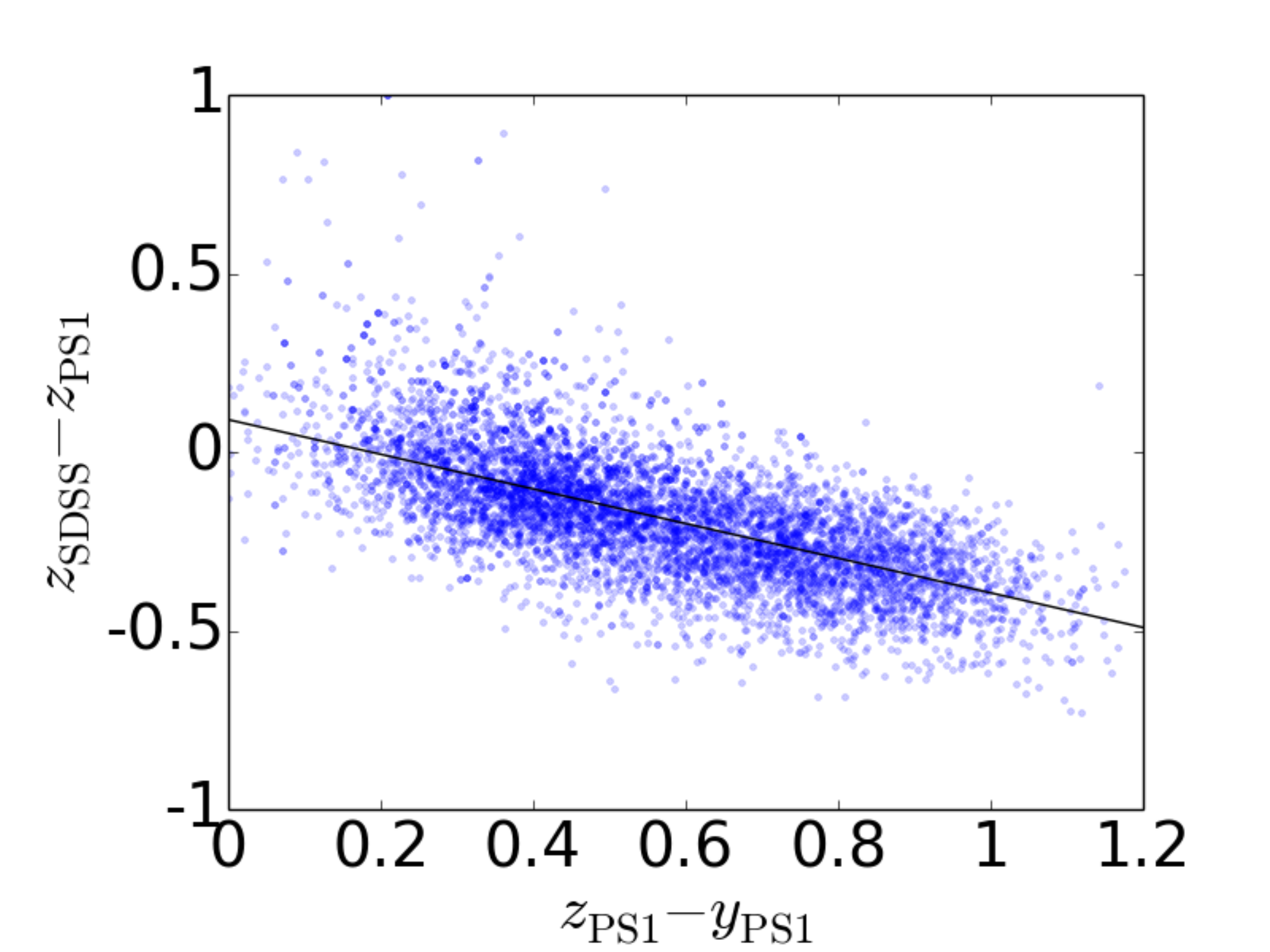}
\caption{\rm{The $z_{\rm{SDSS}}-z_{\rm{PS1}}$ differences of $i$-dropouts versus their $z_{\rm{PS1}}-y_{\rm{PS1}}$ colors. The line was defined using a linear minimum absolute deviations fit, and we use it to compare individual $z_{\rm{PS1}}$'s to $z_{\rm{SDSS}}$'s.} }
\label{fig:zzzy}\end{figure}

To select long term variable objects, we would naturally wish to use the SDSS-PS1 $z$ magnitude difference. Our filter transformations in Eq.\ \ref{eq:spcon}, however, are not designed to work with $i$-dropouts, which are bound to have extreme colors, so we must derive our own SDSS-PS1 filter corrections. In Fig.\ \ref{fig:zzzy}, we fit $\Delta z = z_{\rm{SDSS}}-z_{\rm{PS1}}$ versus $zy = z_{\rm{PS1}}-y_{\rm{PS1}}$ as a line, $\Delta z = a + b\ zy$, by minimizing the absolute deviations: 
\begin{eqnarray}
S&=&\sum_i \left|\frac{\Delta z_i - (a+b\ zy_i)}{\sigma_i}\right|,
\end{eqnarray}
yielding
\begin{eqnarray}
a&=&0.141,\nonumber\\
b&=&-0.525.\nonumber
\end{eqnarray}
Minimizing the absolute deviations is more robust to outliers than a typical $\chi^2$ method. With this linear fit, we can define an expected $z_{\rm{SDSS}}$ given PS1 colors:
\begin{equation}
z_{\rm{SDSS}}* = z_{\rm{PS1}}+ a+b(z_{\rm{PS1}}-y_{\rm{PS1}}).
\end{equation}
$z_{\rm{SDSS}}*-z_{\rm{SDSS}}$ is 0 for a typical $i$ dropout in  our sample. With this correction, we can define a 2D KDE parameter space analogous to the first two dimensions of our main selection KDE in Eq.\ \ref{eq:xyz}
\begin{eqnarray}
S_1 &=& |z_{\rm{SDSS}}*-z_{\rm{SDSS}}|,\label{eq:xy}\\
\rm{Var}_{\rm{PS1}} &=& \rm{Variance}_{\rm{PS1}}-\rm{Err}_{\rm{PS1}}^2(n_{\rm{PS1}}-1),\nonumber\\
\rm{Var}_{zy\ \rm{PS1}} &=& 0.5(\rm{Var}_{z\ \rm{PS1}}+ \rm{Var}_{y\ \rm{PS1}}),\nonumber\\
S_2 &=& \rm{sign}(\rm{Var}_{zy\ \rm{PS1}}) |\rm{Var}_{zy\ \rm{PS1}}|^{1/2}.\nonumber
\end{eqnarray}

\begin{figure}[ht]
\includegraphics[width=0.98\columnwidth]{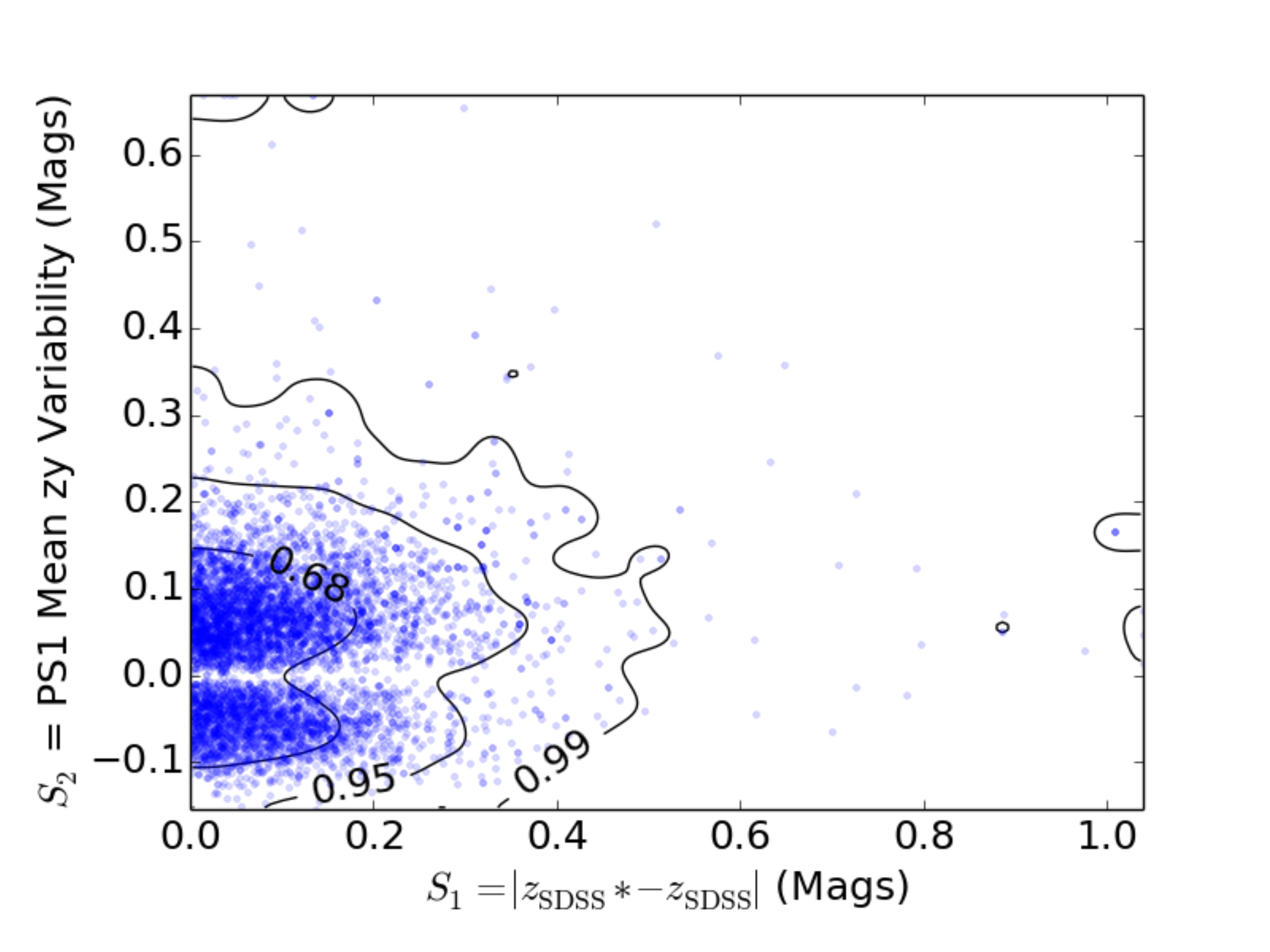}
\caption{\rm{The locations of the $i$-dropouts in the 2D KDE space defined in Eq. \ref{eq:xy}. We selected variable targets from outside the 95\% contour.} }
\label{fig:contoursi}\end{figure}
Figure \ref{fig:contoursi} shows our 2D KDE variable $i$-dropout selection space. The objects which satisfy the criteria in Eq.\ \ref{eq:idreq} tend to be at the faint end of their selection space with magnitude errors near the 0.1 magnitude limit. Their distribution is correspondingly more broad than that of the sources in Figure \ref{fig:contoursi}. We lack a large sample of confirmed variable $i$-dropouts and therefore cannot produce a training set as we did for the main population. Instead, we select the 5\% outliers in variability space. There is a small population of sources with negative PS1 variability (in which the standard deviation is less than what one would expect from the error bars as described in Section \ref{sect:varmeas}) and only moderate PS1-SDSS difference. To avoid "rewarding" sources for having negative PS1 variability, an area of parameter space that is rare, but not particularly likely to indicate true variability, we eliminate sources that satisfy
\begin{equation}
S_1 < 0.6,\ S_2 < 0,
\end{equation}
where $S_1$ and $S_2$ are defined in Eq.\ \ref{eq:xy}. In total, 221 $i$-dropouts satisfy our selection criteria. Of these, only 73 pass our visual inspection and are included in the TDSS target list. 

Only seven previously discovered $z \approx 6$ quasars \citep{FAN++01,FAN++06,MORG++12,BANA++14} satisfy our initial selection criteria, and only one passes our variability threshold. This is not entirely surprising as cosmological time dilation will significantly reduce any observed variability from these quasars. In addition, \citet{MACL++10}, \citet{MORG++14} and others have found significant anticorrelation between quasar variability and luminosity, and $z \approx 6$ quasars detected by SDSS are necessarily extremely luminous.

\section{Photometric Classification of All TDSS Targets}\label{sect:phot}

The algorithm described in Sections \ref{sect:varmeas} and \ref{sect:prioritization} produces a target list that includes 242{,}513 objects. This list has approximately 10\% more sources than we will be able to target spectroscopically due to a combination of extra area and extra density. Nevertheless, the fractions of different classes of objects in this list should closely resemble the final spectroscopic sample. While we do not make any explicit use of color in our variable object selection, classifying our objects by color will allow us to anticipate our final results. We do not attempt to correct for Milky Way redenning in our photometry, because colors do not directly influence our selection. Since the TDSS area is at high Galactic latitude, this only introduces a small error on our color measurements.

\begin{figure}[ht]
\includegraphics[width=0.98\columnwidth]{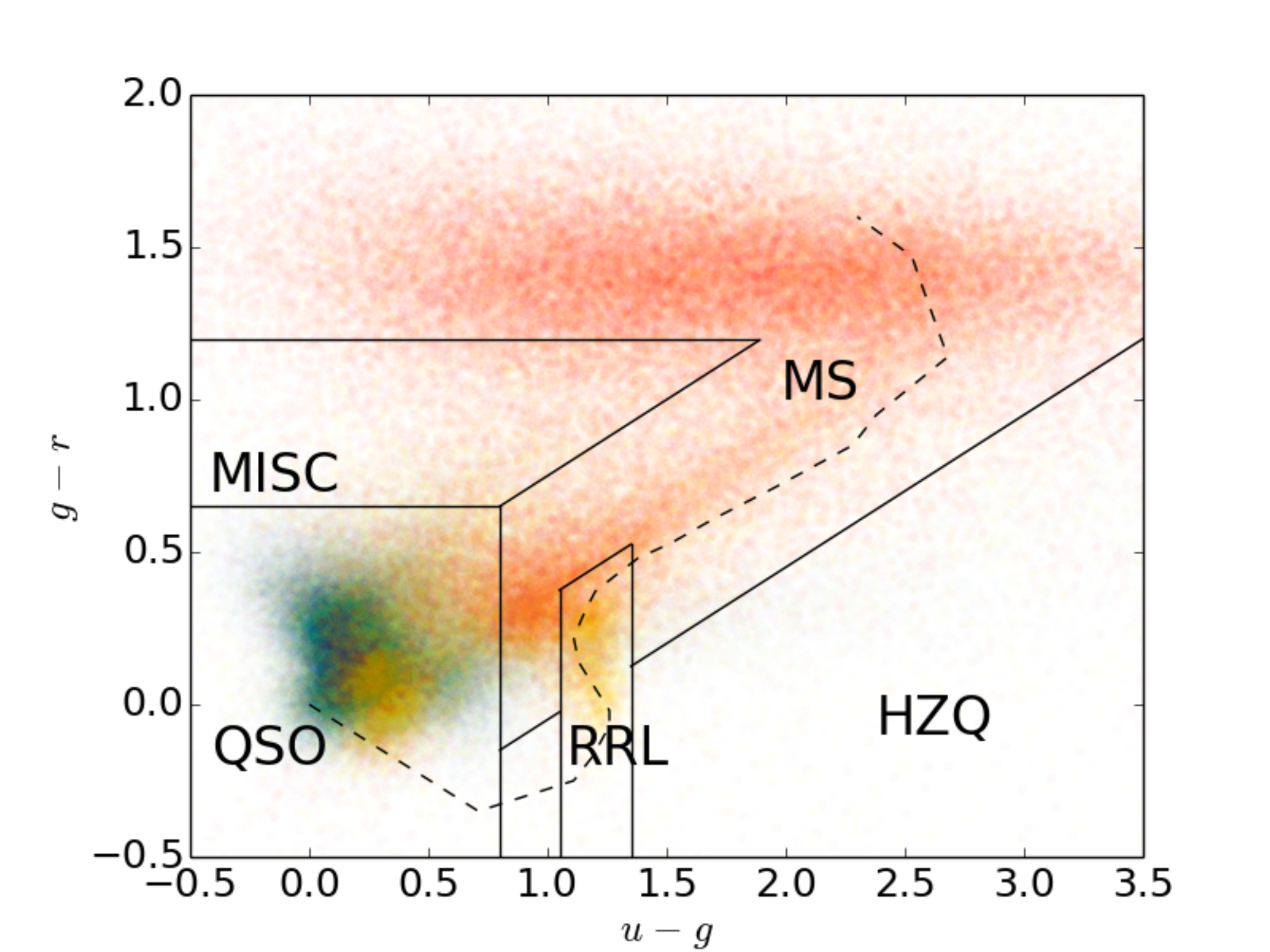}
\includegraphics[width=0.98\columnwidth]{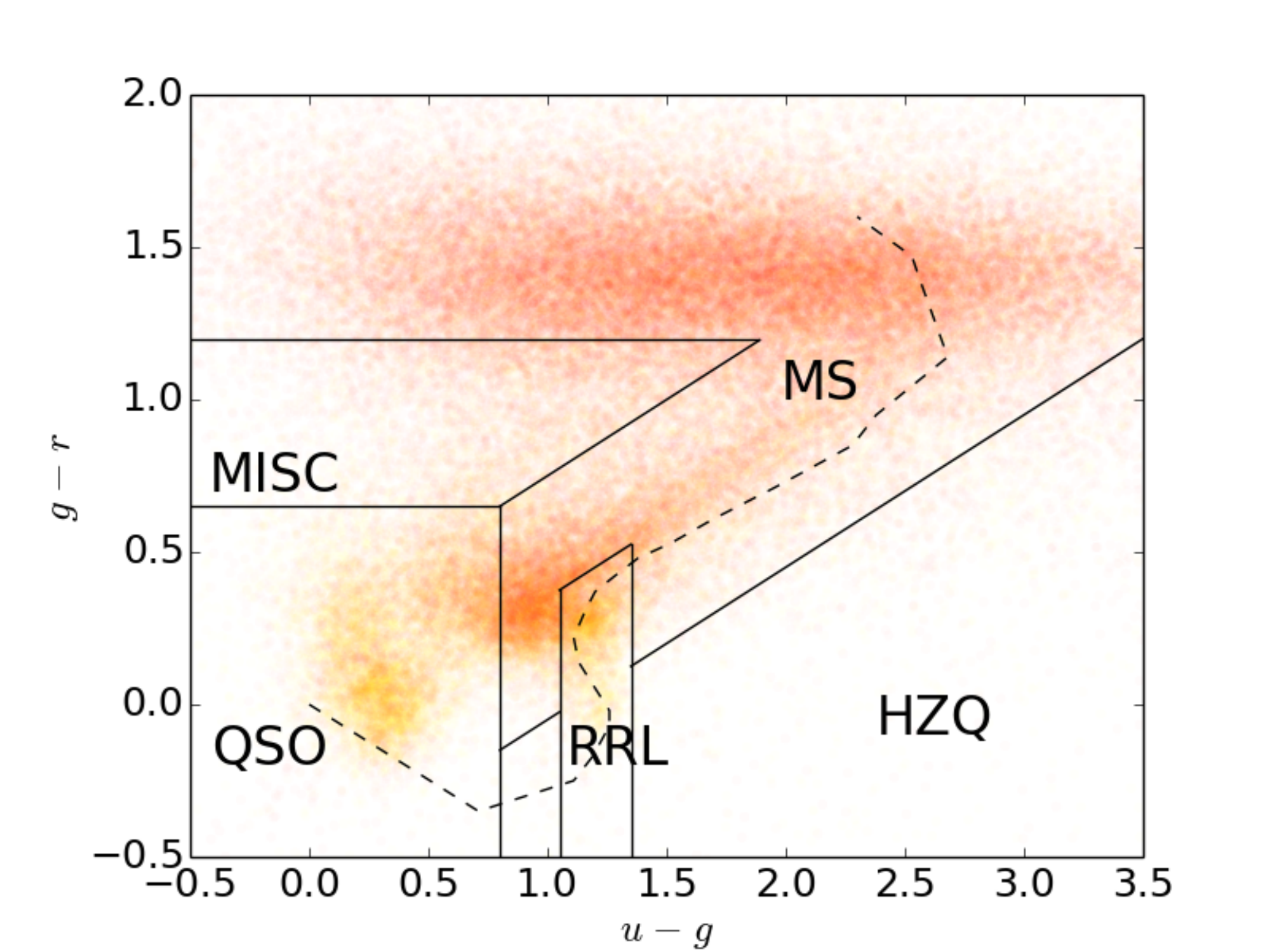}
\caption{\rm{The SDSS $g-r$ versus $u - g$ distribution of all TDSS targets (top) and TDSS-only targets after we remove CORE quasars and objects with previous SDSS spectroscopy (bottom). Low (high) priority eBOSS color-selected quasars are in green (blue). Low (high) priority non-quasars are in red (yellow). The QSO, MS, RRL and HZQ regions are the areas of color space that contain most quasars, main sequence stars, RR Lyrae stars and high-redshift ($z > 2.5$) quasars, respectively. The dotted line represents the stellar main sequence. The horizontal blur at $g - r > 1.5$ is due to objects no being detected in the $u$ band and being assigned an essentially random `Luptitude'. Similarly, marginal $u$ detections are biased to lower Luptitudes and our $u-g$ distribution is shifted lightly to the left of the fiducial main sequence line. } }
\label{fig:grug}\end{figure}

In Fig.\ \ref{fig:grug}, we show the SDSS $g-r$ versus $u - g$ distribution of all TDSS targets. Here, we include all objects with previous spectroscopy as well as those objects we share with the eBOSS CORE quasar group. The extended horizontal cloud at $g-r > 1.2$ is mostly due to objects with essentially zero flux in the $u$ band. These objects have very large error bars in $u - g$. We define regions of color-space on the plot (using SDSS colors):
\begin{eqnarray}
\rm{QSO}:\ u-g &<& 0.8,\ g-r < 0.65,\label{eq:photcat}\\
\rm{RRL}:\ u-g &<& 1.35,\ u-g > 1.05,\nonumber\\
g-r &<& 0.5 (u-g)-.15,\nonumber\\
\rm{MS}:\ g-r &>& 1.2\ \rm{or}\nonumber\\
u-g &>& 0.8,\ g-r < 0.5(u-g)+0.25,\nonumber\\
g-r &>& 0.5(u-g)-0.55,\ \rm{not\ RRL},\nonumber\\
\rm{HZQ}:\ u-g &>& 0.8,\ g-r < 0.5(u-g)-0.55,\ \rm{not\ RRL},\nonumber\\ 
\rm{MISC}:\ g-r &>& 0.6,\ g-r < 1.2,\ g-r > 0.5(u-g)+0.25.\nonumber
\end{eqnarray} 
Our categories are named to indicate the primary type of expected variable object in each region, but no category will be absolutely pure. The QSO fiducial color region is mostly quasars and other AGN and is identical to that defined by our quasar criteria in Eq.\ \ref{eq:qsobox}. RRL contains RR Lyrae stars and other variable F stars. MS contains the bulk of the main sequence. HZQ is the region where high-redshift ($z > 2.5$) quasars typically reside. There is no dominant astrophysical identity of the MISC (miscellaneous) sources, but various white dwarf binary systems are included. Consistent with previous variability studies, the region with the most targets is the QSO region (59.0\%) followed by the MS (31.2\%) as shown in Table \ref{tab:categories}. To estimate our total number of quasar candidates, we add our QSO and HZQ objects and subtract 10\% to obtain 135{,}000. We take 90\% of the sum of our other three categories to estimate 85{,}000 stellar variables.  

\begin{table}
\begin{tabular}{ccccc}
        \hline
& \multicolumn{2}{c}{All Targets} & \multicolumn{2}{c}{TDSS-only Targets}\\
        \hline
Category &  N$_{\rm{objects}}$    &   \% of Total  & N$_{\rm{objects}}$    &   \% of Total  \\
        \hline
MS & 75754 & 31.2 & 67922 & 71.5\\
QSO & 143052 & 59.0 & 12754 & 13.4\\
RRL & 7358 & 3.0 & 4384 & 4.6\\
HZQ & 6948 & 2.9 & 3059 & 3.2\\
MISC & 9401 & 3.9  & 6889 & 7.3\\
        \hline
\end{tabular}
\caption{\rm{Numbers and fractions of different broad color-based categories as shown in Fig.\ \ref{fig:grug} in our total sample and our TDSS-only sample.}}\label{tab:categories}
\end{table}

We can make more sophisticated color classifications of particular classes of objects. Using eBOSS CORE quasar color-based photometric classification and previous SDSS spectroscopy, we can alternately define "CORE quasars" as those objects for which
\begin{eqnarray}
P(\rm{qso}) &>& 0.5\ \rm{or} \label{eq:qso}\\
\rm{Class_{SDSS}} &=& \rm{QSO}.\nonumber
\end{eqnarray}
$P(\rm{qso})$ is provided by the eBOSS CORE quasar team (Myers et al. 2015, \textit{in preparation}) which uses the $XDQSOz$ algorithm \citep{BOVY++12} and $\rm{Class_{SDSS}}$ is the SDSS spectral class from previous spectroscopy. These criteria do not include the small number of potential quasars selected exclusively by the PTF variability quasar search. The eBOSS quasar classifier actually only applies to $z > 0.9$ quasars, but most lower redshift quasars are either swept up into this classifier or already have previous spectra. In Fig.\ \ref{fig:grug}, the 134{,}289 CORE quasars (as now defined by Eq.\ \ref{eq:qso}) are shown in green and blue, with the highest priority variable objects in blue. The 108{,}224 objects not satisfying Eq.\ \ref{eq:qso} are shown in red and yellow with the highest priority objects shown in yellow.  

Reassuringly, the bulk of our quasars are centered around $u-g = 0.2$,\ $g-r=0.2$, the known center of the quasar locus. There is also a high density of points along the main sequence, although there is more than the $\approx 0.1$ magnitude of scatter we would expect from our statistical error bars. This just indicates that many of our variable stars have somewhat unusual colors and is to be expected for variables (e.g. unresolved binaries or stars with particularly active photospheres). One notable subpopulation of this plot is the "blue cloud" of 7{,}548 sources not classified as quasars by Eq.\ \ref{eq:qso} in the region defined by
\begin{eqnarray}
0.5 &<& u-g < 1.0\\
0.1 &<& g-r < 0.5.\label{eq:bluecloud}\nonumber
\end{eqnarray} 
This cloud extends off the left of the main sequence and while photometrically blue, is colored red in our plot. A large fraction of these objects are likely to be $z \approx 2.8$ quasars. This is a well-known region in color space where quasars begin to overlap with the main sequence and the color selection used to produce the eBOSS CORE quasar sample is insufficient to distinguish the two. The addition of variability information has likely allowed us to break the color degeneracy and may be used in the future to extend the redshift range of quasar samples. Note that the coloring in Fig.\ \ref{fig:grug} is effectively opaque in high density regions with non-quasars being plotted over quasars. Underneath the "blue cloud" there are also 11{,}160 objects identified as quasars by the criteria in Eq.\ \ref{eq:qso} "underneath" the "blue cloud" in the top figure.  

\begin{figure}[ht]
\includegraphics[width=0.98\columnwidth]{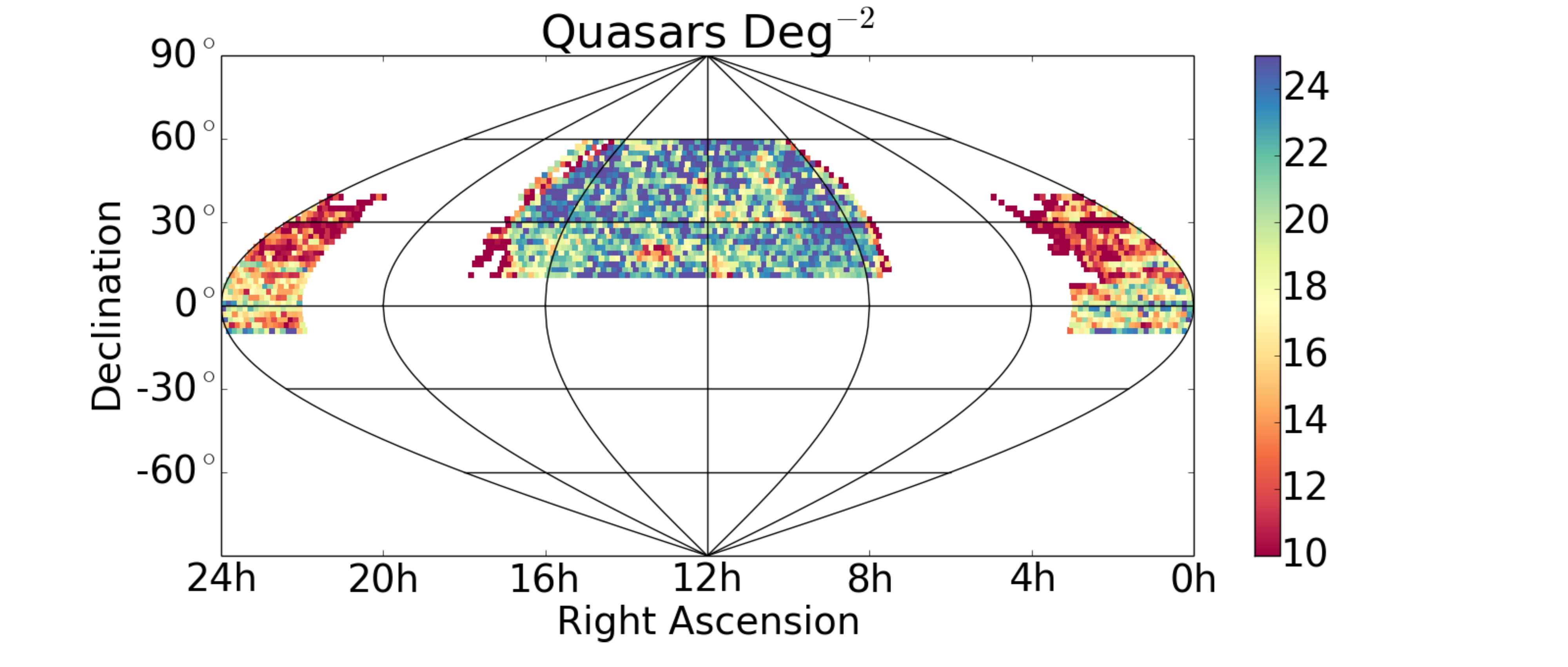}
\includegraphics[width=0.98\columnwidth]{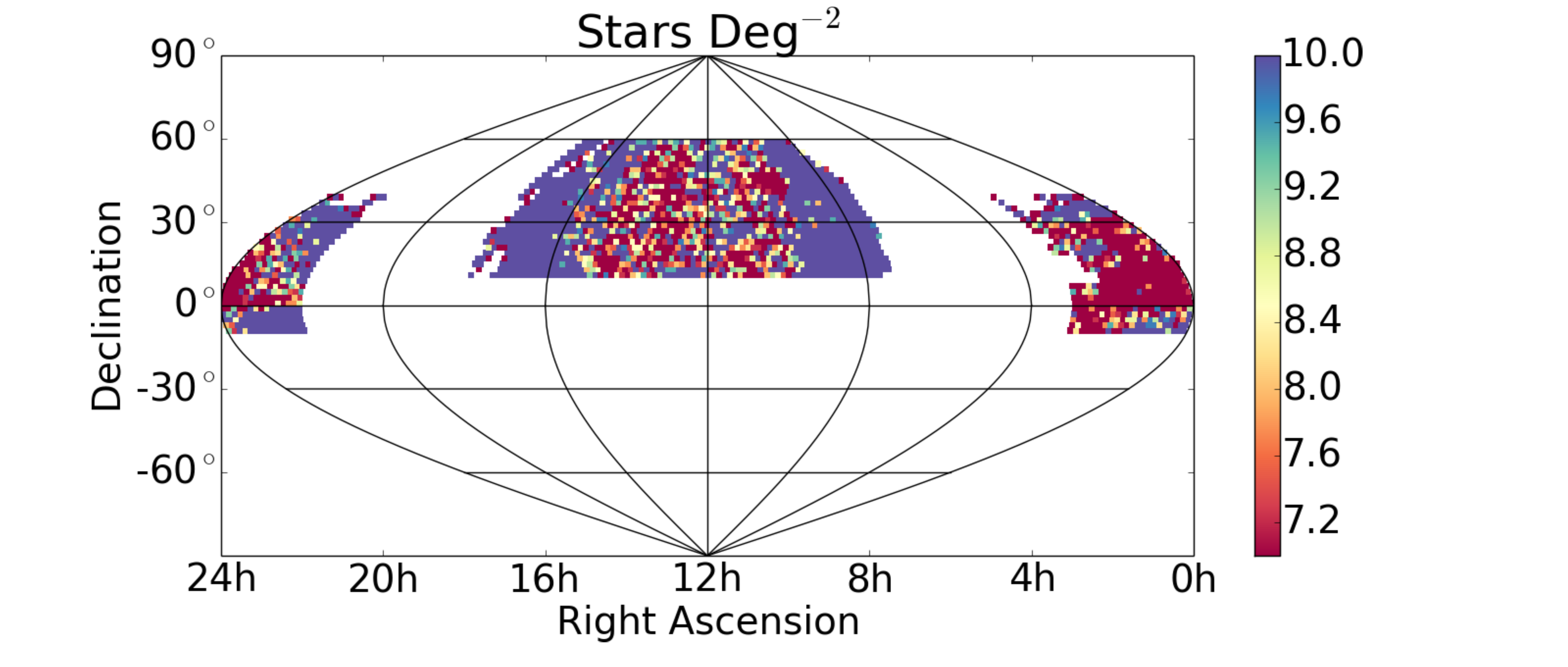}
\caption{\rm{The estimated density of quasars deg$^{-2}$ (top) and stars deg$^{-2}$ (bottom) in our target list across the sky. The "ratty" edges are in very dusty regions that are not actually part of our sample.} }
\label{fig:density}\end{figure}

In Fig.\ \ref{fig:density} we show separately the estimated density of quasars and stars deg$^{-2}$ across our target sample. In this plot, we define quasars via the simple color box in Eq.\ \ref{eq:qsobox}. Our map does not perfectly match the eBOSS area and some extra areas near the edges contain significant (unaccounted for) dust that limits depth and reddens quasars out of our color box. This reddening, combined with geometric incompleteness near the edges of our survey, lower densities near the edge of our field. Beyond these small underdense edges that will not be included in the final survey, our targets are uniformly distributed, not displaying the strong Galactic density variation of the sky plots in Fig.\ \ref{fig:maps}. 

\begin{figure}[ht]
\includegraphics[width=0.98\columnwidth]{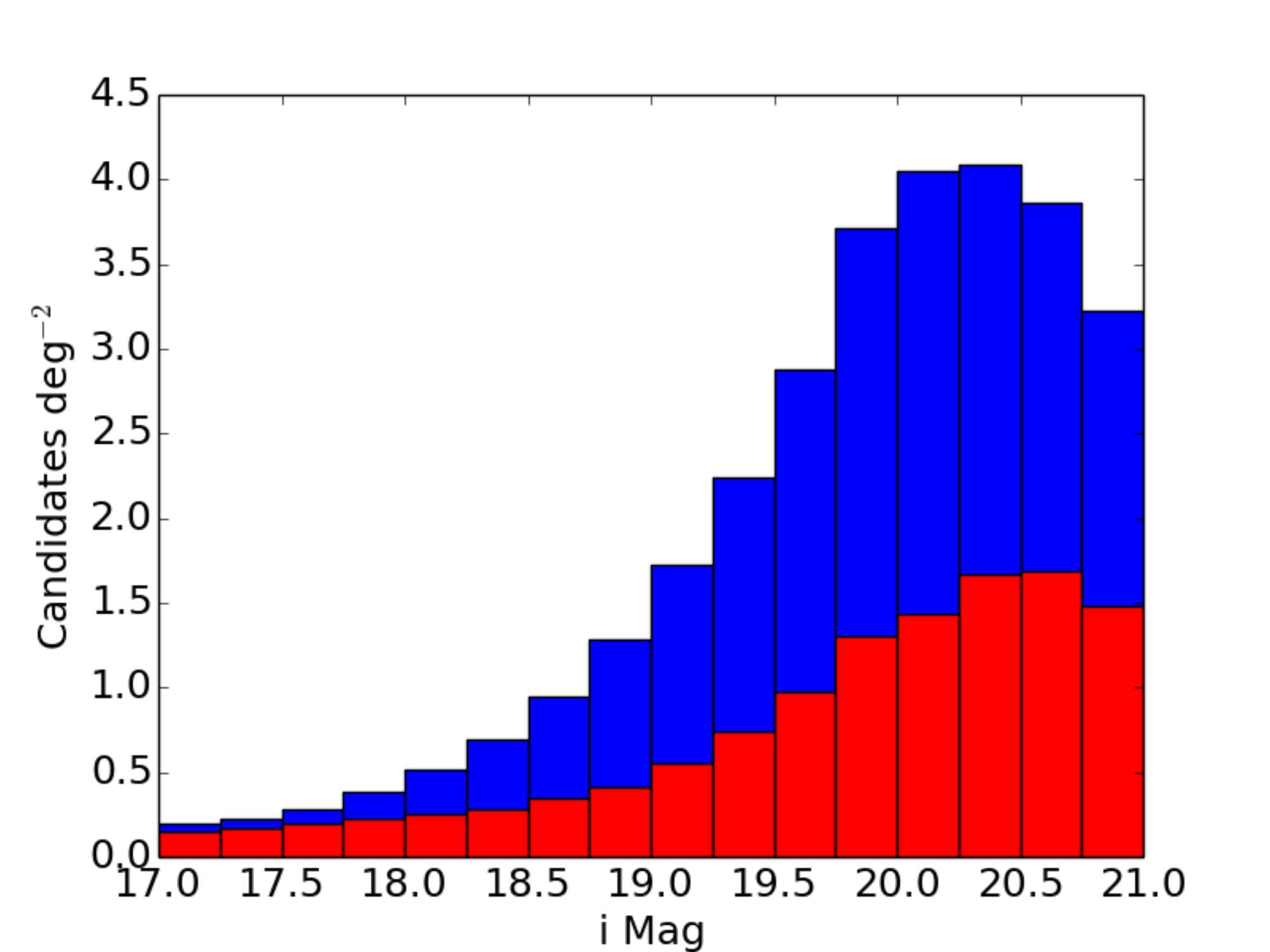}
\caption{\rm{The magnitude distribution of all targets (blue) and TDSS-only targets (red). } }
\label{fig:magdist}\end{figure}

Fig.\ \ref{fig:magdist} shows the magnitude distribution of the TDSS targets. In general, we would expect unbiased magnitude distributions to increase exponentially at fainter magnitudes. Instead, our magnitude distribution peaks at $i = 20.25$. This is a price we pay to ensure high purity. The requirement of detections in multiple filters, our accounting for error bars in Eq.\ \ref{eq:req} and the increased variability requirements at fainter magnitudes as shown in Fig.\ \ref{fig:contours} all decrease the target density at fainter magnitudes. Our TDSS-only targets have a  larger tail on the bright end than the CORE quasars and objects with previous spectra. This result can be explained by the fact that our variable objects are mostly stars (see Table \ref{tab:s82res}), and stars are more concentrated at brighter magnitudes relative to quasars. 

\subsection{The Quasar Population}\label{sect:quasars}

We can probe our likely quasar targets in significantly more detail using a combination of previous spectroscopy and photometry. We are particularly interested in seeing if we are strongly biased towards selecting quasars in a particular color or redshift region. If this were the case, it might indicate that our filter transformations in Eq.\ \ref{eq:spcon} were failing catastrophically in that region. Fortunately, as we show below, the only redshift and color biases are subtle and expected.

\begin{figure}[ht]
\includegraphics[width=0.98\columnwidth]{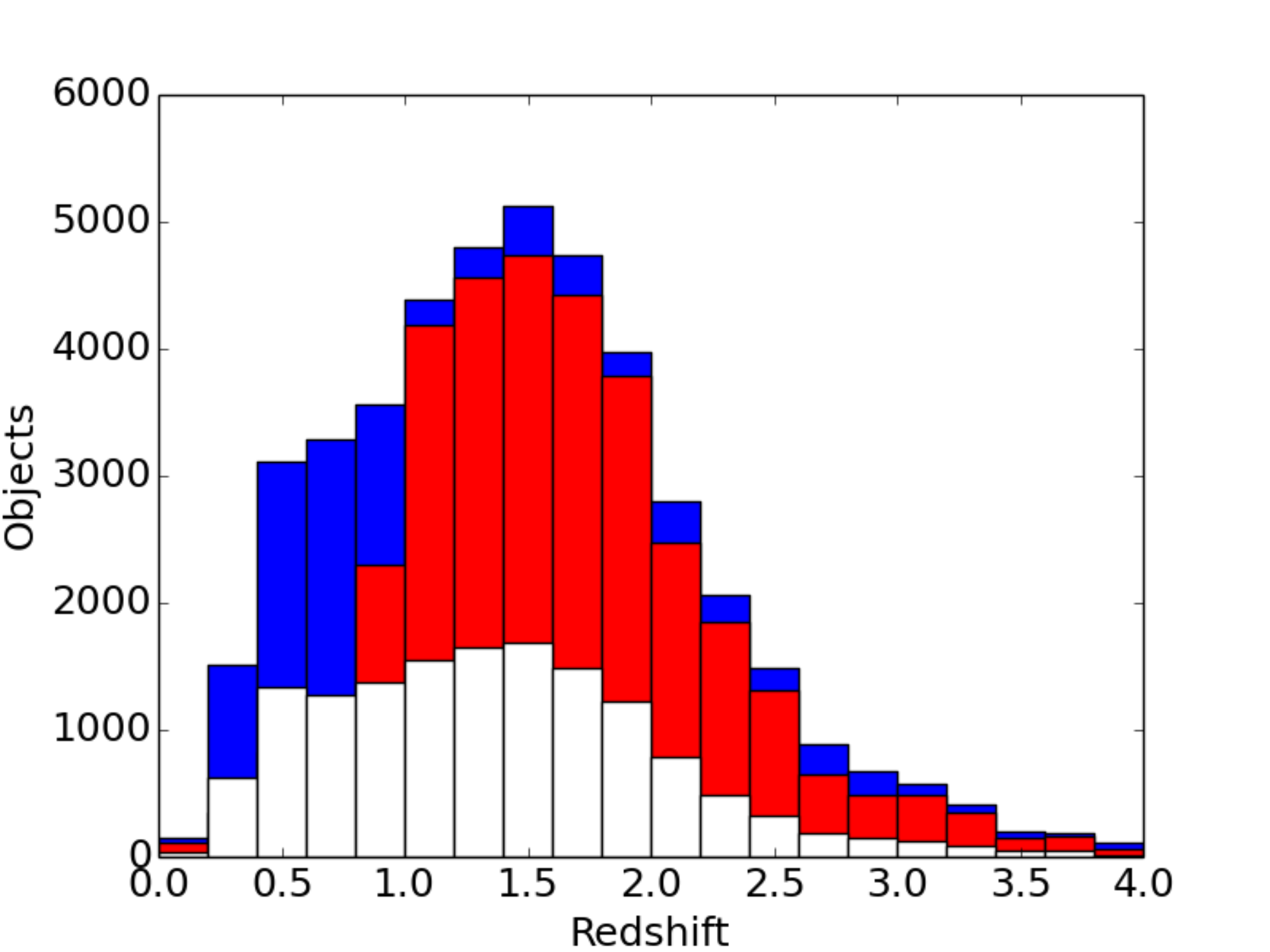}
\includegraphics[width=0.98\columnwidth]{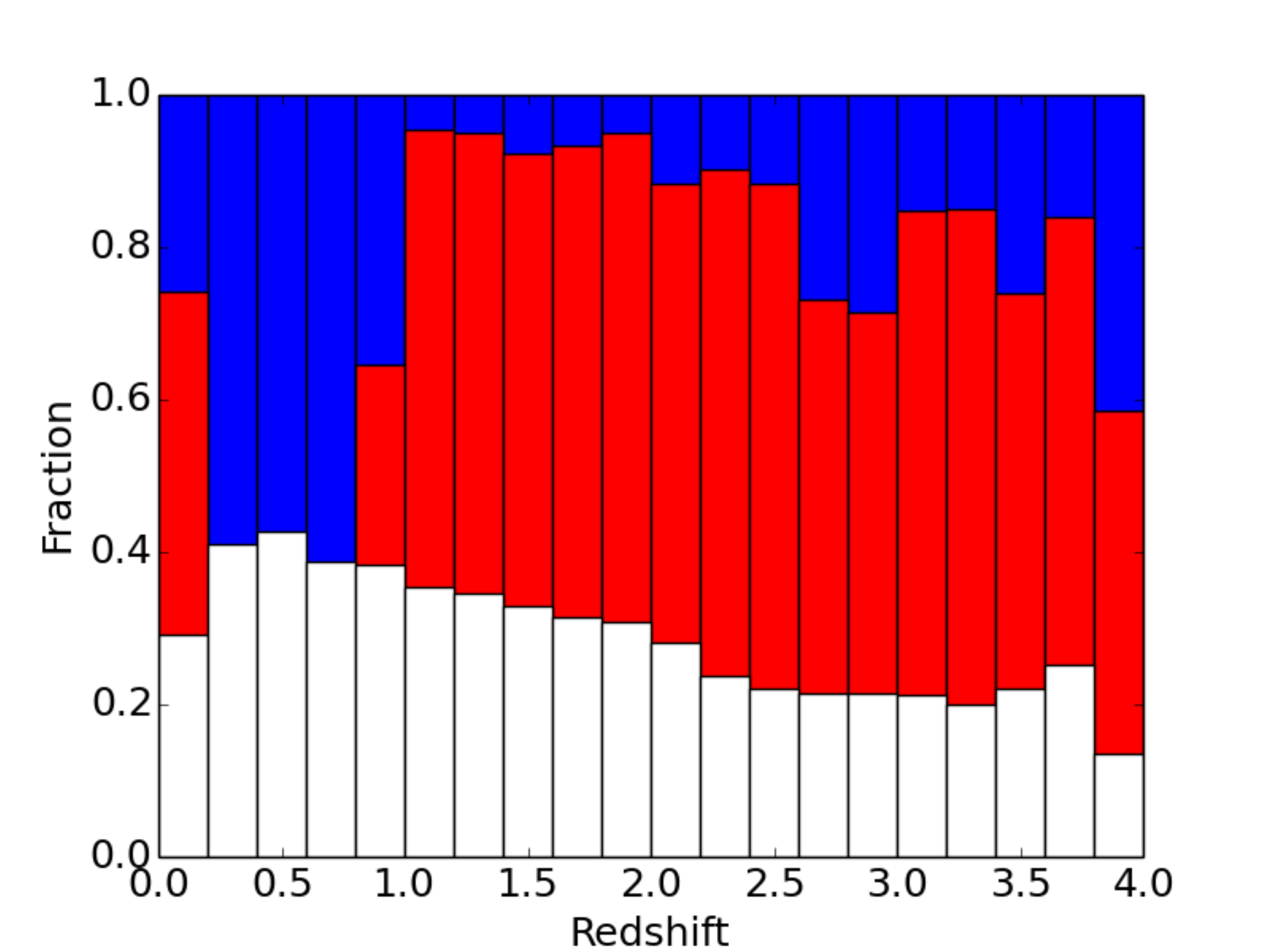}
\caption{\rm{The redshift distribution of quasars with previous SDSS spectroscopy (top). The histograms show all unresolved, $17.8 < i_{\rm{SDSS}} < 19.1$ spectroscopic quasars in the TDSS area (blue), spectroscopic quasars that have an $XDQSOz$ probability $P_{\rm{qso}} > 0.5$ according to the CORE quasar team (red) and quasars that make our final target list (white). The bottom panel shows the fraction of each population as a fraction of the total spectroscopic quasar population. Note that the $XDQSOz$ probability used to select eBOSS CORE quasars intentionally excludes $z < 0.9$ quasars from their sample. } }
\label{fig:redshift}\end{figure}

Fig.\ \ref{fig:redshift} shows the redshift distribution of three categories of spectroscopic quasars: all the unresolved, $17.8 < i_{\rm{SDSS}} <19.1$ SDSS spectroscopic quasars in the TDSS footprint, those with $P_{\rm{qso}} > 0.5$ according to the eBOSS CORE quasar sample and those that make our target list. We chose these limits because the eBOSS CORE bright limit is 17.8, and the previous SDSS spectroscopic faint limit (for the main $z < 2.5$ quasar population) is approximately 19.1. To be clear, these quasars all have previous SDSS spectroscopy and will not generally be reobserved in TDSS. The eBOSS team excludes $z < 0.9$ quasars from their sample. In general, TDSS recovers 30\% of all spectroscopically-confirmed quasars across a broad range of redshift. There are no sharp gaps or spikes that indicate that quasars at particular redshifts are being over-selected or under-selected due to Eq.\ \ref{eq:spcon} or other effects. 

The bottom panel of Fig.\ \ref{fig:redshift} compares the selection efficiency of the CORE quasar sample and TDSS. TDSS underselects $z < 0.2$ objects spectroscopically classified as quasars. Most lower redshift objects with SDSS spectral classification of `QSO' are in fact lower luminosity active galaxies whose emission is not dominated by the central black hole. This is indicated by the fact that $0.2 < z <2.5$ quasars from the plot have mean (median) $u-g$ color of 0.22 (0.25), whereas the $z < 0.2$ quasars in this plot have mean (median) $u-g$ color of 0.95 (0.53). This extra redness is indicative of significant host galaxy flux contamination. Both the CORE quasar sample and the TDSS sample have a decreasing selection efficiency with increasing redshift. For the CORE quasar sample, this effect arises because quasars have less distinct colors at $z > 2.5$, particularly at $z \approx 2.8$ where quasars have similar optical colors to main sequence stars. The TDSS roll-off in efficiency is more gradual and is likely due to the fact that higher redshift quasars vary more slowly due to cosmological time dilation as well as their high luminosities and implied large black hole masses. 

\begin{figure}[ht]
\includegraphics[width=0.98\columnwidth]{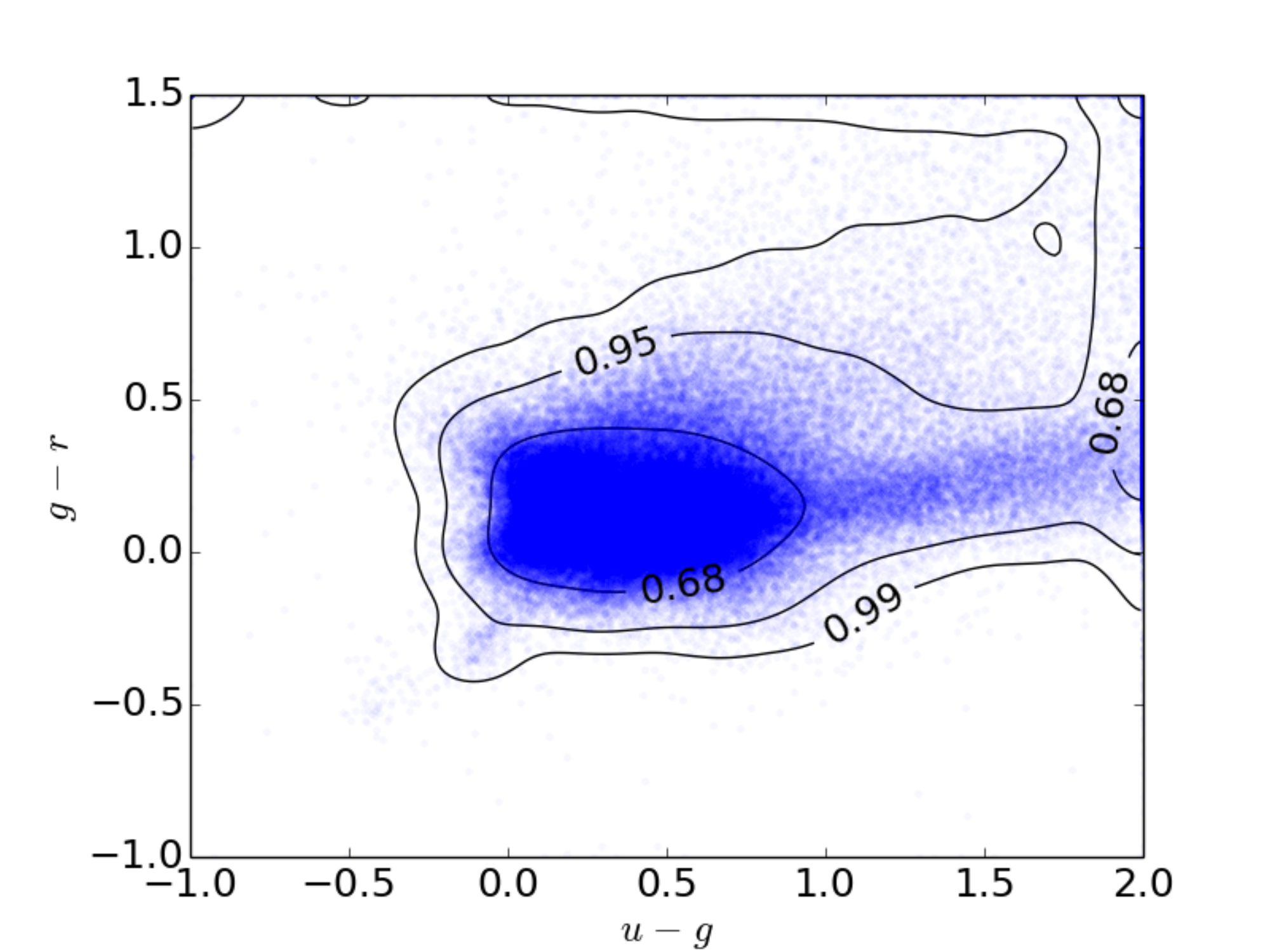}
\includegraphics[width=0.98\columnwidth]{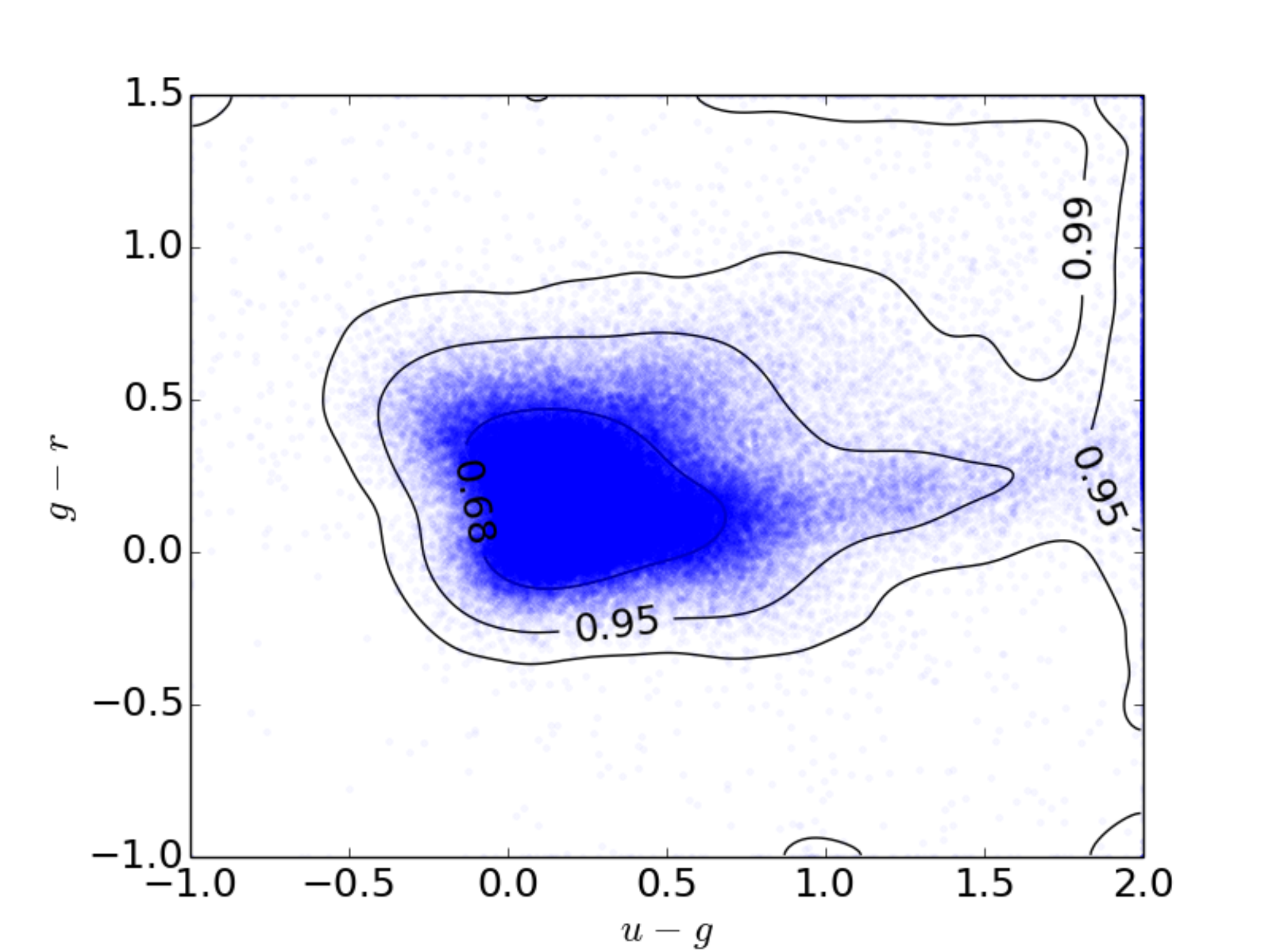}
\includegraphics[width=0.98\columnwidth]{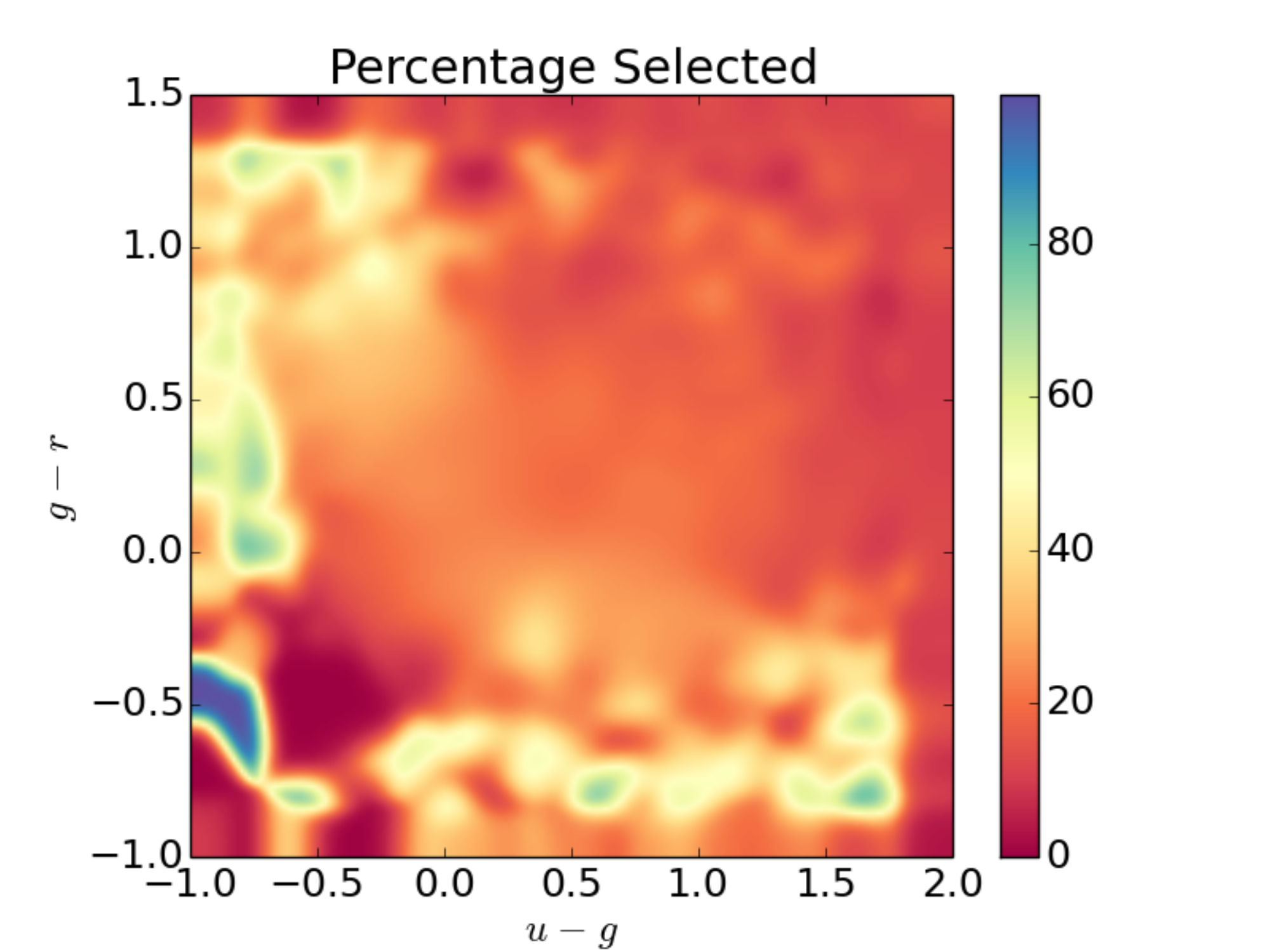}
\caption{\rm{The SDSS $g-r$ versus $u - g$ distribution of all quasars for all $17.8 < i < 21.0$ quasars in the eBOSS area (top) and the subset of those selected by TDSS  (middle). The bottom panel shows the ratio of the two populations.} }
\label{fig:qsogrug}\end{figure}

Fig.\ \ref{fig:qsogrug} compares $g-r$ versus $u - g$ for all $17.8 < i < 21.0$, $P_{\rm{qso}} > 0.5$ CORE quasars and $17.8 < i < 21.0$ spectroscopically identified quasars as well as the subset of those quasars selected by TDSS. The distributions are qualitatively nearly identical. Fig.\ \ref{fig:qsogrug} (bottom) shows the ratio of the two populations across color space. Across the main quasar locus, TDSS recovers 20-30\% of the CORE and spectroscopic quasars. In Fig.\ \ref{fig:qsogrug} (top) there is a faint peninsula of CORE quasar targets stretching from $u-g = 0,\ g - r = -0.2$ to $u-g = -0.5,\ g - r = -0.5$  that are not selected by TDSS in Fig.\ \ref{fig:qsogrug} (middle). These objects are likely to be white dwarfs. Excluding this area, TDSS shows a broad tendency to be more complete at the blue end in both the $u - g$ and $g-r$ axes, although there is a low completeness region in the lower left hand corner of Fig.\ \ref{fig:qsogrug} (bottom) that may be due to small number statistics. This preference for blue objects may partly stem from our decreasing completeness at higher redshift shown in in Fig.\ \ref{fig:redshift}. For a given $i$, we will also generally be more sensitive to variability for blue objects that are bright in $r$ and $g$. So our $i$ magnitude limit may lead to an implicit blue source selection bias.

It is not surprising that the CORE quasar team is significantly more complete at selecting quasars than we are. Their selection is focused on quasars, and it is roughly 4 times larger than our sample. But is should be noted in the analysis above, we do not (and cannot) evaluate the fraction of quasars selected by their variability with TDSS that are missed by conventional color selection. Some poorly constrained fraction of quasars are reddened by dust or otherwise have non-standard colors, and the spectra from TDSS will allow us to study how well many of these quasars we can select from their variability. 

\subsection{The Stellar Population}\label{sect:stars}

Using spectroscopy and eBOSS color-based quasar selection, we can statistically remove most quasars from our sample and investigate the colors of our stellar targets. Again, TDSS does not select stellar targets with color classification, so we expect our targets will span a large range of stellar types and colors. 

\begin{figure}[ht]
\includegraphics[width=0.98\columnwidth]{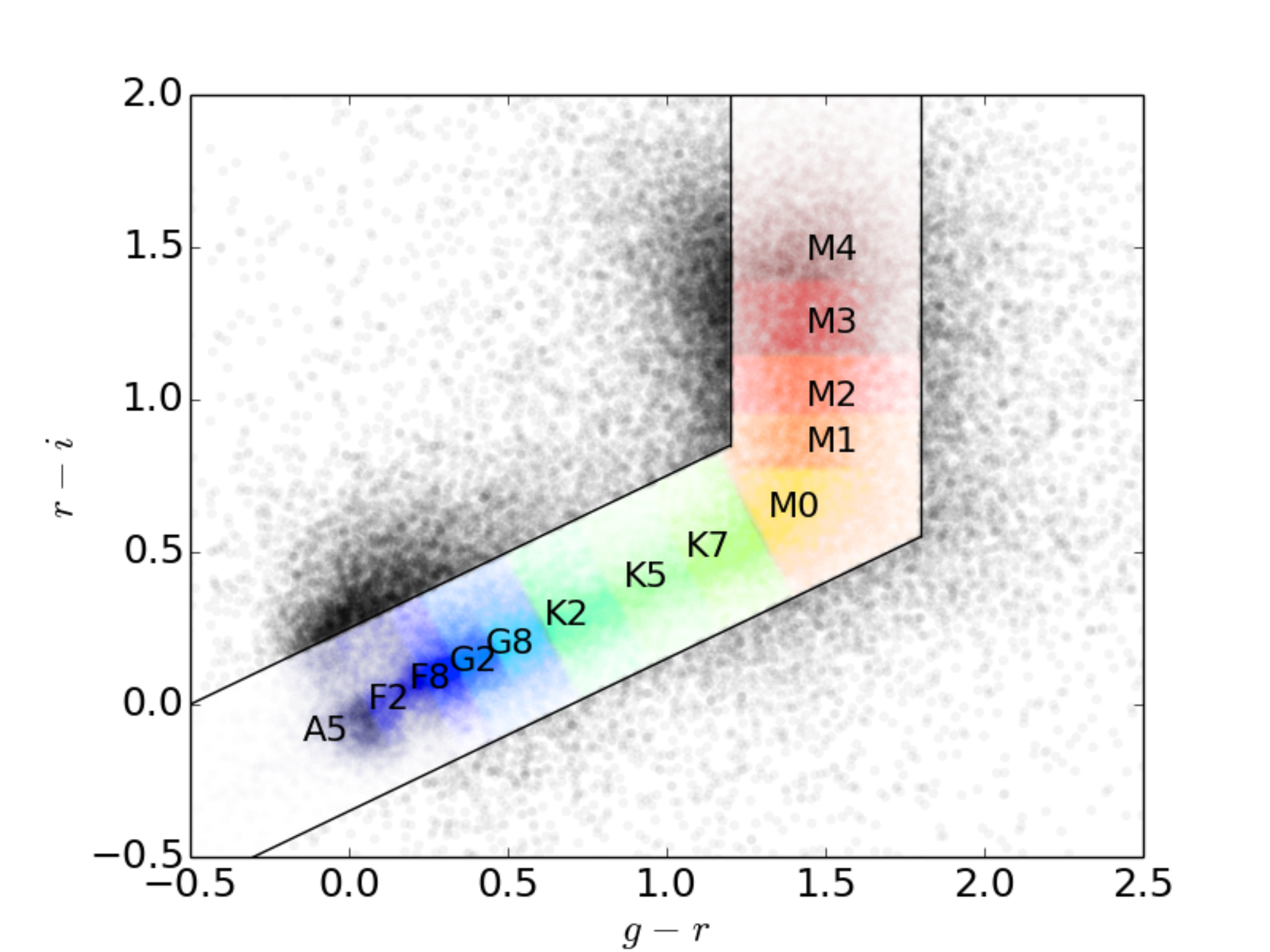}
\caption{\rm{The SDSS $r-i$ versus $g - r$ distribution of all TDSS non-quasars (mostly stars). We approximate the main sequence and label and color-code different stellar types. } }
\label{fig:rigr}\end{figure}

Fig.\ \ref{fig:rigr} shows the $r-i$ versus $g - r$ color distribution of all sources after removing the objects defined as quasars in Eq.\ \ref{eq:qso}. Statistically, we expect the vast majority of remaining objects to be stars. We match the objects to the SDSS main sequence from \citet{K&H07} which we approximate as
\begin{eqnarray}
r-i &=& 0.5(g-r)-0.05,\ \rm{for}\ g-r < 1.45,\label{eq:ms}\\
r-i &>& 0.675,\ \rm{for}\ g-r = 1.45.\nonumber
\end{eqnarray}
This is just a diagonal line which approximates the A through M0 stars and a vertical line that matches the colors of M1 and later stars. We classify our stars into categories defined in Table \ref{tab:stellartypes}. These categories are chosen to be spaced at roughly 0.2 magnitude intervals in $g-r,\ r-i$ so that they are meaningful distinctions for a sample with error bars of just under $0.1$ magnitudes. We set the location of the median subclass of star in Table \ref{tab:stellartypes} to the nearest point on the the main sequence approximation in Eq.\ \ref{eq:ms}. We then match each star to the nearest stellar category median. The results are shown by the coloring in Fig.\ \ref{fig:rigr}. We exclude stars that do not satisfy 
\begin{eqnarray}
r-i &>& 0.5(g-r)-0.35,\\
g-r &<& 1.8,\nonumber\\
r-i &<&  0.5(g-r)+0.25\ \rm{or}\ r-i > 1.2\nonumber
\label{eq:notms}\end{eqnarray}
for tabulation purposes, these stars are called "Not MS" in Table \ref{tab:stellarpercentages} and are colored in greyscale in Fig.\ \ref{fig:rigr}.

\begin{table}
\begin{tabular}{cccccc}
        \hline
Stellar Class & Median Class & $g-r$ & $r-i$ &  $(g-r)_{\rm{line}}\rm$ & $(r-i)_{\rm{line}}$\\
        \hline
OBA     & A5 & -0.02 & -0.17 & -0.06 & -0.08 \\
Early F & F2 & 0.22 & -0.01 & 0.19 & 0.05 \\
Late F  & F8 & 0.31 & 0.03 & 0.28 & 0.09 \\
Early G & G2 & 0.42 & 0.11 & 0.40 & 0.15 \\
Late G  & G8 & 0.53 & 0.18 & 0.52 & 0.21 \\
Early K & K2 & 0.71 & 0.29 & 0.70 & 0.30 \\
Mid K   & K5 & 0.95 & 0.44 & 0.96 & 0.43 \\
Late K  & K7 & 1.14 & 0.55 & 1.15 & 0.53 \\
M0      & M0 & 1.40 & 0.67 & 1.45 & 0.67 \\
M1      & M1 & 1.47 & 0.88 & 1.45 & 0.88 \\
M2      & M2 & 1.48 & 1.03 & 1.45 & 1.03 \\
M3      & M3 & 1.48 & 1.27 & 1.45 & 1.27 \\
M4+     & M4 & 1.48 & 1.51 & 1.45 & 1.51 \\
        \hline
\end{tabular}
\caption{\rm{The different stellar categories shown in Fig.\ \ref{fig:rigr}. We show the description, the median stellar subclass, the actual location of that subclass in  $g-r,\ r-i$ space from \citet{K&H07} and our approximation of this point on the main sequence approximation defined in Eq.\ \ref{eq:ms}.}}\label{tab:stellartypes}
\end{table}

Table \ref{tab:stellarpercentages} lists the numbers and percentages of different stellar types shown in Fig.\ \ref{fig:rigr}. It also presents the numbers and percentages of different stellar types after removing the "blue cloud" stars described in Eq.\ \ref{eq:bluecloud}. These "blue cloud stars", if they are not actually quasars, are most likely F and G type stars. 

There are two notable trends in Table \ref{tab:stellarpercentages}. First, 17.1\% of all objects are classified as "Not Main Sequence". This large fraction is perhaps not surprising since many of our variable targets will be interacting or eclipsing binaries, stars undergoing intense chromospheric activity or will otherwise have colors not consistent with simple stellar physics. Additionally, the fractions of variables are fairly constant across our stellar categories, ranging from 4.2\% to 8.8\%. There is no obvious reason for this to be the case. But it is convenient, as it will allow the study of a broad range of targets. Understanding why the fraction of stellar variables is constant in $r-i,\ g-r$ space will likely be a significant topic of interest for TDSS as spectra are analyzed.

\begin{table}
\begin{tabular}{ccccc}
        \hline
Stellar Class & $N$ & $P$ & $N_{\rm{no\ bc}}$ &  $P_{\rm{no\ bc}}$ \\
        \hline
OBA     & 4421 & 4.2 & 4406 & 4.5 \\
Early F & 6711 & 6.3 & 5240 & 5.4 \\
Late F  & 6657 & 6.3 & 3951 & 4.0 \\
Early G & 6574 & 6.2 & 4006 & 4.1 \\
Late G  & 6293 & 5.9 & 5507 & 5.6 \\
Early K & 6407 & 6.0 & 6405 & 6.5 \\
Mid K   & 5014 & 4.7 & 5014 & 5.1 \\
Late K  & 5857 & 5.5 & 5857 & 6.0 \\
M0      & 8455 & 8.0 & 8455 & 8.6 \\
M1      & 5894 & 5.5 & 5894 & 6.0 \\
M2      & 7061 & 6.6 & 7061 & 7.2 \\
M3      & 9380 & 8.8 & 9380 & 9.6 \\
M4+     & 9390 & 8.8 & 9390 & 9.6 \\
        \hline
MS       & 88114 & 82.9 & 80566 & 82.4 \\
Not MS   & 18190 & 17.1 & 17236 & 17.6 \\
        \hline
Previous SDSS Spectra\\
        \hline
Star & 1742 & 1.6 & 1646 & 1.7 \\
Galaxy   & 196 & 0.2 & 167 & 0.2 \\
        \hline
\end{tabular}
\caption{\rm{The number and percentage of targets in the TDSS candidate list from different stellar classes/subclasses after removing all quasars (as defined by Eq.\ \ref{eq:qso}). $N$ is the number of non-quasar targets of each type. P is the percentage of our total non-quasar targets from each stellar type. N$_{\rm{no\ bc}}$ and P$_{\rm{no\ bc}}$ are the analogous quantities for targets after objects in the "blue cloud" (Eq.\ \ref{eq:bluecloud}) are also excluded. The first 13 rows add up to the main sequence (MS) line, and the total is of course 100\%.}}\label{tab:stellarpercentages}
\end{table}

Our stellar candidates are distributed much more uniformly across the main sequence than those presented in the Catalina Surveys Periodic Variable Star Catalog \citep{DRAK++14} and the analogous catalog from LINEAR \citep{PALA++13}. Specifically, a much larger fraction of our sources are redder K and M stars. The CSS and LINEAR teams require a period measurement for inclusion in their catalogs and are thus particularly sensitive to RR-Lyrae and other (mostly blue) pulsating variables with short periods. Since we do not require a period measurement, our sample includes many eclipsing binaries whose period is difficult to measure due to their low duty cycle. Eclipsing binaries occur across a wide range of stellar masses, so should be distributed rather uniformly across the main sequence. We also expect to find various flaring stars, especially towards the red end of the mains sequence, which may not be periodic at all. 

\subsection{The Hypervariable Population}\label{sect:hypercolor}

\begin{figure}[ht]
\includegraphics[width=0.98\columnwidth]{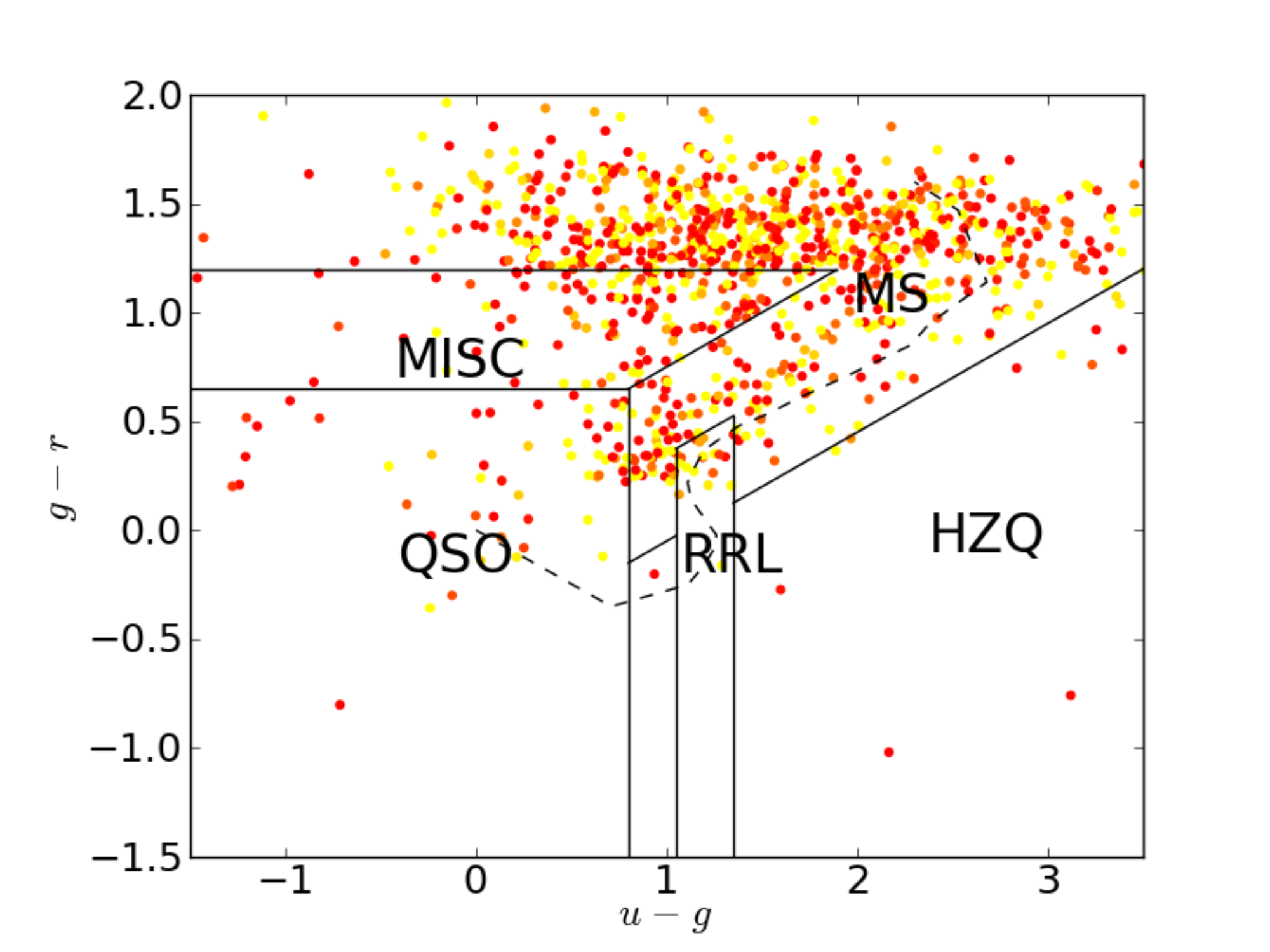}
\includegraphics[width=0.98\columnwidth]{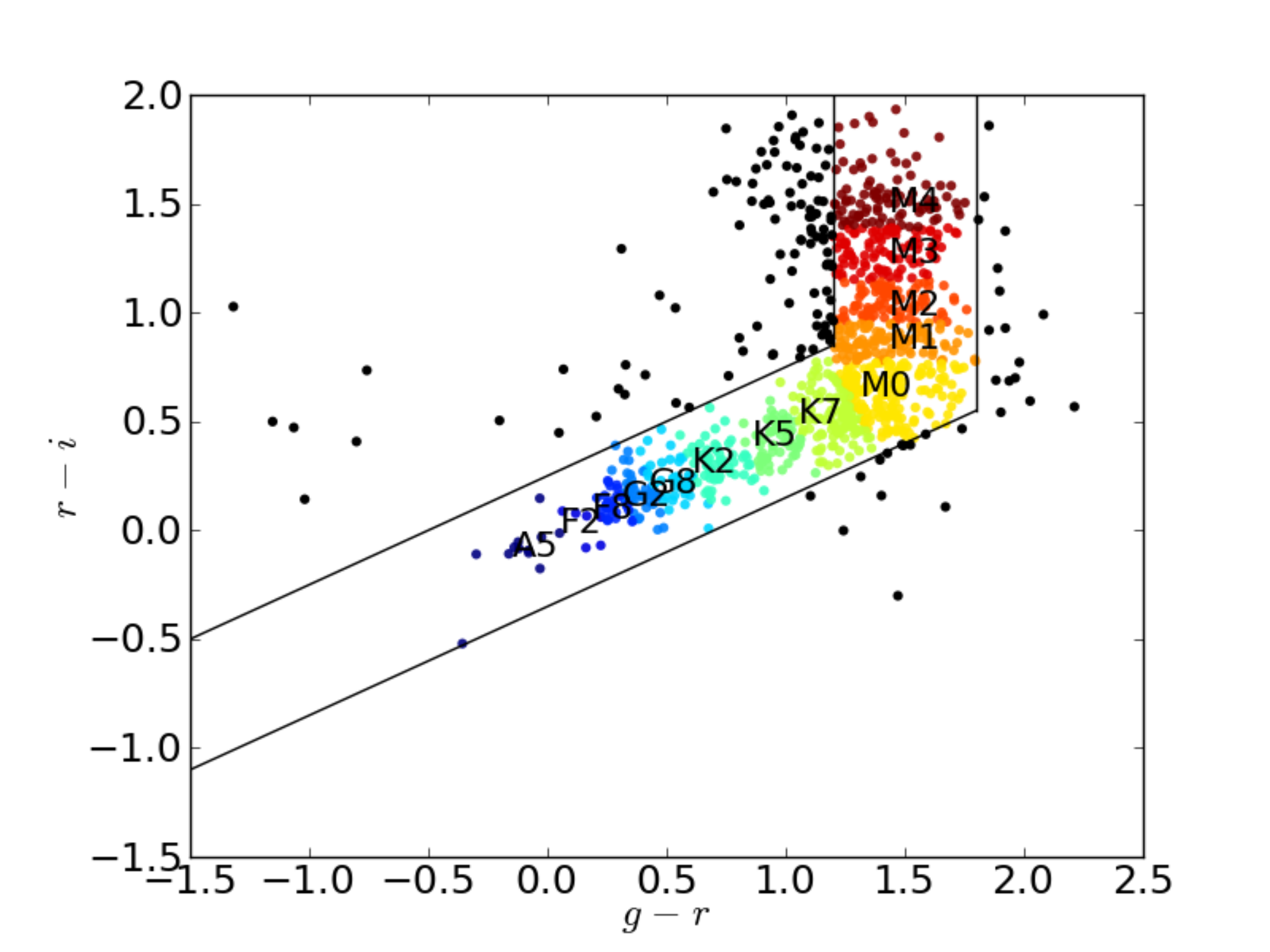}
\caption{\rm{The SDSS $g-r$ versus $u - g$ distribution of all TDSS hypervariables (top) and the  $r-i$ versus $g-r$ distribution of all TDSS hypervariables (bottom). In the top panel, low (high) priority objects are in red (yellow) and the QSO, MS, RRL and HZQ regions are the areas of color space that contain most quasars, main sequence stars, RR Lyrae stars and high-redshift quasars, respectively. In the bottom panel, we show the approximate positions of main sequence classifications. } }
\label{fig:hypercolor}\end{figure}

As mentioned in section \ref{sect:hyper}, 1{,}108 of our sources are hypervariables with 2 or more magnitudes of variability, $V$ (see Eq. \ref{eq:V}). In Fig.\ \ref{fig:hypercolor}, these variables have an unusual distribution of colors, with almost none near the quasar locus. These hypervariables are also significantly redder than our main population, suggesting that many of these stars may be cataclysmic variables, Mira variables or long-period variables. 

We expect the hypervariables to be some of the most interesting objects in our survey and plan on examining this hypervariable population as well as the high variability stellar and quasar populations (mentioned as FES projects in the introduction). Specifically, we will examine the light curves from PS1 and shallower surveys like the Catalina Sky Survey, the Palomar Transient Factory and LINEAR (when available) and see how these relate to our early spectral identifications.
\section{TDSS Selection Fraction as a Function of Color}\label{sect:fraction}

\begin{figure}[ht]
\includegraphics[width=0.98\columnwidth]{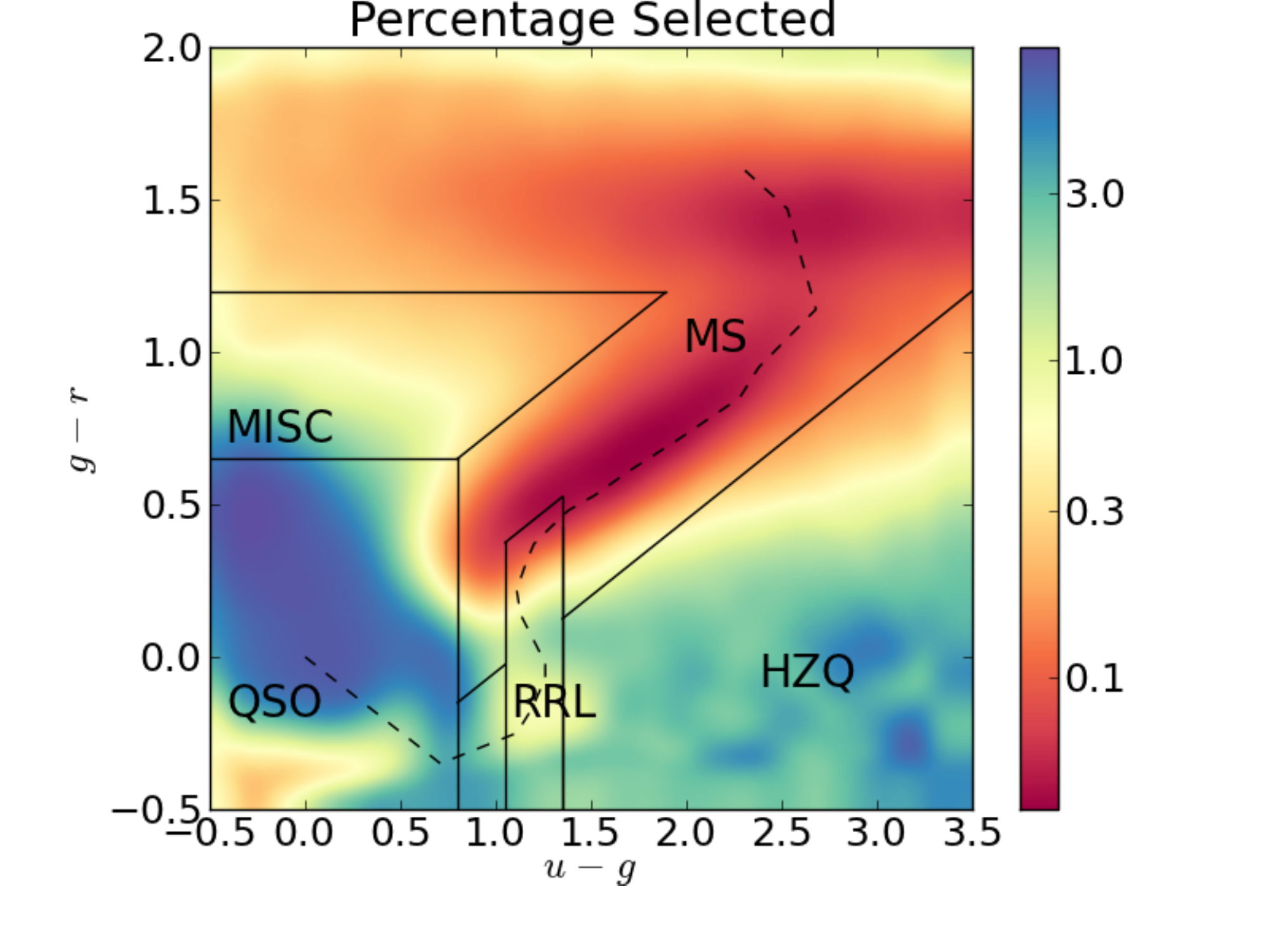}
\includegraphics[width=0.98\columnwidth]{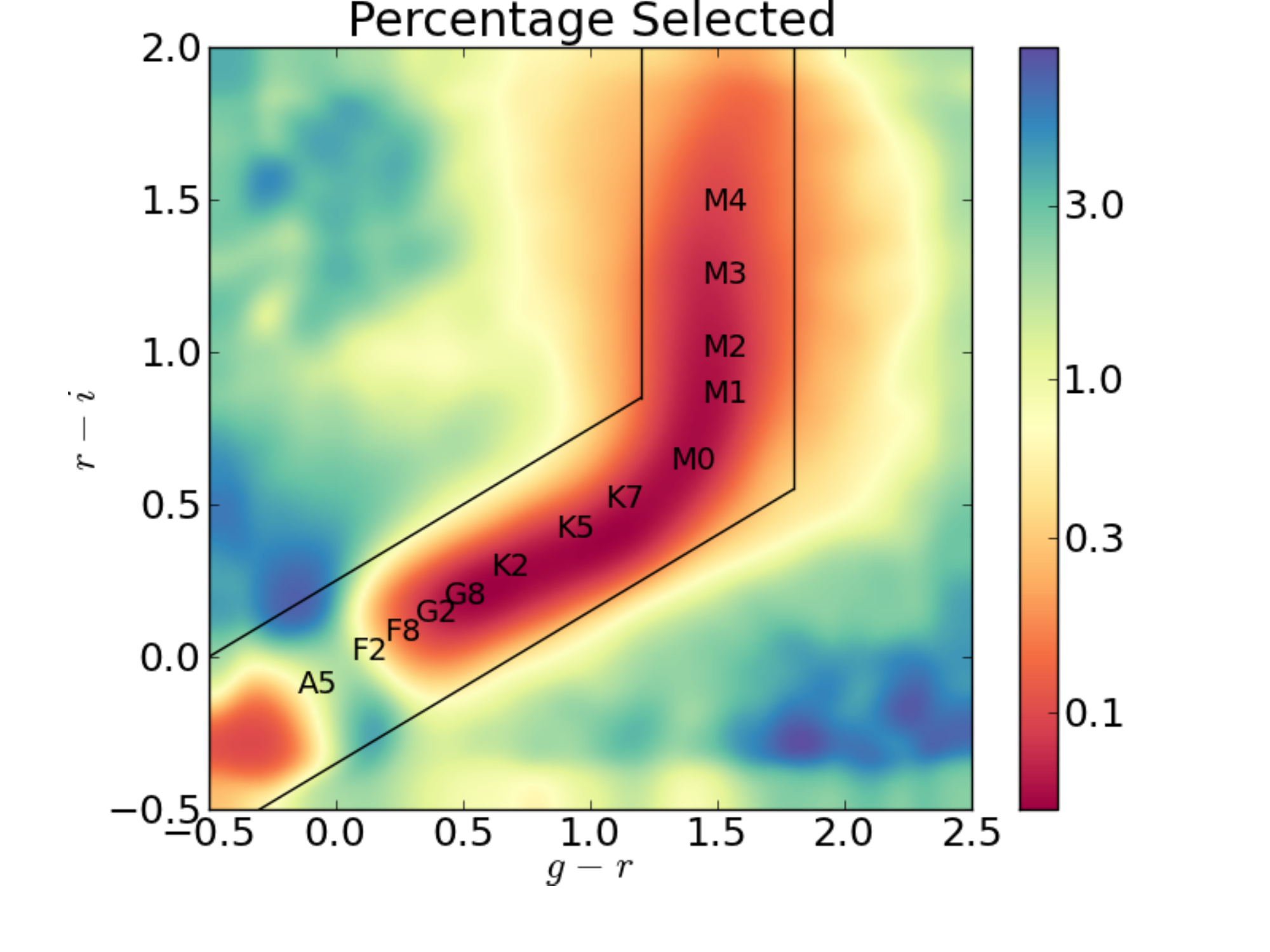}
\caption{\rm{The percentage of SDSS objects which satisfy Eq.\ \ref{eq:dataq} that we select as TDSS targets as a function of $g-r$ and $u - g$ (top). The same percentage as a function of $r-i$ and $g - r$ (bottom). The variable object categories from Fig.\ \ref{fig:grug} and the main sequence categories from Fig.\ \ref{fig:rigr} are also shown.}}
\label{fig:selectpercent}\end{figure}

We can learn more about the TDSS selection algorithm by inverting the analysis in Section \ref{sect:phot} and determining what percentage of objects with particular colors are selected as targets. Fig.\ \ref{fig:selectpercent} displays the selection percentage in the $g-r$, $u-g$ space from Fig.\ \ref{fig:grug} and the $r-i$, $g-r$ space from Fig.\ \ref{fig:rigr}. In this plot and in the accompanying tables below, we compare the total number of TDSS targets to the total number of objects in the TDSS footprint that pass our data quality cuts in Eq.\ \ref{eq:dataq}. In broad strokes, the selection percentage is extremely low (0.3\%) along the main sequence and much higher (above 10\%) in areas of color space in which quasars or other more exotic astrophysical objects are expected to reside. 

Table \ref{tab:categoriespercent} tabulates the fraction of sources selected as variable objects in the categories in Fig.\ \ref{fig:selectpercent} (top) and Eq.\ \ref{eq:photcat}. We only select 0.28\% of objects on the main sequence, excluding the RR Lyrae box from which we select 0.61\% of objects. Within the (very broad) quasar box (which includes many nonvariable, blue stars), we select 11.9\%, although we select approximately 30\% of quasars with previous SDSS spectra as noted in Section \ref{sect:phot}. We select 1.61\% and 1.06\% of sources in the HZQ and MISC regions, respectively. These off-main sequence regions include variable subclasses like cataclysmic variables and white-dwarf main sequence binaries in addition to high-redshift quasars.

\begin{table}
\begin{tabular}{cccc}
        \hline
Category &  N$_{\rm{targets}}$    & N$_{\rm{total\ objects}}$    &   \% Selected  \\
        \hline
MS & 75{,}754 & 27{,}079{,}176 & 0.28\\
QSO & 143{,}052 & 1{,}201{,}995 & 11.90\\
RRL & 7{,}358 & 1{,}204{,}246 & 0.61\\
HZQ & 6{,}948 & 430{,}329 & 1.61\\
MISC & 9{,}401 & 890{,}721 & 1.06\\
        \hline
\end{tabular}
\caption{\rm{Total number of targets, total number of objects and percentage-selected of different broad color-based categories as shown in Fig.\ \ref{fig:grug} in our total TDSS sample.}}\label{tab:categoriespercent}
\end{table}

Table \ref{tab:stellarpercent} tabulates the fraction of sources we select as variable objects from the categories in Fig.\ \ref{fig:selectpercent} (middle) and from section \ref{sect:stars} after likely quasars are removed. We also present our results after removing the ambiguous "blue cloud" region from Eq.\ \ref{eq:bluecloud} in the right half of the table. Along the main sequence, we preferentially select OBA stars (4.4\%) and F stars (3.31\%) over redder stars (0.2\%-0.5\%). Perhaps some of these early-type (blue) stars are the unusually colored quasars that remain after excluding our color-selected quasar sample, but the huge difference in selection percentage between early-type and late-type stars suggests that a relatively large fraction of early-type stars are early-type variables, including pulsators such as RR Lyrae stars. 

\begin{table*}
\begin{tabular}{ccccccc}
        \hline
Stellar Class & N$_{\rm{targets}}$ & N$_{\rm{total\ objects}}$  & \% Selected & N$_{\rm{targets\ no\ bc}}$  & N$_{\rm{total\ objects\ no\ bc}}$ & \% Selected$_{\rm{no\ bc}}$\\
        \hline
OBA     & 4{,}421 & 100{,}368 & 4.40 & 4{,}406 & 100{,}219 & 4.40 \\
Early F & 6{,}711 & 202{,}730 & 3.31 & 5{,}240 & 88{,}923 & 5.89 \\
Late F  & 6{,}657 & 1{,}407{,}242 & 0.47 & 3{,}951 & 455{,}731 & 0.87 \\
Early G & 6{,}574 & 2{,}408{,}184 & 0.27 & 4{,}006 & 1{,}479{,}662 & 0.27 \\
Late G  & 6{,}293 & 2{,}810{,}731 & 0.22 & 5{,}507 & 2{,}668{,}135 & 0.21 \\
Early K & 6{,}407 & 2{,}755{,}323 & 0.23 & 6{,}405 & 2{,}755{,}301 & 0.23 \\
Mid K   & 5{,}014 & 2{,}147{,}150 & 0.23 & 5{,}014 & 2{,}147{,}150 & 0.23 \\
Late K  & 5{,}857 & 2{,}565{,}386 & 0.23 & 5{,}857 & 2{,}565{,}386 & 0.23 \\
M0      & 8{,}455 & 3{,}465{,}678 & 0.24 & 8{,}455 & 3{,}465{,}678 & 0.24 \\
M1      & 5{,}894 & 2{,}616{,}959 & 0.23 & 5{,}894 & 2{,}616{,}959 & 0.23 \\
M2      & 7{,}061 & 2{,}944{,}265 & 0.24 & 7{,}061 & 2{,}944{,}265 & 0.24 \\
M3      & 9{,}380 & 3{,}443{,}992 & 0.27 & 9{,}380 & 3{,}443{,}992 & 0.27 \\
M4+     & 9{,}390 & 2{,}365{,}245 & 0.40 & 9{,}390 & 2{,}365{,}245 & 0.40 \\
        \hline
MS      & 88{,}114 & 29{,}233{,}253 & 0.30 & 80{,}566 & 2{,}7096{,}646 & 0.30 \\
NMS     & 18{,}190 & 1{,}011{,}878 & 1.80 & 17{,}236 & 997{,}638 & 1.73 \\
        \hline
Previous SDSS Spectra\\
        \hline
Star & 1{,}742 & 219{,}463 & 0.79 & 1{,}646 & 158{,}830 & 1.04 \\
Galaxy & 196 & 6{,}981 & 2.81 & 167 & 6{,}005 & 2.78 \\
        \hline
\end{tabular}
\caption{\rm{The numbers and percentages of targets selected from different stellar classes/subclasses after removing all quasars (as defined by Eq.\ \ref{eq:qso}). N$_{\rm{targets}}$ is the number of non-quasar targets of each selected while  N$_{\rm{total}}$  is the total number of non-quasar objects that pass our data quality requirements. The \% Selected columns is the percentage of objects that we select in our total sample. We also show the analogous quantities for targets after objects in the "blue cloud" (Eq.\ \ref{eq:bluecloud}) are also excluded (subscripted "no bc").}}\label{tab:stellarpercent}
\end{table*}

\section{Stripe 82 and CSS Targets with Previous Spectroscopy or Variability Classifications}\label{sect:spec}

As a final probe into the TDSS sample, we run our algorithm on SDSS and PS1 data across the high Galactic latitude, $315^\circ < \rm{RA} < 60^\circ$, area of SDSS Stripe 82 and cross-match our results with samples of objects with previous spectroscopy or variability classification. Both spectroscopy and known variable objects are significantly more dense in Stripe 82 than in the larger SDSS or eBOSS areas, so this dataset provides a relatively complete and homogeneous sample. We slightly modify our selection algorithm by using 2.5$^\circ$ $\times$ 2.5$^\circ$ pixels with 62 TDSS-only targets per pixel since Stripe 82 is 2.5$^\circ$ wide. We then cross-match these targets (including shared CORE quasar targets) with $17 < i < 21$ point sources that have previous public SDSS spectroscopy and also cross-match our sample with known variable objects. We use a set of 173 ellipsoidal/eclipsing binaries from \citet{BHAT12}, 235 RR Lyrae from \citet{SESA++10} and 91 other low mass periodic sources from \citet{BECK++11}. We also cross-match our complete target list with the union of the Catalina Sky Survey (CSS) periodic variables from \citet{DRAK++14} and RR Lyrae variables from  \citet{DRAK++13} and \citet{TORR++15}. This union contains 68{,}956 stellar variables, 5{,}978 of which satisfy the minimum data quality requirement from Eq. \ref{eq:dataq} and are in the TDSS area. Both our spectroscopic and variable object samples are the results of multiple different surveys with acute and intentional biases rather than a single statistically complete sample. The relative fractions of different sources that we detect are thus only suggestive of how our techniques will select various subclasses of variable objects.

\begin{table}
\begin{tabular}{cccccc}
        \hline
Spec Class & $N_{S82}$ & $\rho_{S82}$ & $N_{S82\ TDSS}$ & $\rho_{S82\ TDSS}$ & TDSS\%\\ 
       \hline
AGN & 24{,}315 & 47.44 & 6{,}788 & 13.24 & 27.92\\
        \hline
AGN Broadline & 18{,}999 & 37.07 & 5{,}727 & 11.17 & 30.14\\
AGN Non-Broadline & 5{,}316 & 10.37 & 1{,}061 & 2.07 & 19.96\\
        \hline
Star & 62{,}147 & 121.26 & 358 & 0.70 & 0.58\\
        \hline
OBA & 6{,}080 & 11.86 & 160 & 0.31 & 2.63\\
Early F & 10{,}151 & 19.81 & 55 & 0.11 & 0.54\\
Late F & 6{,}895 & 13.45 & 26 & 0.05 & 0.38\\
Early G & 3{,}469 & 6.77 & 3 & 0.01 & 0.09\\
Late G & 410 & 0.80 & 3 & 0.01 & 0.73\\
Early K & 8{,}789 & 17.15 & 20 & 0.04 & 0.23\\
Mid K & 432 & 0.84 & 3 & 0.01 & 0.69\\
Late K & 3{,}177 & 6.20 & 4 & 0.01 & 0.13\\
M0 & 3{,}200 & 6.24 & 1 & 0.00 & 0.03\\
M1 & 2{,}696 & 5.26 & 10 & 0.02 & 0.37\\
M2 & 3{,}746 & 7.31 & 3 & 0.01 & 0.08\\
M3 & 4{,}485 & 8.75 & 8 & 0.02 & 0.18\\
M4+ & 6{,}274 & 12.24 & 28 & 0.05 & 0.45\\
L, T & 556 & 1.08 & 18 & 0.04 & 3.24\\
Carbon Star & 117 & 0.23 & 4 & 0.01 & 3.42\\
CV & 253 & 0.49 & 8 & 0.02 & 3.16\\
WD & 1{,}417 & 2.76 & 4 & 0.01 & 0.28\\
        \hline
Galaxy & 1{,}448 & 2.83 & 58 & 0.11 & 4.01\\
        \hline
\end{tabular}
\caption{\rm{A summary of SDSS spectroscopic pipeline classes and subclasses of all $315^\circ < \rm{RA} < 60^\circ$, $17 < i < 21$ Stripe 82 point sources with spectroscopy. These columns are the number and density deg$^{-2}$ of each type of object, the number and density deg$^{-2}$ of each type of object that is selected by TDSS and the percentage of these objects that would be selected by TDSS. Many L, T, carbon star and CV classifications are suspect.}}\label{tab:specclass}
\end{table}

Table \ref{tab:specclass} shows the numbers of objects of different spectroscopic types that pass our selection cut. We use SDSS spectroscopic pipeline classes (`quasar', `star' or `galaxy') and subclasses (of which there are many) rather than performing independent spectroscopic analysis.  We combine all objects with spectroscopic type `quasar' into the AGN category and classify them as either `AGN Broadline' or `AGN Non-Broadline'. As expected, we select a significantly higher fraction of Broadline AGN. Many `Non-Broadline' AGN are starburst galaxies or Seyfert type 2 galaxies in which the potentially variable central black hole is less dominant in the overall emission. 

We only select 0.58\% of objects with stellar spectra. This is also expected as most stars, unlike quasars, are not inherently variable. Conversely, only 358 of our approximately 2{,}400 stellar targets (15\%) in Stripe 82 have previous spectra. The fact that 85\% of our stellar targets are new, even in Stripe 82, an area with a disproportionately high density of spectra, emphasizes how large and unique the TDSS stellar sample is.  

For convenience, we have bundled our stellar spectroscopic subclasses into the same photometric color subclasses we use in Tables \ref{tab:stellartypes} and \ref{tab:stellarpercentages} with additional categories for L and T dwarfs, carbon stars, cataclysmic variables and white dwarfs. Roughly half of the stars selected have OBA type colors. This population is highly weighted toward the `A' end, and many of these stars are likely RR Lyrae or anomalous Cepheid variables. The list of stars with previous SDSS spectra is heavily biased towards OBA stars. Only 4.2\% of our non-quasar targets are OBA targets. We also tend to select a relatively high percentage of L and T stars (3.24\%) as well as carbon stars (3.42\%), which are likely in binaries \citep{GREE13}.  We only select 3.16\% of cataclysmic variables, objects that by definition have large variability amplitudes, but relatively short duty cycles. The L, T, carbon star and cataclysmic variable selection fractions are all suspect as a large number of objects are misidentified with these intrinsically rare classifications in the SDSS spectroscopic pipeline. In practice, objects identified by TDSS with these rare classifications may require additional observations to classify them with certainty. We select 4\% of unresolved objects with galaxy spectra. These are probably intermediate AGN not recognized as quasars by the SDSS algorithm due to relatively weak emission lines, AGN with resolved galaxy flux that SDSS misclassified morphologically or occasionally supernova hosts. 

\begin{table}
\begin{tabular}{cccccc}
        \hline
Var Class & $N_{\rm{S82}}$ & $\rho_{\rm{S82}}$ & $N_{\rm{S82\ TDSS}}$ & $\rho_{\rm{S82\ TDSS}}$ & TDSS\%\\ 
       \hline
Binaries & 173 & 0.34 & 26  & 0.05 & 15.03\\
RR Lyrae & 235 & 0.46 & 120 & 0.23 & 51.06\\
Other Periodic & 91 & 0.18 & 10 & 0.02 & 10.99\\
        \hline
\end{tabular}
\caption{\rm{The classes of selected $315^\circ < \rm{RA} < 60^\circ$ Stripe 82 $17 < i < 21$ variable point sources. These columns are the number and density deg$^{-2}$ of each type of object, the number and density deg$^{-2}$ of each type of object that is selected by TDSS and the percentage of these objects that would be selected by TDSS. }}\label{tab:varclass}
\end{table}

Table \ref{tab:varclass}  lists the fractions of previously identified Stripe 82 variable objects we detect. We only detect 15\% of the \citet{BHAT12} binaries. Binaries typically produce the $\approx 0.2$ magnitudes of variability we require for targets only when they are nearly fully eclipsing and thus have a relatively low duty cycle compared to the more constantly dynamic pulsators. More than half (51\%) of the \citet{SESA++10} RR Lyrae sample makes our cut. In fact, 156 of 235 (66\%) of their RR Lyrae stars pass our ($E > 45.4$) RR Lyrae cut, with 15\% being removed by our random downsampling in areas with more than 10 targets deg$^{-2}$. If the density of selected RR Lyrae stars here were applied over the whole sky, we would expect to find 1700 RR Lyrae stars. Additionally, our broad variability selector should identify many RR Lyrae stars whose light curves are too faint to be precisely classified as RR Lyrae stars. It is likely that our estimate in Table \ref{tab:categories} of 4{,}384 TDSS-only RR Lyrae targets made solely from photometry is not more than a factor of two too high. We only detect 11\% of other periodic stars, likely due to their relatively small variability amplitudes. 

\begin{table}
\begin{tabular}{cccc}
        \hline
Var Class & Num$_{\rm{CSS}}$ & Num$_{\rm{CSS\ TDSS}}$ &  TDSS\%\\
        \hline
W-Ursae Majoris&              1{,}982 &                     550 &   27.75 \\
Algol Eclipsing&               364 &                      47 &   12.91 \\
$\beta$ Lyrae &                27 &                       7 &   25.93 \\
RR Lyrae &              3{,}494 &                    1{,}867 &   53.43 \\
Blazhko   &                 3 &                       3 &  100.00 \\
RS Canum Venaticorum&                29 &                       7 &   24.14 \\
Anomalous Cepheid&                 3 &                       2 &   66.67 \\
Cepheid-II&                11 &                       3 &   27.27 \\
High Amplitude $\delta$ Scuti&                21 &                       7 &   33.33 \\
Long-Period Variables&                 7 &                       3 &   42.86 \\
Rotating Ellipsoidal&                18 &                       5 &   27.78 \\
Post Common Envelope Binary&                17 &                       6 &   35.29 \\
        \hline
All       &              5{,}978 &                    2{,}507 &   41.94   \\
        \hline
\end{tabular}
\caption{\rm{The classes of selected periodic variable point sources from the Catalina Sky Survey. The columns are the number of each type of object in the TDSS area, the number detect by TDSS and the percentage of these objects that would be selected by TDSS. }}\label{tab:cssclass}
\end{table}

We can perform a more in depth analysis for many of our sources over the full TDSS area by cross-matching with known periodic variable objects from CSS. CSS is significantly shallower that PS1 (typical limiting magnitude of $V = 19.7$), and the CSS sources with measurable periodicity are biased towards the brighter end of the survey. Our sample of 5{,}978 CSS periodic variables analyzed by TDSS is heavily biased towards the bright end of the survey with 3{,}963 $i < 18$ and 5{,}621 $i < 19$ objects, respectively. Table \ref{tab:cssclass} shows the numbers and percentages of CSS periodic variables selected by TDSS. The categories are those used by \citet{DRAK++14}. Our results here are similar to those in Stripe 82. In particular, we recover 53\% of RR Lyrae and generally recover a large fraction of the pulsating stars (RR Lyrae, Blazhko stars, Cepheid variables, $\delta$ Scuti stars and Long period variables) which tend to have high amplitudes and duty cycles. As a reminder, we are randomly downsampling by 30\%, so we should not exceed 70\% completeness for a large population. We generally recover a smaller fraction of binary systems (W-Ursae Majoris, Algol Eclipsing, $\beta$ Lyrae, RS Canum Venaticorum and Post Common Envelope Binaries) which tend to have lower duty cycles and amplitudes (although the categories here have relatively high amplitude).

As TDSS spectra are processed, we plan to compare our spectral identification of brighter TDSS-identified variable objects to those derived from higher cadence light curve analysis from other time domain imaging surveys (particularly the Catalina Sky Survey, the Palomar Transient Factory and LINEAR). Photometric classification of the stellar population may be supported through a machine-learning approach to the photometric time-series light curves. For example, the artificial neural-network based Eclipsing Binary Factory (EBF) pipeline \citep{PAEG++14,PARV++14} has been used to automatically identify and sub-classify eclipsing binary stars in the \textit{Kepler} field as eclipsing contact, eclipsing semi-detached, and eclipsing detached systems with a low false positive rate. These EBF sub-classifications are accompanied by a confidence level (i.e., posterior classification probability) for each target as a given variable type (e.g., Eclipsing Binary, Cepheid, $\delta$ Scuti, RR Lyrae). This EBF-generated confidence may then be used as quantitative corroboration for the spectral classification of TDSS stellar variable targets, and extrapolated cautiously to fainter targets.

\section{Conclusions}\label{sect:conc}

TDSS promises to open a new window into the nature of astrophysical variable objects. Obtaining 220{,}000 $R \approx 2{,}000$,  optical spectra will make TDSS a massive and unique spectroscopic survey of variable objects. Just as important as the scale of the TDSS sample is its breadth. By adopting a general variability metric and not selecting for specific types of variable objects in color space, TDSS will not only acquire spectra of 135{,}000 variable quasars, but it will also obtain spectra of 85{,}000 stellar targets including perhaps 4{,}000 RR Lyrae stars and 1{,}108 hypervariables (including blazars, CVs or other flaring stars), hundreds of carbon stars and multitudes of other variables yet to be determined. The TDSS stellar spectra have little overlap with previous SDSS stellar spectra and should prove to be a truly unique sample. 

This survey is facilitated by the combination of SDSS and PS1 photometry. SDSS and PS1 both produce 10\% level photometry out to $i = 21$ in the $griz$ filters across an overlapping area of 14{,}400 deg$^2$, including the entire 7{,}500 deg$^2$ eBOSS area. The combination of an SDSS-PS1 photometry difference, spanning 6-10 years, and PS1-only variation, with time scales of hours to years, efficiently selects both long term variable objects (quasars) and shorter term variable objects (most variable stars). After flagging and rejecting sources with unreliable photometry using sensible database queries, we use a Kernel Density Estimator and a Stripe 82 training set to produce a sample that we estimate to be 95\% pure, based on Stripe 82 variability measurements. We suspect that our final sample will have even higher purity since some Stripe 82 non-variables may have simply been dormant during the epochs of Stripe 82 imaging but active during those of PS1. In addition, we increase purity further with visual image inspection. While the vast majority of our sample is selected in a relatively unbiased manner, we deliberately select 1{,}108 hypervariables (which vary by more than 2 magnitudes) and 73 $i$-dropouts to ensure that these potentially interesting objects are not excluded from our sample. 

While precise and complete identification of variable objects is impossible with basic photometric colors, we analyze our sample in $u-g,\ g-r,\ r-i$ color space to characterize our sample in broad strokes. The majority of our sample (59\%) resides in the traditional $z < 2.5$ quasar color region. However, after removing our overlap with the eBOSS CORE quasar sample and previous spectroscopy, only 13.4\% of our TDSS-only targets reside in this region, while 76.1\% of them lie along or near the main sequence (including 4.1\% which are in the F-star region where most RR Lyrae lie). Our stellar population is spread out relatively evenly with 37.7\% of our non-quasar sample being M stars, 40.9\% being FGK stars, 4.2\% being (intrinsically rare) OBA stars and 17.1\% being outside our main sequence classifying scheme. This target diversity was a natural result of selecting objects based on their variability without explicit regard for their colors. Inverting this analysis, we select 11.9\% of objects within a broad quasar color box while we only select 0.28\% of main sequence stars. Within the main sequence, we select 4.4\% of OBA stars, 3.31\% of F stars and roughly 0.25\% of all other stars.

We anticipate that the breadth of the TDSS sample will lead to a wide variety of applications. Our work here suggests variability will help improve quasar selection in redshift regimes where photometric color selection is difficult ($z \approx 2.8$) and distinguish white dwarfs from quasars. More interestingly, variability can help us identify quasars that are reddened by dust, have weakened emission lines or otherwise have unusual colors that mask them from conventional quasar searches. TDSS will also produce a relatively pure and complete quasar sample with respect to variability allowing a study of how quasar properties change with variability in a statistically robust way. Determining how the concentration of different types of stellar variables changes across the Milky Way will be a major survey goal of TDSS. TDSS also promises to produce the largest sample of outer Milky Way RR Lyrae spectra and will thus probe the outer halo with new precision. TDSS should also significantly expand our samples of cataclysmic variables and variable carbon stars, although confident identification may require additional observations, particularly for objects that are not in a quiet state when observed by TDSS. Finally, as the first truly large scale spectroscopic survey to access a broad range of variable types, TDSS serves as a pathfinder for future variability surveys like LSST, allowing both a statistical spectroscopic characterization of the variable object population and the identification of rare or extreme examples only found in large variable samples.

\section{Acknowledgments}

Funding for the Sloan Digital Sky Survey IV has been provided by the Alfred P. Sloan Foundation and the Participating Institutions. SDSS-IV acknowledges support and resources from the Center for High-Performance Computing at the University of Utah. The SDSS web site is www.sdss.org.

SDSS-IV is managed by the Astrophysical Research Consortium for the Participating Institutions of the SDSS Collaboration including the Carnegie Institution for Science, Carnegie Mellon University, the Chilean Participation Group, Harvard-Smithsonian Center for Astrophysics, Instituto de Astrof\'isica de Canarias, The Johns Hopkins University, Kavli Institute for the Physics and Mathematics of the Universe (IPMU) / University of Tokyo, Lawrence Berkeley National Laboratory, Leibniz Institut f\"ur Astrophysik Potsdam (AIP), Max-Planck-Institut f\"ur Astrophysik (MPA Garching), Max-Planck-Institut f\"ur Extraterrestrische Physik (MPE), Max-Planck-Institut f\"ur Astronomie (MPIA Heidelberg), National Astronomical Observatory of China, New Mexico State University, New York University, The Ohio State University, Pennsylvania State University, Shanghai Astronomical Observatory, United Kingdom Participation Group, Universidad Nacional Aut\'onoma de M\'exico, University of Arizona, University of Colorado Boulder, University of Portsmouth, University of Utah, University of Washington, University of Wisconsin, Vanderbilt University, and Yale University.

The PS1 Surveys have been made possible through contributions of the Institute for Astronomy, the University of Hawaii, the Pan-STARRS Project Office, the Max-Planck Society, and its participating institutes, the Max Planck Institute for Astronomy, Heidelberg, and the Max Planck Institute for Extraterrestrial Physics, Garching, The Johns Hopkins University, Durham University, the University of Edinburgh, Queen's University Belfast, the Harvard-Smithsonian Center for Astrophysics, and the Las Cumbres Observatory Global Telescope Network, Incorporated, the National Central University of Taiwan, the National Aeronautics and Space Administration under Grant No. NNX08AR22G issued through the Planetary Science Division of the NASA Science Mission Directorate, the National Science Foundation under Grant No. AST-1238877, the University of Maryland, and Eotvos Lorand University (ELTE).

We thank Don York for many discussions, spanning a number of years, related to the combined SDSS and PS scientific potential, and we thank Tim Heckman for his support linking the two surveys through TDSS.

We gratefully acknowledge help with candidate visual inspection provided by Jerica Green (SAO) and Caroline  Scott (Imperial).

\bibliography{ms}

\appendix

\section{Comparison of KDE to Boosted Decision Tree}\label{sect:comp}

In order to investigate whether more complex techniques that utilize a greater variety of variability features can offer significant improvement over our variability-based KDE approach, we compared the KDE results with those obtained using a Stochastic Gradient Boosted Decision Tree (SGBDT) technique \citep{FRIE01,FRIE02}. Gradient boosting is one of the most powerful and commonly used machine learning techniques, and among its advantages are that it is highly flexible and fairly robust against overfitting. The basic idea behind gradient boosting is to build up a classifier (or regression function) as a linear combination of many weak classifiers. In most applications, including ours, the weak classifiers are shallow binary decision trees. One can think of the technique as modeling the logarithm of the probability that an object is a variable object, given the set of input variability features, as a basis expansion in a set of shallow decision trees, where each decision tree is derived sequentially from the training data. In the stochastic implementation that we used, the decision trees are derived sequentially using a random subsample of the training data, which improves the prediction error by reducing variance in the estimator through averaging. In addition to the median(SDSS-PS1), median(Var) and median(mag) features used in our standard selection algorithm, we add $\chi_{\rm{c\ red}}^2$, $Q_{\rm{tot}}$, $v$ and median($\sigma$). Here, $\chi_{\rm{c\ red}}^2$ is the reduced $\chi^2$ of our PS1 $\griz$ magnitudes assuming a constant for each of the $\griz$ filters. $Q_{\rm{tot}}$ is the average of $Q_{75}-Q_{25}$ across $griz$ filters, where $Q_{75}$ and $Q_{25}$ are, respectively, the 75th and 25th percentile PS1 measurement in each filter. The quantity $v$ is a four filter white noise amplitude described in \citet{MORG++14}. Median($\sigma$) is the median PS1 standard deviation across the $griz$ filters.

We used the stochastic gradient boosting algorithm implemented by the Python scikit-learn package \footnote[0]{\url{http://scikit-learn.org}}. There are a few tuning parameters in this algorithm. The first is the fraction of the training data that is used in each subsample when deriving each weak classifier. We set this parameter to 0.5, a recommended default value. Another tuning parameter is the learning rate, which controls the amount of shrinkage employed. A higher learning rate means that less shrinkage is applied to each of the base classifiers (shallow decision trees), and the model is built up faster. We adopt the default value of 0.1. The number of decision trees to use in the sum is chosen to be 84, found by minimizing the `out-of-bag' error; the out-of-bag error is the error as evaluated by that subsample of the training set that was not used to build the next weak classifier. Finally, the maximum allowed depth of each decision tree in the sum was chosen to be 3, found to minimize the test error, where we withheld 25\% of the Stripe 82 data set as test data and used the remaining 75\% to train the algorithm. Ultimately, the SGBDT assigns every object in our $135^\circ < RA < 150^\circ,\ 45^\circ < DEC < 60^\circ$ test set (as well as our training variable object and standard sets) a probability of being a variable object. This quantity is analogous to the $E$ quantity (and related probability) defined in Section \ref{sect:varmeas} for our KDE. 

The SGBDT also provides a relative measure of the importance of each feature in classifying variable objects. The most important feature was found to be median$(|SDSS-PS1|)$, followed by median(Var) and median($\sigma$). These three features contained approximately 60\% of the total feature importance measure. 

\begin{table*}
\begin{tabular}{cccccccccc}
        \hline
$N_{\rm{tar\ 20}}$ & $N_{\rm{tar\ test}}$ & $N_{\rm{QSO}}$ & $N_{*}$ & $N_{\rm{lovar}}$ & $N_{\rm{CORE}}$ & $N_{\rm{prev}}$ & $N_{\rm{tot}}$ & $P_{\rm{tar}}$ & $P_{\rm{tot}}$ \\
        \hline
60 & 67.8 & 2.9 & 11.7 & 53.1 & 14.1 & 15.7 & 97.6 & 36.8 & 45.6 \\
50 & 56.5 & 2.6 & 13.1 & 40.7 & 13.5 & 14.9 & 84.9 & 41.9 & 52.0 \\
40 & 45.4 & 2.3 & 14.6 & 28.6 & 13.0 & 14.0 & 72.4 & 49.1 & 60.6 \\
30 & 35.2 & 1.9 & 15.5 & 17.8 & 12.1 & 13.0 & 60.4 & 58.2 & 70.4 \\
20 & 23.7 & 1.5 & 14.3 & 7.9 & 10.7 & 11.4 & 45.7 & 71.2 & 82.7 \\
10 & 11.3 & 0.9 & 9.4 & 1.1 & 8.2 & 8.3 & 27.8 & 90.8 & 96.0 \\
        \hline
\end{tabular}
\caption{\rm{The Stochastic Gradient Boosted Decision Tree analog of Table \ref{tab:s82res}. Estimated target counts and purities from Stripe 82 tests at different variability cutoffs. All counts are in units of deg$^{-2}$. All purities are percentages. $N_{\rm{tar\ 20}}$ is the number of targets in the 20th percentile pixel for a given threshold while $N_{\rm{tar\ test}}$ is the number of targets in our test field. $N_{\rm{QSO}}$, $N_{*}$ and $N_{\rm{lovar}}$ are the estimated numbers of TDSS-unique quasars, stars and low-variability objects, respectively. $N_{\rm{CORE}}$ and $N_{\rm{prev}}$ are the estimated numbers of objects we share with the CORE quasar sample or have previous SDSS spectroscopy. $N_{\rm{tot}}$ is the total number of candidates.  $P_{\rm{tar}}$ and $P_{\rm{tot}}$ are the estimated purities of our TDSS-only targets and our total targets, respectively.}}\label{tab:bdts82res}
\end{table*}

In Table \ref{tab:bdts82res}, we show the SGBDT analog of Table \ref{tab:s82res}. As in Section \ref{sect:varmeas}, we set thresholds in our SGBDT $P_{\rm{var}}$ so that 10, 20... 60 TDSS-only targets deg$^{-2}$ pass the threshold in our test set. We can then count the number of variable objects and standards that pass these thresholds and calculate purities and other quantities with the same procedures described in Section \ref{sect:purity}. At the crucial density of 10 TDSS-only targets deg$^{-2}$ (the density of our actual target list), the SGBDT sample is slightly more pure than our KDE sample (90.8\% versus 86.4\% in  $P_{tar}$). However, our KDE performs significantly better at finding CORE quasars and objects with previous SDSS spectra and identifies 9.1 additional objects deg$^{-2}$. Since we are interested in the total sample that passes our threshold, this feature is a decisive advantage for the KDE. We also conceptually prefer using the KDE method which uses a few robust quantities that may be more homogeneous across our sample.

\end{document}